\newif\iflong
\newif\ifshort

\longtrue

\iflong 
\else
\shorttrue
\fi

\documentclass[a4paper]{comsoc2020}
\usepackage{amsmath,amssymb,graphicx,amsthm,dsfont,mathtools}

\usepackage[utf8]{inputenc}
\usepackage[T1]{fontenc}

\usepackage{cutwin}

\usepackage{xspace}
\usepackage[pagebackref]{hyperref}
\hypersetup{%
  backref=true, 
  pagebackref=true, 
  hypertexnames=true,
  colorlinks=true,citecolor=green!35!black,linkcolor=red!60!black%
} 
\usepackage{amsthm}
\usepackage{xifthen}
\usepackage{paralist}

\usepackage[numbers]{natbib}

\usepackage{tikz}
\usetikzlibrary{decorations,arrows,petri,topaths,backgrounds,shapes,positioning,fit,calc,decorations.pathreplacing,patterns,intersections}

\tikzstyle{agent} = [draw=black, circle, fill=black,  inner sep=1.2pt]
\tikzstyle{agentU} = [draw=black, circle, fill=black,  inner sep=1.2pt, line width=0.8pt]
\tikzstyle{agentW} = [draw=orange, rectangle, fill=orange, inner sep=1.5pt, line width=0.8pt]

\usepackage[textsize=tiny,textwidth=1.5cm,linecolor=green!70!black, backgroundcolor=green!10, bordercolor=black,disable]{todonotes}

\usepackage[noend,
ruled,linesnumbered]{algorithm2e}

\SetAlFnt{\small}
\SetAlCapFnt{\small}
\SetAlCapNameFnt{\small}
\SetAlCapHSkip{0pt}

\SetKw{KwAnd}{and}
\SetKw{KwOr}{or}
\SetKw{KwNo}{no}
\SetKw{KwYes}{yes}
\SetKw{KwT}{true}
\SetKw{KwF}{false}
\SetKw{KwST}{s.t.}

\SetCommentSty{mycommfont}
\SetKwComment{Mycomment}{$\triangleright$\ }{}

\usepackage[capitalize]{cleveref}

\newtheorem{example}{Example}
\newtheorem{lemma}{Lemma}
\newtheorem{observation}{Observation}

\newtheorem{theorem}{Theorem}
\newtheorem{claim}{Claim}

\crefname{table}{Table}{Tables}
\crefname{figure}{Figure}{Figures}
\crefname{theorem}{Theorem}{Theorems}
\crefname{corollary}{Corollary}{Corollaries}
\crefname{observation}{Observation}{Observations}
\crefname{lemma}{Lemma}{Lemmas}
\crefname{example}{Example}{Examples}
\crefname{reduction}{Reduction}{Reductions}
\crefname{construction}{Construction}{Constructions}
\crefname{subsection}{Subsection}{Subsections}
\crefname{section}{Section}{Sections}
\crefname{claim}{Claim}{Claims}
\creflabelformat{enumi}{(#2#1#3)}

\theoremstyle{definition}
\newtheorem{definition}{Definition}
\crefname{definition}{Definition}{Definitions}

\newcommand{\decprob}[3]{
   \begin{center}%
    \begin{minipage}{0.92\linewidth}%
      \textsc{#1}\\[0.2ex]
      \textbf{Input:} #2\\[0.2ex]
      \textbf{Question:} #3
    \end{minipage}%
  \end{center}
}

\newcommand{\CESR}{\textsc{Coalitional Exchange-Stable Roommates}\xspace}
\newcommand{\CESRs}{\text{\normalfont CESR}\xspace}

\newcommand{\CESM}{\textsc{Coalitional Exchange-Stable Marriage}\xspace}
\newcommand{\CESMs}{\text{\normalfont CESM}\xspace}

\newcommand{\ESR}{\textsc{Exchange-Stable Roommates}\xspace}
\newcommand{\ESRs}{\text{\normalfont ESR}\xspace}

\newcommand{\ESM}{\textsc{Exchange-Stable Marriage}\xspace}
\newcommand{\ESMs}{\text{\normalfont ESM}\xspace}

\newcommand{\dCESR}[1][]{{\ifthenelse{\equal{#1}{}}{$d$}{$#1$}\textsc{-Coalitional Exchange-Stable Roommates}}\xspace}
\newcommand{\dCESRs}[1][]{{\ifthenelse{\equal{#1}{}}{$d$}{$#1$}-{\normalfont CESR}}\xspace}

\newcommand{\dCESM}[1][]{{\ifthenelse{\equal{#1}{}}{$d$}{$#1$}\textsc{-Coalitional Exchange-Stable Marriage}}\xspace}
\newcommand{\dCESMs}[1][]{{\ifthenelse{\equal{#1}{}}{$d$}{$#1$}-{\normalfont CESM}}\xspace}

\newcommand{\dESR}[1][]{{\ifthenelse{\equal{#1}{}}{$d$}{$#1$}\textsc{\nobreakdash-\hspace{0pt}Exchange-Stable Roommates}}\xspace}
\newcommand{\dESRs}[1][]{{\ifthenelse{\equal{#1}{}}{$d$}{$#1$}\nobreakdash-\text{\normalfont ESR}}\xspace}

\newcommand{\dESM}[1][]{{\ifthenelse{\equal{#1}{}}{$d$}{$#1$}\textsc{\nobreakdash-\hspace{0pt}Exchange-Stable Marriage}}\xspace}
\newcommand{\dESMs}[1][]{{\ifthenelse{\equal{#1}{}}{$d$}{$#1$}\nobreakdash-\text{\normalfont ESM}}\xspace}

\newcommand{\LESM}{\textsc{Path to Exchange-Stable Marriage}\xspace}
\newcommand{\LESMs}{\textsc{P\nobreakdash-ESM}\xspace}
\newcommand{\LESMsnonfragile}{\textsc{P-ESM}\xspace}

\newcommand{\rsat}{\textsc{R\nobreakdash-$3$SAT}\xspace}
\newcommand{\rrsat}{\textsc{(2,2)\nobreakdash-$3$SAT}\xspace}

\newcommand{\IS}[0]{\textsc{Independent Set}}

\newcommand{\intervalI}[2]{[#2]}

\newcommand{\pair}[2]{(#1 , #2)}
\newcommand{\set}[1]{\{ #1 \}}

\newcommand{\ppp}{{\cal P}}

\newcommand{\myemph}[1]{{\color{green!25!black}\emph{#1}}}

\definecolor{darkgreen}{rgb}{0.01,0.6,0.1}
\definecolor{darkblue}{rgb}{0,0,0.4}
\definecolor{winered}{rgb}{0.6,0.1,0.1}
\definecolor{doncolor}{RGB}{78,154,0}
\definecolor{falsecolor}{RGB}{0,55,255}
\definecolor{truecolor}{RGB}{164,0,0}
\definecolor{lightblue}{rgb}{0.527,0.805,0.977}

\newcommand{\dde}[1][]{{\ifthenelse{\equal{#1}{}}{$d$}{$#1$}-dimensional Euclidean}\xspace}
\newcommand{\dder}[1][]{{\ifthenelse{\equal{#1}{}}{$d$}{$#1$}-dimensional Euclidean representation}\xspace}

\newcommand{\dEuclid}[1][]{{\ifthenelse{\equal{#1}{}}{$d$}{$#1$}-{\color{blue!40!black}Euclidean}}\xspace}
\newcommand{\dManhattan}[1][]{{\ifthenelse{\equal{#1}{}}{$d$}{$#1$}-{\color{red!40!black}Manhattan}}\xspace}

\newcommand{\ebp}{exchange-blocking pair\xspace}
\newcommand{\ebc}{exchange-blocking coalition\xspace}

\newcommand{\ebps}{ebp\xspace}
\newcommand{\ebcs}{ebc\xspace}

\newcommand{\es}{exchange-stable\xspace}
\newcommand{\ces}{coalitional exchange-stable\xspace}

\newcommand{\esty}{exchange-stability\xspace}
\newcommand{\cesty}{coalitional exchange-stability\xspace}

\newcommand{\Esty}{Exchange-Stability\xspace}

\newcommand{\enn}{\ensuremath{{\hat{n}}}}
\newcommand{\emm}{\ensuremath{{\hat{m}}}}
\newcommand{\lit}{\ensuremath{\mathsf{lit}}}
\newcommand{\UU}{\ensuremath{{U}}}
\newcommand{\WW}{\ensuremath{{W}}}

\newcommand{\truematch}{\ensuremath{N^1}}
\newcommand{\falsematch}{\ensuremath{N^2}}
\newcommand{\donmatch}{\ensuremath{N^D}}

\newcommand{\mysucc}{\ensuremath{\succ}}

\newcommand{\oc}{\ensuremath{\mathsf{o}}}
\newcommand{\ii}{\ensuremath{\mathsf{s}}}

\newcommand{\cboxed}[2]{{\colorlet{oldcolor}{.}
  \color{#1}\boxed{\color{oldcolor}#2}}}
\newcommand{\tboxed}[1]{\cboxed{truecolor!80}{#1}}
\newcommand{\fboxed}[1]{\cboxed{falsecolor!80}{#1}}
\newcommand{\dboxed}[1]{\cboxed{doncolor!80}{#1}}

\newcommand{\tr}{\ensuremath{\mathsf{t}}}
\newcommand{\fa}{\ensuremath{\mathsf{f}}}

\newcommand{\rest}{\ensuremath{\mathsf{R}}}
\newcommand{\lin}{\ensuremath{\mathsf{L}}}

\newcommand{\myparagraph}[1]{
  \noindent
  \underline{\emph{#1}}
}

\newcommand{\crosstable}{\ensuremath{D_c}}
\newcommand{\horitable}{\ensuremath{D_h}}

\newcommand{\sixcat}{
    \foreach \i / \nn /  \y / \x / \p / \typ in {
      u0/{u_0}/0/0/above/U, w0/{w_0}/2/0/below/U,
      u1/{u_1}/0/-2/above/U, w1/{w_1}/2/-2/below/U,
      u4/{u_{h-1}}/0/-8/above/U, w4/{w_{h-1}}/2/-8/below/U%
    } {
      \node[agent\typ] (\i) at (\x*\xs, \y*\ys) {};
      \node[\p = 0pt of \i] (l\i) {\small $\nn$};
    }

    \foreach \i / \nn /  \y / \x / \p / \typ in {%
      u2/{u_2}/0/-4/above/U, w2/{w_2}/2/-4/below/U, %
      u3/{u_3}/0/-6/above/U, w3/{w_3}/2/-6/below/U%
    }{
      \node[white, inner sep=1pt] (\i) at (\x*\xs, \y*\ys) {};
    }   

    \foreach  \i / \nn /  \y / \x / \p / \typ in {%
      u00/{}/-.2/2/above/U, w00/{}/2.2/2/below/U, %
      u40/{}/-.2/-10/above/U, w40/{}/2.2/-10/below/U}{
      \node[agent\typ] (\i) at (\x*\xs, \y*\ys) {};
    }
    
    \foreach \i / \p in {0/above,2/below} {
      \node[\p=1pt] at (-5*\xs, \i*\ys) {$\cdots$};
    } 

    \foreach \i in {0,1,4} {
      \path[draw] (u\i) edge (w\i);
    }

    \foreach \i / \j in {0/1,1/2,3/4} {
      \path[draw] (u\i) edge (w\j);
      \path[draw] (u\j) edge (w\i);
    }
    
    \foreach \i / \j in {u0/u00,w0/w00,u4/u40/,w4/w40} {
      \path[draw] (\i) edge (\j);
    }
}
\tikzset{ttrue/.style={color=truecolor!50!white}}
\tikzset{ffalse/.style={color=falsecolor!50!white}}
\tikzset{dont/.style={color=doncolor!50!white}}
\tikzset{trueline/.style =   {line width= 3pt, ttrue}}
\tikzset{falseline/.style =   {line width= 3pt, ffalse}}
\tikzset{dontline/.style =   {line width= 3pt, dont}}
  \def \xss {8ex}
  \def \ys {.5}
  \def \xs {1.3}

\newcommand{\switchgadget}{
    \def \xss {8ex}
    \def \ys {.5}
    \def \xs {.9}
    \foreach \i / \nn /  \y  / \x / \p / \typ in {
      alpha/{\alpha}/-2/0/above/U, beta/{\beta}/-.5/0/below/W,
      delta/{\delta}/-2/-4/above/W, gamma/{\gamma}/-.5/-4/below/U,
      a0/{a^0}/0/-.5/below/U, a1/{a^1}/-2/-1/above/U, a2/{a^2}/-2/-1.5/above/U, a3/{a^3}/-2/-2/above/U, a4/{a^4}/-2/-2.5/above/U, a5/{a^5}/-2/-3/above/U, a6/{a^6}/-2.5/-3.5/above/U,
      b0/{b^0}/-2.5/-.5/above/W, b1/{b^1}/-.5/-1/below/W, b2/{b^2}/-.5/-1.5/below/W, b3/{b^3}/-.5/-2/below/W, b4/{b^4}/-.5/-2.5/below/W, b5/{b^5}/-.5/-3/below/W, b6/{b^6}/0/-3.5/below/W%
    }
    {
      \node[agent\typ] at (-\x*\xs, -\y*\ys) (\i) {};
      \node[\p = 0pt of \i] {$\nn$};
    }

    \foreach \s / \t in {alpha/b0, beta/a0,a6/delta,b6/gamma} {
      \path[draw] (\s) edge (\t);
    }
    \foreach \i / \j in {0/1,1/0,1/2,2/1,2/3,3/2,3/4,4/3,4/5,5/4,5/6,6/5} {
      \path[draw] (a\i) edge (b\j);
    }
    \foreach \i in {1,2,3,4,5} {
      \path[draw] (a\i) edge (b\i);
    }
  }

  \newcommand{\alphaagent}[2]{
      \foreach \i / \nn /  \y  / \x / \p / \typ in {
        alpha/{#1}/-2/0/left/U} { 
        \node[agent\typ] at (-\x*\xs, -\y*\ys) (#2) {};
        \node[\p = 0pt of #2] {\small \nn};
    }
  }
  \newcommand{\deltaagent}[2]{
    \foreach \i / \nn /  \y  / \x / \p / \typ in {
      delta/{#1}/-2/-4/right/W} {
      \node[agent\typ] at (-\x*\xs, -\y*\ys) (#2) {};
      \node[\p = 0pt of #2] {\small \nn};
    }
    }
  
  \newcommand{\switchgadgetV}[6]{
   \foreach \i / \nn /  \y  / \x / \p / \typ in {
      beta/{e}/-.5/0/above/W,
     gamma/{f}/-.5/-4/above/U%
    }
    {
      \node[agent\typ] at (-\x*\xs, -\y*\ys) (\i) {};
      \node[\p = 0pt of \i, inner sep=1pt] {\small $\nn^{#1}_{#2}$};
    }

    \foreach \i / \nn /  \y  / \x / \p / \typ in {
      a0/{a^0}/0/-.5/below/U, a1/{a^1}/-2/-1/above/U, a2/{a^2}/-2/-1.5/above/U, a3/{a^3}/-2/-2/above/U, a4/{a^4}/-2/-2.5/above/U, a5/{a^5}/-2/-3/above/U, a6/{a^6}/-2.5/-3.5/above/U,  
      b0/{b^0}/-2.5/-.5/above/W, b1/{b^1}/-.5/-1/below/W, b2/{b^2}/-.5/-1.5/below/W, b3/{b^3}/-.5/-2/below/W, b4/{b^4}/-.5/-2.5/below/W, b5/{b^5}/-.5/-3/below/W, b6/{b^6}/0/-3.5/below/W%
    }
    {
      \node[agent\typ] at (-\x*\xs, -\y*\ys) (\i) {};
      \node[\p = 0pt of \i, inner sep=0pt] {\small $\nn_{#1,#2}$};
    }

    \node[agentU] at (a6) (#6) {};

    \foreach \s / \t in {#5/b0, beta/a0, b6/gamma} {
      \path[draw] (\s) edge (\t);
    }
    \foreach \i / \j in {0/1,1/0,1/2,2/1,2/3,3/2,3/4,4/3,4/5,5/4,5/6,6/5} {
      \path[draw] (a\i) edge (b\j);
    }
    \foreach \i in {1,2,3,4,5} {
      \path[draw] (a\i) edge (b\i);
    }
  }

  \newcommand{\clausegadget}[7]{
    \foreach \i / \nn /  \x  / \y / \p / \typ in {
      c#1/{c_#1}/#6/-2/left/U, d#2/{d_#2}/#7/-2/right/W%
    }{
      \node[agent\typ] at (\x*\xs, \y*\ys) (\i) {};
      \node[\p = 0pt of \i, inner sep=1pt] {\small $\nn$};
    }
    \path[draw] (c#1) edge[bend right=-#5] (#3);
    \path[draw] (d#2) edge[bend right=#5] (#4);
  }

\iflong
\title{On (Coalitional) Exchange-Stable Matching}
\else
\title{On (Coalitional) Exchange-Stable Matching}
\fi
\author{Jiehua Chen, Adrian Chmurovic, Fabian Jogl, and Manuel Sorge\\
  TU Wien, Austria\\
  \texttt{\{jiehua.chen, manuel.sorge\}@tuwien.ac.at}}
\begin{document}

\begin{abstract}
  \looseness=-1
  We study \myemph{(coalitional) exchange stability}, which Alcalde~[Economic Design, 1995] introduced as an alternative solution concept for matching markets involving property rights, such as assigning persons to two-bed rooms.
  Here, a matching of a given \textsc{Stable Marriage} or \textsc{Stable Roommates} instance is called \myemph{\ces} if it does not admit any \myemph{\ebc}, that is, %
  a subset~$S$ of agents in which everyone prefers the partner of some other agent in~$S$.
  The matching is \myemph{\es} if it does not admit any \myemph{\ebp}, that is, an \ebc{} of size two.

  We investigate the computational and parameterized complexity of the \CESM (resp.\ \textsc{Coalitional Exchange Roommates}) problem, which is to decide whether a \textsc{Stable Marriage} (resp.\ \textsc{Stable Roommates}) instance admits a \ces{} matching.
  Our findings  resolve an open question and confirm the conjecture of Cechl{\'a}rov{\'a} and Manlove~[Discrete Applied Mathematics, 2005] that \CESM{} is NP-hard even for complete preferences without ties.
  We also study bounded-length preference lists and a local-search variant of deciding whether a given matching can reach an \es{} one after at most $k$~\myemph{swaps}, where a swap is defined as exchanging the partners of the two agents in an \ebp{}.
\end{abstract}

\section{Introduction}\label{sec:intro}

An instance in a matching market consists of a set of agents that each have preferences over other agents with whom they want to be matched with.
The goal is to find a matching, i.e., a subset of disjoint pairs of agents, which is \myemph{fair}.
A classical notion of fairness is \myemph{stability}~\cite{GaleShapley1962}, meaning that no two agents can form a \myemph{blocking pair}, i.e., they would prefer to be matched with each other rather than with the partner assigned by the matching.
This means that a matching is fair if the agents cannot take local %
action to improve their outcome.
If we assign property rights via the matching, however, then the notion of blocking pairs may not be actionable, as \citet{Alcalde95} observed:
For example, if the matching represents an assignment of persons to two-bed rooms, then two persons in a blocking pair may not be able to deviate from the assignment because they may not find a new room that they could share.
Instead, we may consider the matching to be \emph{fair} if no two agents form an \myemph{\ebp{}}, that is, they would prefer to have each other's partner rather than to have the partner given by the matching~\cite{Alcalde95}.
In other words, they would like to \emph{exchange} their partners.
Note that such an exchange would be straightforward in the room-assignment problem mentioned before.
We refer to \citet{Alcalde95}, Cechl{\'{a}}rov{\'{a}}~\cite{Cec2002}, and Cechl{\'{a}}rov{\'{a}} and Manlove~\cite{CecMan2005} for more discussion and examples of markets involving property rights.

\looseness=-1
If a matching does not admit an \ebp, then we say the matching is \myemph{\es}.
If we also want to exclude the possibility that several agents may collude to favorably exchange partners, 
then we arrive at \myemph{\cesty}~\cite{Alcalde95}, a concept that is more stringent than the \esty.
In contrast to classical stability and \esty{} for perfect matchings (i.e., everyone is matched),
it is not hard to observe that \cesty implies \myemph{Pareto-optimality}, another fairness concept which asserts that no other matching can make at least one agent better-off without making some other agent worse-off (see also \citet{AbrMan2004}).
Note that, in contrast, the classical Gale/Shapley stability and Pareto-optimality are incompatible to each other. 

Cechl{\'{a}}rov{\'{a}} and Manlove~\cite{CecMan2005} showed that the problem of deciding whether an \es{} matching exists is NP\nobreakdash-hard,
even for the marriage case (where the agents are partitioned into two subsets of equal size such that each agent of either subset has preferences over the agents of the other subset) with complete preferences but without ties.
They left open whether the NP-hardness transfers to the case with \cesty,
but observed NP-containment.

\paragraph{Our contributions.}
We study the algorithmic complexity of problems revolving around (coalitional) \esty{}.
In particular, we establish a first NP-hardness result for deciding \cesty, confirming a conjecture of Cechl{\'{a}}rov{\'{a}} and Manlove~\cite{CecMan2005}.
The NP-hardness reduction is based on a novel \myemph{switch-gadget} wherein each preference list contains at most three agents.
Utilizing this, we can carefully complete the preferences so as to prove the desired NP-hardness.
We then investigate the impact of the maximum length~$d$ of a preference list.
We find that NP-hardness for both \esty{} and \cesty{} starts already when $d=3$, while it is fairly easy to see that the problem becomes polynomial-time solvable for $d=2$.
For $d=3$, we obtain a fixed-parameter algorithm for \esty{} regarding a parameter which is related to the number of switch-gadgets.

\looseness=-1
Finally, we look at a problem variant, called \LESM~(\LESMs), for uncoordinated (or decentralized) matching markets.
Starting from an initial matching, in each iteration the two agents in an \ebp{} may exchange their partners.
An interesting question regarding the behavior of the agents in uncoordinated markets is whether such iterative exchange actions can reach a stable state, i.e., \esty,
and how hard is it to decide.
It is fairly straight-forward to verify that if the number~$k$ of exchanges is bounded by a constant,
then
\iflong
we can decide in polynomial-time whether an \es{} matching is reachable since there are only polynomially many possible sequences of exchanges to be checked.
\fi
From the parameterized complexity point of view, we obtain an XP algorithm for~$k$, i.e., the exponent in the polynomial running time depends on~$k$.
We further show that this dependency is unlikely to be removed by showing W[1]-hardness with respect to~$k$.

\paragraph{Related work.}
Alcalde~\cite{Alcalde95} introduced (coalitional) exchange stability and discussed restricted preference domains where (coalitional) exchange stability is guaranteed to exist.
Abizada~\cite{Abizada2019} showed a weaker condition (on the preference domain) to guarantee the existence of exchange stability.
Cechl{\'{a}}rov{\'{a}} and Manlove~\cite{CecMan2005} proved that it is NP-complete to decide whether an \es{} matching exists, even for the marriage case with complete preferences without ties.
Aziz and Goldwasser~\cite{AzizGoldwasser2017} introduced several relaxed notions of \cesty{} and discussed their relations.

\newcommand{\PDIV}{\textsc{Path-to-Stability via Divorces}}
\newcommand{\PDIVs}{\textsc{PSD}}
The \LESMs{} problem is inspired by the \PDIV\ 
(\PDIVs) problem, originally introduced by Knuth~\cite{Knuth1976}, see also Bir{\'o} and Norman~\cite{biro_analysis_2013} for more background.
Very recently, Chen~\cite{Chen2020} showed that \PDIVs\ is NP-hard and W[1]-hard when parameterized by the number of divorces.
\LESMs\ can also be considered as a local search problem and is a special case of the \textsc{Local Search Exchange-Stable Seat Arrangement}~(\textsc{Local-STA}) problem, introduced by Bodlaender et al.~\cite{BodEtAl2020}:
Given a a set of agents, each having cardinal preferences (i.e., real values) over the other agents, an undirected graph~$G$ with the same number of vertices as agents, and an initial assignment (bijection) of the agents to the vertices in~$G$, is it possible to swap two agents' assignments iteratively so as to reach an \es{} assignment?
Herein an assignment is called \myemph{\es{}} if no two agents can each have a higher sum of cardinal preferences over the other's neighboring agents.
\LESMs{} is a restricted variant of~\textsc{Local-STA}, where $G$ consists of disjoint edges
and the agents have ordinal preferences.
 Bodlaender et al.~\cite{BodEtAl2020arxiv} showed that~\textsc{Local-STA} is W[1]-hard wrt.\ the number~$k$ of swaps.
Their reduction relies on the fact that the given graph contains cliques and stars, and the preferences of the agents may contain ties.
Our results for~\LESMs\ that \textsc{Local-STA} is W[1]-hard even if the given graph consists of disjoint edges and the preferences do not have ties.
\iflong

Finally, we mention that \citet{Irving2008StabilityEs} and \citet{MCS07} studied the complexity of computing stable matchings in the marriage setting with preference lists, requiring additionally that the matching be man-exchange stable, that is, no two men form an exchange-blocking pair.
The resulting problem is NP-hard, even if the men's preference lists are of length at most $c$ for each $c \geq 4$~\cite{MCS07}, but it becomes polynomial-time solvable when the men's preference lists have length at most three~\cite{Irving2008StabilityEs}.
This contrasts with \CESM\ which, as we show, remains remains NP-hard even when all preference lists have length at most three.

\paragraph{Organization.}
In \cref{sec:defi}, we introduce relevant concepts and notation, and define our central problems.
In \cref{sec:ces}, we investigate the complexity of deciding (coalitional) \esty{}, both when the preferences are complete and when the preferences length are bounded.
In \cref{sec:d-ces}, we provide algorithms for profiles with preference length bounded by three. 
In \cref{sec:local}, we turn to the local search variant of reaching \esty{}.
\section{Basic Definitions and Observations}
\label{sec:defi}
\looseness=-1
For each natural number $t$, we denote the set~$\{1, 2, \ldots, t\}$ by $[t]$.

Let $V=\{1,2,\ldots, 2n\}$ be a set of~$2n$~agents.
Each agent~$i\in V$ has a nonempty subset of agents~$V_i\subseteq V$ which he finds \myemph{acceptable} as a partner and has a \myemph{strict preference list~$\succ_i$} on~$V_i$ (i.e., a linear order on~$V_i$).
The \myemph{length} of preference list~$\succ_i$ is defined as the number of acceptable agents of~$i$, i.e.,
$|V_i|$.
Here, $x \succ_i y$ means that $i$ \myemph{prefers} $x$ to~$y$.

We assume that the acceptability relation among the agents is \myemph{symmetric}, i.e.,
for each two agents~$x$ and $y$ it holds that $x$ is acceptable to $y$ if and only if $y$ is acceptable to~$x$.
For two agents~$x$ and $y$, we call $x$ \myemph{most acceptable} to~$y$ if $x$ is a maximal element in the preference list of~$y$.
For notational convenience, we write \myemph{$X\succ Y$} to indicate that for each pair of agents~$x \in X$ and $y \in Y$ it holds that $x\succ y$.

\looseness=-1
A \myemph{preference profile}~$\ppp$ is a tuple $(V, (\succ_i)_{i\in V})$ consisting of an agent set $V$ and a collection~$(\succ_i)_{i\in V}$ of preference lists for all agents~$i\in V$.
For a graph~$G$, by $V(G)$ and $G(G)$ we refer to its vertex set and edge set, respectively.
Given a vertex~$v\in V(G)$, by $N_G(v)$ and $d_G(v)$ we refer to the neighborhood of~$v$ and degree of $v$ in~$G$, respectively.
To a preference profile~$\ppp$ with agent set~$V$ we assign an \myemph{acceptability graph}~$G(\ppp)$ which has $V$ as its vertex set and two agents are connected by an edge if they find each other acceptable.
A preference profile~$\ppp$ may have the following properties:
\begin{compactitem}[--]
  \item It is \myemph{bipartite}, if the agent set~$V$ can be partitioned into two agent sets~$U$ and $W$ of size~$n$ each, such that each agent from one set has a preference list over a subset of the agents from the other set.
  \item It has \myemph{complete} preferences if the underlying acceptability graph~$G(\ppp)$ is a complete graph
  or a complete bipartite graph on two disjoint sets of vertices of equal size; otherwise it has \myemph{incomplete} preferences.
\iflong \end{compactitem} \fi
Profile $\ppp$ has \myemph{bounded length~$d$} if each preference list in~$\ppp$ has length at most~$d$.

\paragraph{(Coalitional) exchange-stable matchings.}
A \myemph{matching~$M$} for a given profile~$\ppp$ is a subset of disjoint edges from the underlying acceptability graph~$G(\ppp)$.
Given a matching~$M$ for~$\ppp$, and two agents~$x$ and $y$,
if it holds that $\{x,y\}\in M$, then we use \myemph{$M(x)$} (resp.\ \myemph{$M(y)$}) to refer to $y$ (resp.\ $x$),
and we say that $x$ and $y$ are their respective assigned \myemph{partners} under~$M$ and that they are \myemph{matched} to each other;
otherwise we say that $\{x,y\}$ is an \myemph{unmatched pair} under~$M$.
If an agent~$x$ is \emph{not} assigned any partner by $M$, then we say that $x$ is \myemph{unmatched by $M$} and we put \(M(x)= x\).
We assume that each agent~$x$ prefers to be matched than remaining unmatched.
To formalize this, we will always say that $x$ prefers all acceptable agents from~$V_x$ to himself~$x$.

A matching~$M$ is \myemph{perfect} if every agent is assigned a partner.
It is \myemph{maximal} if for each unmatched pair~$\{x,y\}\in E(G(\ppp))\setminus M$ it holds that $x$ or $y$ is matched under~$M$.
For two agents~$x,y$, we say that \myemph{$x$ envies $y$ under~$M$} if $x$ prefers the partner of $y$, i.e., $M(y)$, to his partner~$M(x)$.
We omit the ``under $M$'' if it is clear from the context.

\looseness=-1
We say that matching~$M$ admits an \myemph{\ebc} (in short \myemph{\ebcs}) if there exists a sequence~\myemph{$\rho=(x_0,x_1,\ldots,x_{r-1})$} of $r$~agents, $r\ge 2$, such that each agent~$x_i$ envies her successor~$x_{i+1}$ in~$\rho$ (index~$i+1$ taken modulo~$r$). 
The \myemph{size} of an \ebcs{} is defined as the number of agents in the sequence.
An \myemph{\ebp} (in short \myemph{\ebps}) is an {\ebcs} of size two.
A matching~$M$ of~$\ppp$ is \myemph{\es} (resp.\ \myemph{\ces}) if it does not admit any \ebps{} (resp.\ \ebcs{}.)
Note that an \ces\ matching is \es.

For an illustration, let us consider the following example.

\renewcommand{\windowpagestuff}{
  \begin{align*}
    \begin{array}{@{}l@{\;}l@{}c@{}l@{\;}l@{}}
      x\colon & a\succ \fboxed{b} \succ \tboxed{c}, &\;\;\;& a\colon & y \succ x \succ \fboxed{\tboxed{z}},\\ 
      y\colon & \tboxed{b}\succ a \succ \fboxed{c}, && b\colon & \fboxed{x} \succ \tboxed{y} \succ z,\\ 
      z\colon & \fboxed{\tboxed{a}}\succ c \succ b, && c\colon & \tboxed{x} \succ \fboxed{y} \succ z.\\[4ex]
    \end{array}
  \end{align*}
}
\opencutright
\begin{example}\label{ex:ces-es}
  \begin{cutout}{1}{0.52\linewidth}{-2pt}{5}
    The following bipartite preference profile~$\ppp$ with agent sets~$U=\{x,y,z\}$ and $W=\{a,b,c\}$ admits $2$ (coalitional) \es{} matchings~$M_1$ and $M_2$ with $M_1=\{\{x,c\}, \{y,b\}, \{z, a\}\}$ (marked in \textcolor{truecolor}{red} boxes)
    and $M_2=\{\{x,b\}, \{y,c\}, \{z,a\}\}$~(marked in \textcolor{falsecolor}{blue} boxes).
    Matching~$M_3$ with $M_3=\{\{x,c\}, \{y,a\}, \{z, b\}\}$ is not \es{} and hence not \ces{} since for instance $(y,z)$ is an \ebp{} of~$M_3$.
  \end{cutout}
\end{example}

\noindent As already observed by Cechl{\'{a}}rov{\'{a}} and Manlove~\cite{CecMan2005}, \es (or \ces) matchings may not exist, even for bipartite profiles with complete preferences.
\iflong

\renewcommand{\windowpagestuff}{
  \begin{align*}
    \begin{array}{@{}l@{\;}l@{}c@{}l@{}l@{}}
      x\colon & a\succ b \succ {c}, &\;\;\;& a\colon & y \succ z \succ x,\\ 
      y\colon & {b}\succ c \succ {a}, && b\colon & z \succ x \succ y,\\ 
      z\colon & c \succ a \succ b, && c \colon & {x} \succ {y} \succ z.\\
    \end{array}
  \end{align*}
}
\opencutright
\begin{example}\label{ex:ces-es2}
  \begin{cutout}{1}{0.52\linewidth}{-2pt}{3}
    The following bipartite preference profile~$\ppp$ with agent sets~$U=\{x,y,z\}$ and $W=\{a,b,c\}$ admits one \es{} matching~$M$ with $M=\{\{x,b\}, \{y,c\}, \{z, a\}\}$.
    However, it is not \ces{} as for instance~$(x,y,z)$ is an \ebcs.
  \end{cutout}
\end{example}
\fi%
\iflong
By assumption, if a matching~$M$ is \ces, then it is maximal.

\begin{lemma}\label{lem:props-es}
  Every \es{} matching is maximal.
\end{lemma}

\begin{proof}
  Let $\ppp$ be a preference profile and $M$ be an \es{} matching of~$\ppp$.

  Suppose, for the sake of contradiction, that there exists an unmatched pair~$\{x,y\}\in E(G(\ppp))\setminus M$ with $M(x)=x$ and $M(y)=y$.
  Since $x$ and $y$ are acceptable to each other, by our assumption that each agent prefers to be matched than remaining unmatched,
  we obtain that $x$ prefers $y$ to~$M(x)$, and $y$ prefers $x$ to~$M(y)$.
  In other words, $(x,y)$ is an \ebp{} of $M$, a contradiction.
\end{proof}

Moreover, \ebcs{} can only occur in cycles:
\begin{lemma}
  \label{lm:coal_means_cycle}
  Let $\ppp$ be a preference profile and let $M$ be a maximal matching of~$\ppp$.
  Then, for each \ebcs{}~$\rho$ of $M$ with $\rho=(x_0,\cdots,x_{r-1})$ it holds that $\{x_0,M(x_0),\cdots,x_{r-1}, M(x_{r-1})\}$ forms a (not necessarily induced) cycle in~$G(\ppp)$.
\end{lemma}
\begin{proof}
  Let $\ppp,M,\rho$ be as defined in the statement with $\rho=(x_0,\cdots,x_{r-1})$.
  By definition each agent~$x_i$, $0\le i\le r-1$, envies his successor~$x_{i+1}$ in~$\rho$ ($i+1$ taken modulo $r$).
  In other words, $\{x_i,M(x_{i+1})\}\in E(G(\ppp))$ holds for each $i\in \{0,\cdots, r-1\}$.
  Since $M$ is maximal, for each $x_i\in \rho$ it must also hold that agent~$x_i$ or agent~$M(x_{i+1})$ must be matched, i.e., $x_i\neq M(x_i)$ or $x_{i+1}\neq M(x_{i+1})$.
  Define the following pieces for all~$i\in \{0,\cdots,r-1\}$: 
  $(s_i)\coloneqq (M(x_i),x_i)$ if $\{x_i,M(x_i)\}\in M$; $(s_i)=(x_i)$ otherwise.
  Then, it is straightforward to verify that $(s_0,s_1,\cdots,s_{r-1},M(x_0))$ is a cycle in~$G(\ppp)$.
\end{proof}

\else%
Every \ces\ matching is  maximal.
\fi

\paragraph{Central problem definitions.}
We are interested in the computational complexity of deciding whether a given profile admits a \ces{} matching.
\decprob{\CESR~(\CESRs)}
{A preference profile~$\ppp$.}
{Does $\ppp$ admit a \ces{} matching?}

\noindent The bipartite restriction of \CESRs, called \myemph{\CESM}~(\myemph{\CESMs}), has as input a \emph{bipartite} preference profile. %
\myemph{\ESR}~(\myemph{\ESRs}) and \myemph{\ESM}~(\myemph{\ESMs}) are defined analogously.

We are also interested in the case when the preferences have bounded length.
In this case, not every \ces{} (or \es) matching is perfect.
In keeping with the literature~\cite{Cec2002,CecMan2005}, we focus on the perfect case.

\decprob{\dCESR~(\dCESRs)}
{A preference profile~$\ppp$ with preferences of bounded length~$d$.}
{Does $\ppp$ admit a \ces{} and \emph{perfect} matching?}

\noindent We analogously define the bipartite restriction \myemph{\dCESM~(\dCESMs)},
and the \es{} variants~\myemph{\dESR}~(\myemph{\dESRs}) and \myemph{\dESM}~(\myemph{\dESMs}).
Note that the above problems are contained in NP~\cite{CecMan2005}.

Finally, we investigate a local search variant regarding \esty.
To this end, given two matchings~$M$ and $N$ of the same profile~$\ppp$,
we say that $M$ is \myemph{one-swap reachable} from~$N$ if there exists an \ebp{}~$(x,y)$ of $N$
such that
$M = (N \setminus \{\{x,N(x)\}, \{y,N(y)\}\}) \cup \{\{x,y\}, \{N(x), N(y)\}\}$.
Accordingly, we say that $M$ is \myemph{$k$-swaps reachable} from~$N$ if there exists a sequence~$(M_0,M_{1},\cdots, M_{k'})$ of $k'$ \emph{matchings} of profile~$\ppp$ such that 
\begin{inparaenum}[(a)]
  \item $k'\le k$,  $M_0=N$, $M_{k'}=M$, and
  \item for each~$i \in [k']$, $M_i$ is one-swap reachable from~$M_{i-1}$.
\end{inparaenum}
\iflong

For an illustration, let us consider \cref{ex:ces-es} again.

\begin{example}\label{ex:lesm}
  Consider matching~$M_3$ from \cref{ex:ces-es}, which admits two \ebps{s}~$(x,z)$ and $(y,z)$.
  If we let $y$ and $z$ exchange their partners, then we reach~$M_1$, which is (coalitional) \es{}.  
  If we let $x$ and $z$ exchange their partners, however, then we reach matching~$M_4=\{\{x,b\}, \{y,a\}, \{z,c\}\}$, which will never reach an \es{} matching due to the following:
  $(x,y)$ is a unique \ebp{} of~$M_4$.
  If we let $x$ and $y$ exchange partners under~$M_4$, then we obtain matching~$M_5$ with $M_5=\{\{x,a\}, \{y,b\}, \{z,c\}\}$, and it admits a unique \ebp{}~$(a,b)$.
  If we let $a$ and $b$ exchange partners under~$M_5$, then we obtain matching~$M_4$, again.  
\end{example}
\fi
\noindent
\iflong
The local search problem variant that we are interested in is defined as follows:
\else
The local search problem variant is defined as follows:
\fi

\decprob{\LESM~(\LESMs)}{
  A bipartite preference profile~$\ppp$, a matching~$M_0$ of $\ppp$, and an integer~$k$.
}{Does $\ppp$ admit an \es{} matching~$M$ which is $k$-swap reachable from~$M_0$?}

\section{Deciding (Coalitional) \Esty{} is NP-complete}\label{sec:ces}

\citet{CecMan2005} proved NP-completeness for deciding whether a profile with complete and strict preferences admits an \es{} matching, by reducing from the NP-complete \rsat{} problem, where each clause has at most three literals and each literal appears at most two times~\cite{GJ79}. It is, however, not immediate how to adapt \citeauthor{CecMan2005}'s proof to show hardness for \cesty{} since their constructed \es{} matching is not always \ces.
\looseness=-1
To obtain a hardness reduction for \ces{},
we first study the case when the preferences have length bounded by three,
and show that \dCESM[3] is NP-hard, even for strict preferences.
The idea is different than that by~\citeauthor{CecMan2005}.
To simplify the reduction, we will reduce from an NP-complete variant of \rsat.
\decprob{\rrsat}
{A Boolean formula~$\phi(X)$ with variable set~$X$ in 3CNF (i.e., a set of clauses each containing at most $3$ literals) such that no clause contains both the positive and the negated literal of the same variable, and each literal appears \emph{exactly} two times.}
{Is $\phi$ satisfiable?}
\fi

\begin{lemma}\label{rrsat;np-c}
  \rrsat{} is NP-complete.
\end{lemma}

\begin{proof}
  Clearly, the problem belongs to NP.
  We provide a reduction from the NP-complete \rsat, where
  each clause has at most three literals
  and each literal appears at most two times~\cite{GJ79}.

  First, we assume that no variable appears only negatively or only positively: if this
  were the case for some variable, then we could set it to true or false and simplify the formula.
  Second, we assume that no clause contains both the positive and negated literal of the same variable more as otherwise it is always satisfiable and we can delete it from the formula.
  
  Now, for each literal~$\lit_i$ (either $x_i$ or $\overline{x}_i$) which appears only once,
  add two new variables~$a_i$ and $b_i$,
  and four new clauses~$(\lit_i, a_i, \overline{b}_i)$, $(a_i, \overline{b}_i)$, $(\overline{a}_i, b_i)$,
  and $(\overline{a}_i, b_i)$.

  It is straight-forward to see that in the constructed instance no clause contains the positive and negated literal of the same variable and each literal appears exactly two times.
  Moreover, the original instance is a yes-instance if and only if the newly constructed instance is a yes-instance for \rrsat since the newly added variables only affect the newly added clauses which can be satisfied by setting all newly added variables to true.
\end{proof}

A crucial ingredient for our reduction is the following \myemph{switch-gadget} which enforces that each \es{} matching results in a valid truth assignment.
The gadget and its properties are summarized in the following lemma. 

\iflong
\begin{figure}[t!]\centering
  \begin{tikzpicture}[>=stealth]
  \switchgadget
  \node[above = 4ex of a3,text=truecolor] {$\truematch$};
    \begin{scope}[on background layer]
      \foreach \i / \j in {alpha/b0, a0/b1, a1/b2, a2/b3, a3/b4, a4/b5, a5/b6, a6/delta}
      {
        \draw[trueline] (\i) -- (\j);
      }
    \end{scope}
 \end{tikzpicture}
\begin{tikzpicture}[>=stealth]
    \switchgadget
    \node[above = 4ex of a3, text=falsecolor] {$\falsematch$};
    \begin{scope}[on background layer]
      \foreach \i / \j in {a0/beta, a1/b0, a2/b1, a3/b2, a4/b3, a5/b4, a6/b5, b6/gamma}
      {
        \draw[falseline] (\i) -- (\j);
      }
    \end{scope}
  \end{tikzpicture}
\begin{tikzpicture}[>=stealth]
  \switchgadget
    \node[above = 4ex of a3, text=doncolor] {$\donmatch$};
    \begin{scope}[on background layer]
      \foreach \i / \j in {a0/beta, a1/b2, a2/b1, a3/b3, a4/b5, a5/b4, a6/delta, b6/gamma,b0/alpha}
      {
        \draw[dontline] (\i) -- (\j);
      }
    \end{scope}
  \end{tikzpicture}
  \caption{The three possible \es{} matchings discussed in \cref{lem:switch}.}
  \label{fig:switch-match}
\end{figure}
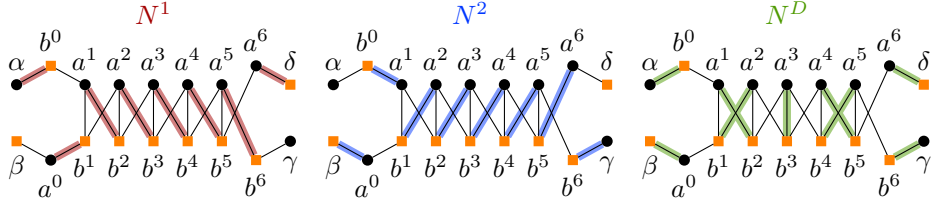
\fi

\begin{lemma}\label{lem:switch}
  Let $\ppp$ be a bipartite preference profile on agent sets~$U$ and $W$.
  Let $A=\{a^z\mid z\in \{0,1,\ldots,6\}\}$ and $B=\{b^z\mid z\in \{0,1,\ldots,6\}\}$ be two disjoint sets of agents,
  and let $Q=\{\alpha, \beta, \gamma, \delta\}$ be four further distinct agents with
  $A\cup \{\alpha, \gamma\}\subseteq U$ and $B\cup \{\beta, \delta\}\subseteq W$.
  The preferences of the agents from $A$ and $B$ are as follows; the preferences of the other agents are arbitrary but fixed.
  \ifshort

  {\centering \begin{tabular}{l@{\;}lcl@{\;}l}
    $ \tikz{\node[agentU] {};}~ a^0\colon$ &  $\tboxed{b^1} \succ \dboxed{\fboxed{\beta}}$,  & \qquad \qquad  & $\tikz{\node[agentW] {};}~b^0\colon$ &$\fboxed{a^1} \succ \dboxed{\tboxed{\alpha}}$,\\
    $\tikz{\node[agentU] {};}~ a^1\colon$ &$\fboxed{b^0} \succ \dboxed{\tboxed{b^2}} \succ b^1$, &   \qquad  &$\tikz{\node[agentW] {};}~ b^1\colon$ & $\tboxed{a^0} \succ  \dboxed{\fboxed{a^2}} \succ a^1$,\\
      $\tikz{\node[agentU] {};}~ a^2\colon$ &  $\tboxed{b^3} \succ \dboxed{\fboxed{b^1}} \succ b^2$,&  \qquad    &$\tikz{\node[agentW] {};}~b^2\colon$ & $a^2 \succ  \fboxed{a^3} \succ \dboxed{\tboxed{a^1}}$,\\
      $\tikz{\node[agentU] {};}~ a^3\colon$  & $\fboxed{b^2} \succ \dboxed{b^3} \succ \tboxed{b^4}$,&   \qquad &    $\tikz{\node[agentW] {};}~b^3\colon$  &   $\fboxed{a^4} \succ \dboxed{a^3} \succ \tboxed{a^2}$,\\
      $\tikz{\node[agentU] {};}~ a^4\colon$  & $   b^4 \succ \fboxed{b^3} \succ \dboxed{\tboxed{b^5}}$,  &  \qquad   &  $\tikz{\node[agentW] {};}~b^4\colon$  &   $   \tboxed{a^3} \succ  \dboxed{\fboxed{a^5}} \succ a^4$,\\
      $\tikz{\node[agentU] {};}~ a^5\colon$  & $   \tboxed{b^6} \succ \dboxed{\fboxed{b^4}} \succ b^5$,  &  \qquad  &   $\tikz{\node[agentW] {};}~b^5\colon$  &   $    \fboxed{a^6} \succ \dboxed{\tboxed{a^4}} \succ a^5$,\\
      $\tikz{\node[agentU] {};}~ a^6\colon$  & $   \fboxed{b^5} \succ \dboxed{\tboxed{\delta}}$,   & \qquad  &   $\tikz{\node[agentW] {};}~b^6\colon$  & $   \tboxed{a^5} \succ \dboxed{\fboxed{\gamma}}$.
  \end{tabular}
  \par}

\else  \begin{align*}
         \allowdisplaybreaks
    \begin{array}{l@{\;}lcl@{\;}l}
      \tikz{\node[agentU] {};}~ a^0\colon & \tboxed{b^1} \succ \dboxed{\fboxed{\beta}}, & \qquad &  \tikz{\node[agentW] {};}~b^0\colon & \fboxed{a^1} \succ \dboxed{\tboxed{\alpha}},\\
      \tikz{\node[agentU] {};}~ a^1\colon & \fboxed{b^0} \succ \dboxed{\tboxed{b^2}} \succ b^1, & \qquad & \tikz{\node[agentW] {};}~ b^1\colon & \tboxed{a^0} \succ  \dboxed{\fboxed{a^2}} \succ a^1,\\
      \tikz{\node[agentU] {};}~ a^2\colon & \tboxed{b^3} \succ \dboxed{\fboxed{b^1}} \succ b^2, & \qquad &  \tikz{\node[agentW] {};}~b^2\colon & a^2 \succ  \fboxed{a^3} \succ \dboxed{\tboxed{a^1}},\\
      \tikz{\node[agentU] {};}~ a^3\colon & \fboxed{b^2} \succ \dboxed{b^3} \succ \tboxed{b^4}, & \qquad &  \tikz{\node[agentW] {};}~b^3\colon &  \fboxed{a^4} \succ \dboxed{a^3} \succ \tboxed{a^2},\\
      \tikz{\node[agentU] {};}~ a^4\colon & b^4 \succ \fboxed{b^3} \succ \dboxed{\tboxed{b^5}}, & \qquad &  \tikz{\node[agentW] {};}~b^4\colon & \tboxed{a^3} \succ  \dboxed{\fboxed{a^5}} \succ a^4,\\
      \tikz{\node[agentU] {};}~ a^5\colon & \tboxed{b^6} \succ \dboxed{\fboxed{b^4}} \succ b^5, & \qquad &  \tikz{\node[agentW] {};}~b^5\colon &  \fboxed{a^6} \succ \dboxed{\tboxed{a^4}} \succ a^5,\\
      \tikz{\node[agentU] {};}~ a^6\colon & \fboxed{b^5} \succ \dboxed{\tboxed{\delta}}, & \qquad &  \tikz{\node[agentW] {};}~b^6\colon & \tboxed{a^5} \succ \dboxed{\fboxed{\gamma}}.
    \end{array}
       \end{align*}
 \noindent Define the following matchings:
   \begin{align*}
    {\color{truecolor}\truematch} & \coloneqq \{\{\alpha,b^0\},\{a^6,\delta\}\}\cup \{\{a^{z-1},b^{z}\}\mid z\in [6]\},\\
    {\color{falsecolor}\falsematch} &\coloneqq \{\{a^0,\beta\},\{\gamma,b^6\}\}\cup \{\{a^z,b^{z-1}\}\mid z\in [6]\},\\
    {\color{doncolor}\donmatch} &\coloneqq \{\{\alpha,b^0\}, \{a^0,\beta\},
                                  \{a^6,\delta\}, \{\gamma, b^6\}, \{a^1,b^2\}, \{a^2,b^1\}, \{a^3,b^3\}, \{a^4,b^5\},\{a^5,b^4\}\}.
  \end{align*}
  See \cref{fig:switch-match} for an illustration of the gadget and the matchings above.
  \noindent Every \emph{perfect} matching~$M$ of~$\ppp$ satisfies the following,
  \fi
  \ifshort \noindent Every \emph{perfect} matching~$M$ of~$\ppp$ satisfies the following,
  where 
 ${\color{truecolor}\truematch}  \coloneqq \{\{\alpha,b^0\},\{a^6,\delta\}\}\cup \{\{a^{z-1},b^{z}\}\mid z\in [6]\}$, ${\color{falsecolor}\falsematch} \coloneqq \{\{a^0,\beta\},\{\gamma,b^6\}\}\cup \{\{a^z,b^{z-1}\}\mid z\in [6]\}$, ${\color{doncolor}\donmatch} \coloneqq \{\{\alpha,b^0\}, \{a^0,\beta\}, \{a^6,\delta\}$, $\{\gamma, b^6\}, \{a^1,b^2\}, \{a^2,b^1\}, \{a^3,b^3\}, \{a^4,b^5\},\{a^5,b^4\}\}$.
  \fi
  \begin{compactenum}[(1)]
  \item\label{switch:es} If $M$ is \es{}, then
    \begin{inparaenum}[(i)]
    \item\label{esm:x-true}either $N^{1}\subseteq M$, or %
    \item\label{esm:x-false} $N^{2}\subseteq M$, or %
    \item\label{esm:x-dontcare} $N^{D}\subseteq M$. %
    \end{inparaenum}
  \item\label{switch:T} If $\truematch\subseteq M$,
    then every \ebcs{} of~$M$ which involves an agent from~$A$ (resp.~$B$) also involves~$\alpha$ (resp.~$\delta$).
  \item\label{switch:F} If $\falsematch\subseteq M$,
    then every \ebcs{} of~$M$ which involves an agent from~$A$ (resp.~$B$) also involves~$\gamma$ (resp.~$\beta$).
  \item\label{switch:D} If $\donmatch\subseteq M$,
    then every \ebcs{} of~$M$ which involves an agent from~$A$ (resp.~$B$) also involves an agent from~$\{\alpha, \gamma\}$ (resp.\ $\{\beta,\delta\}$).
  \end{compactenum}
\end{lemma}

\iflong
\begin{proof}
  For statement~\eqref{switch:es}, since $M$ is perfect, every agent is matched.
  In particular, $M(a^3)\in \{b^2,b^3,b^4\}$.
  We show that if $M$ is \es{}, then the partner~$M(a^3)$ decides which of the three cases in the first statement holds.

  If $M(a^3) = b^2$, then $M(a^2)\neq b^3$ as otherwise $(b^2,b^3)$ is an \ebp{}.
  Hence, $M(a^2) = b^1$ and $M(b^3)=a^4$ since $b^2$, $a^2$, and $a^3$ are already matched to other agents.
  By the acceptability relations, $M(a^1)=b^0$, $M(a^0) = \beta$, $M(b^4)=a^5$, $M(b^5)=a^6$, and $M(b^6)=\gamma$.
  This implies that $M$ satisfies the condition given in case~\eqref{esm:x-false}.

  If $M(a^3) = b^3$, then $M(a^2)\neq b^2$ and $M(a^4)\neq b^4$ as otherwise $(a^2,a^3)$
  or $(b^3,b^4)$ is an \ebp{}.
  Hence, $M(a^2) = b^1$ and $M(a^4)=b^5$ since $b^3$ is already matched to~$a^3$.
  By the acceptability relations, $M(b^2)=a^1$, $M(b^4) = a^5$, $M(a^0)=\beta$, $M(a^6)=\delta$, $M(b^0)=\alpha$, $M(b^6)=\gamma$.
  This implies that $M$ satisfies the condition given in case~\eqref{esm:x-dontcare}.

  If $M(a^3) = b^4$, then $M(a^4)\neq b^3$ as otherwise $(a^3,a^4)$ is an \ebp{}.
  Hence, $M(a^4) = b^5$ and $M(b^3)=a^2$ since $b^5$ and $a^2$ are the only agents available to~$a^4$ and $b^3$, respectively.
  By the acceptability relations, $M(a^5)=b^6$, $M(a^6) = \delta$,
  $M(b^2)=a^1$,
  $M(b^1)=a^0$,
  and $M(b^0)=\alpha$.
  This implies $M$ satisfies the condition given in case~\eqref{esm:x-true}.
  Together, this completes the proof for Statement~\eqref{switch:es}.

  To show Statements~\eqref{switch:T}--\eqref{switch:D}, %
  we show their contra-positives, by means of a straight-forward verification. %
  For Statement~\eqref{switch:T}, assume that $\truematch\subseteq M$ and let $\rho$ denote an arbitrary \ebcs{} of $M$ which does not involve~$\alpha$.
  
  We aim to show that $\rho$ does not involve any agent from~$A$.
  Suppose, for the sake contradiction, that $\rho$ involves an agent~$x$ with $x\in A$.
  
  Certainly, $\rho$ cannot involve any agent from~$S=\{a^0,a^2,a^5\}$ since each of the agents from~$S$ already receives his most-preferred agent.
  Moreover, $x\neq a^1$ since $a^1$ only envies the partner of~$b^0$, which is $\alpha$ and $\alpha \notin \rho$.
  Consequently, $x \neq a^3$, because $a^3$ envies only $a^1, a^2 \in S \cup \{a^1\}$.
  Thus, $x \neq a^4$ because $a^4$ envies only $a^3$ and $a^2$.
  Finally, $x \neq a^6$ because $a^6$ envies only $a^4$, a contradiction.

  Analogously, assume that $\truematch\subseteq M$ and let $\rho$ denote an arbitrary \ebcs{} of $M$ which does not involve~$\delta$.
  Certainly, $\rho$ cannot involve any agent from~$S=\{b^1,b^4,b^6\}$ since each of the agents from~$S$ already receives his most-preferred agent.
  Moreover, $x\neq b^5$ since $b^5$ only envies the partner of~$a^6$, which is $\delta$,
  but $\delta\notin \rho$.
  Consequently, $x \neq b^3$.
  It follows that, $x \neq b^2$, because $b^2$ envies only $b^3$ and $b^4$.
  Finally, $x \neq b^0$, because $b^0$ envies only $b^2$, a contradiction.
  This completes the proof for Statement~\eqref{switch:T}.

  For Statement~\eqref{switch:F}, assume that $\falsematch\subseteq M$ and let $\rho$ denote an arbitrary \ebcs{} of $M$ which does not involve~$\gamma$.
  Suppose, for the sake contradiction, that $\rho$ involves an agent~$x$ with $x\in A$.
  Certainly, $x\notin \{a^1,a^3,a^6\}$ since each of these agents is matched with his most preferred partner.
  Moreover, $x\neq a^5$ since $\gamma\notin \rho$.
  Consequently, $x\notin \{a^4,a^2,a^0\}$ by an analogous reasoning, a contradiction.

  Analogously, assume that $\falsematch\subseteq M$ and let $\rho$ denote an arbitrary \ebcs{} of $M$ which does not involve~$\beta$.
  Suppose, for the sake contradiction, that $\rho$ involves an agent~$x$ with $x\in B$.
  Certainly, $x\notin \{b^0,b^3,b^5\}$ since each of these agents is matched with his most preferred partner.
  Moreover, $x\neq b^1$ since $\beta \notin \rho$.
  Consequently, $x\notin \{b^2,b^4,b^6\}$ by an analogous reasoning, a contradiction.
  This completes the proof for Statement~\eqref{switch:F}.
   
  For Statement~\eqref{switch:D}, assume that $\donmatch \subseteq M$ and let $\rho$ denote an arbitrary \ebcs{} of $M$ which involves neither~$\alpha$ nor~$\gamma$.
  Suppose, for the sake contradiction, that $\rho$ involves an agent~$x$ with $x\in A$.
  Certainly, $x\notin \{a^1, a^5\}$ since $a^1$ only envies~$\alpha$ and $a^5$ only envies~$\gamma$,
  but neither $\alpha$ nor $\gamma$ is included in~$\rho$.
  Consequently, $x\neq a^3$ since $a^3$  only envies $a^1$.
  Using a similar reasoning, we infer that $x \notin \{a^2, a^4, a^6, a^0\}$.
  Altogether, no agent from~$A$ is involved in~$\rho$.

  Analogously,  assume that $\donmatch\subseteq M$ and let $\rho$ denote an arbitrary \ebcs{} of $M$ which involves neither~$\beta$ nor~$\delta$.
  Certainly, $x\notin \{b^1, b^5\}$ since $b^1$ only envies~$\beta$ and $b^5$ only envies~$\delta$.
  Consequently, $x\neq b^3$ since $b^3$  only envies $b^5$.
  Using a similar reasoning, we infer that $x \notin \{b^2, b^4, b^6, b^0\}$.
  Altogether, no agent from~$B$ is involved in~$\rho$.
  This completes the proof for Statement~\eqref{switch:D}.
\end{proof}
\fi
\noindent
Using \cref{lem:switch}, we can show NP-completeness for bounded preference length.

\begin{theorem}\label{thm:cesm-length=3-nph}
  \dCESMs[3], \dESMs[3], \dCESRs[3], and \dESRs[3] %
  are NP-complete.
\end{theorem}

\begin{proof}
  \looseness=-1
  As already mentioned~\cite{CecMan2005}, by checking for cycles in the envy graph all discussed problems are in~NP.
  \iflong
  For the sake of completeness, we show the NP-containment here.
  Given a preference profile~$\ppp$ and a matching~$M$, we can check in polynomial time whether $M$ is \ces\ for $\ppp$ as follows.
  Compute an auxiliary directed graph $H$ that has the agent set $V$ as vertices and an arc $(u, v)$ if agent $u$ envies $v$ under~$M$.
  Observe that $M$ admits an \ebc\ if and only if there is a directed cycle in $H$.
  Thus, it suffices to check for a directed cycle in~$H$.
  It follows that all problems are in NP.

  \fi
  For the NP-hardness, it is enough to show that \dCESMs[3]\ and \dESMs[3]\ are NP-hard.
  We use the same reduction from \rrsat\ for both.
  Let $(X,C)$ be an instance of \rrsat{} where $X=\{x_1,x_2,\cdots,x_{\enn}\}$ is the set of variables and $\phi=\{C_1,C_2,\cdots,C_{\emm}\}$ the set of clauses.

  We construct a bipartite preference profile on two disjoint agent sets~$U$ and $W$.
  The set~$U$ (resp.\ $W$) will be partitioned into three different agent-groups: the variable-agents, the switch-agents, and the clause-agents.
  The general idea is to use the variable-agents and the clause-agents to determine a truth assignment
  and satisfying literals, respectively.
  Then, we use the switch-agents from \cref{lem:switch} to make sure that the selected truth assignment is consistent with the selected satisfying literals.
  For each~$\lit_i\in X\cup \overline{X}$ that appears in two different clauses~$C_j$ and $C_{k}$ with $j < k$,
  we use \myemph{$\oc_1(\lit_i)$} and \myemph{$\oc_2(\lit_i)$} to refer to the indices~$j$ and $k$; recall that in~$\phi$ each literal appears exactly two times.
  \iflong
  For example, if literal~$x_i$ appears in $C_3$ and $C_5$, then $\oc_1(x_i)=3$ and $\oc_2(x_i)=5$.
  \fi

  \begin{figure}[t]
    \begin{alignat*}{3}
      \allowdisplaybreaks
      \forall i \in [\enn]\colon &~~\,
      \begin{array}{l@{\qquad\quad\;\;}l}
        \tikz{\node[agentU] {};}~  v_i \colon  \fboxed{y_i} \mysucc \tboxed{\overline{y}_i}, &   \tikz{\node[agentW] {};}~ w_i\colon  \fboxed{x_i} \mysucc \tboxed{\overline{x}_i},\\
       \tikz{\node[agentU] {};}~   x_i\colon  \fboxed{w_i} \mysucc \tboxed{b^0_{i, \oc_1(x_i)}}, & \tikz{\node[agentW] {};}~  y_i \colon  \fboxed{v_i} \mysucc \tboxed{a^6_{i,\oc_2(x_i)}},\\
        \tikz{\node[agentU] {};}~  \overline{x}_i\colon  \tboxed{w_i} \mysucc \fboxed{b^0_{i, \oc_1(\overline{x_i})}}, &  \tikz{\node[agentW] {};}~ \overline{y}_i \colon  \tboxed{v_i} \mysucc \fboxed{a^0_{i,\oc_2(\overline{x}_i)}},\\
        \end{array}\\
        \forall j \in [\emm]\colon &  ~~\, \begin{array}{l@{\qquad\quad\qquad\qquad\qquad}l}\tikz{\node[agentU] {};}~ c_j\colon  [E_j], & \tikz{\node[agentW] {};}~ d_j \colon   [F_j],\end{array}\\
         \parbox[t]{.26\textwidth}{$\forall i \in [\enn]$, \mbox{$\forall j\in [\emm]$}  with $\lit_i \in C_j  \colon$} &
      \begin{cases}
          \tikz{\node[agentU] {};}~ f^{i}_j\colon  d_j \mysucc b^6_{i,j} &  \tikz{\node[agentW] {};}~ e^i_{j}\colon  c_j \mysucc a^0_{i,j}\\
          \tikz{\node[agentU] {};}~   a^0_{i,j}\colon  \tboxed{b^1_{i,j}} \mysucc \dboxed{\fboxed{e^i_{j}}}, &    \tikz{\node[agentW] {};}~ b^0_{i,j}\colon  \fboxed{a^1_{i,j}} \mysucc \dboxed{\tboxed{\alpha_{i,j}}},\\
          \tikz{\node[agentU] {};}~  a^1_{i,j}\colon  \fboxed{b^0_{i,j}} \mysucc \dboxed{\tboxed{b^2_{i,j}}} \mysucc b^1_{i,j}, &    \tikz{\node[agentW] {};}~ b^1_{i,j}\colon  \tboxed{a^0_{i,j}} \mysucc  \dboxed{\fboxed{a^2_{i,j}}} \mysucc a^1_{i,j},\\
          \tikz{\node[agentU] {};}~   a^3_{i,j}\colon  \fboxed{b^2_{i,j}} \mysucc \dboxed{b^3_{i,j}} \mysucc \tboxed{b^4_{i,j}}, &    \tikz{\node[agentW] {};}~ b^3_{i,j}\colon   \fboxed{a^4_{i,j}} \mysucc \dboxed{a^3_{i,j}} \mysucc \tboxed{a^2_{i,j}},\\
          \tikz{\node[agentU] {};}~   a^4_{i,j}\colon  b^4_{i,j} \mysucc \fboxed{b^3_{i,j}} \mysucc \dboxed{\tboxed{b^5_{i,j}}}, &   \tikz{\node[agentW] {};}~  b^4_{i,j}\colon  \tboxed{a^3_{i,j}} \mysucc  \dboxed{\fboxed{a^5_{i,j}}} \mysucc a^4_{i,j},\\
          \tikz{\node[agentU] {};}~  a^5_{i,j}\colon  \tboxed{b^6_{i,j}} \mysucc \dboxed{\fboxed{b^4_{i,j}}} \mysucc b^5_{i,j}, &    \tikz{\node[agentW] {};}~ b^5_{i,j}\colon   \fboxed{a^6_{i,j}} \mysucc \dboxed{\tboxed{a^4_{i,j}}} \mysucc a^5_{i,j},\\
            \tikz{\node[agentU] {};}~   a^6_{i,j}\colon  \fboxed{b^5_{i,j}} \mysucc \dboxed{\tboxed{\delta_{i,j}}}, &    \tikz{\node[agentW] {};}~ b^6_{i,j}\colon \tboxed{a^5_{i,j}} \mysucc \dboxed{\fboxed{f^i_{j}}}.
          \end{cases}
      \end{alignat*}
    \caption{The preferences constructed in the proof for \cref{thm:cesm-length=3-nph}.
      Recall that for each literal~$\lit_i\in X\cup \overline{X}$,
      the indices~$\oc_1(\lit_i)$ and $\oc_2(\lit_i)$
      denote the two indices $j < j'$ of the clauses that contain~$\lit_i$.
      For each clause~$C_j\in \phi$, the expression~$[E_j]$ (resp.\ $[F_j]$) denotes an arbitrary but fixed order of the agents in~$E_j$ (resp.~$F_j$). }
    \label{fig:cesm-length=3-nph}
  \end{figure}

  For illustration of the construction below, refer to \cref{fig:cesm-length=3-nph}.
  
  \myparagraph{The variable-agents.} For each variable~$x_i\in X$, introduce $6$~\myemph{variable-agents}~$v_i$, $w_i$, $x_i$, $\overline{x}_i$, $y_i$, $\overline{y}_i$.
  Add~$v_i,x_i,\overline{x}_i$ to~$U$, and $w_i,y_i,\overline{y}_i$ to~$W$.
  For each literal~$\lit_i \in X\cup \overline{X}$ let \myemph{$y(\lit_i)$} denote the corresponding $Y$-variable-agent, that is, $y(x_i) = y_i$ and $y(\overline{x}_i) = \overline{y}_i$.
  Define $\overline{X}\coloneqq \{\overline{x}_i\mid i\in [\enn]\}$, and
  $\overline{Y}\coloneqq \{\overline{y}_i\mid i\in [\enn]\}$.
  
  \myparagraph{The clause-agents.} For each clause~$C_j\in C$, introduce two \myemph{clause-agents}~$c_j,d_j$.
  Further, for each literal~$\lit_i\in C_j$ with $\lit \in \{x,\overline{x}\}$,
  introduce two more \myemph{clause-agents}~$e_j^i, f_j^i$.
  Add $c_j,f_j^i$ to~$U$, and $d_j,e_j^i$ to~$W$.
  \iflong

  \fi
  For each clause~$C_j\in \phi$, define~$E_j\coloneqq \{e_j^{i}\mid \lit_i\in C_j\}$, and
  $F_j\coloneqq \{f_j^{i}\mid \lit_i\in C_j\}$.
  Moreover, define~${E} \coloneqq \bigcup_{C_j\in \phi}E_j$  and~${F} \coloneqq \bigcup_{C_j\in \phi}F_j$.
  
  \myparagraph{The switch-agents.} For each clause~$C_j\in C$, and each literal~$\lit_i \in C_j$
  introduce fourteen \myemph{switch-agents}~$a^z_{i,j},b^z_{i,j}$, $z\in \{0,1,\cdots,6\}$.
  Define $A_{i,j}=\{a^z_{i,j}\mid z\in \{0,1,\ldots,6\}\}$ and  $B_{i,j}=\{b^z_{i,j}\mid z\in \{0,1,\ldots,6\}\}$.
  Add $A_{i,j}$ to~$U$ and $B_{i,j}$ to~$W$.

  In total, we have the following agent sets:
  \ifshort

  \noindent $U \coloneqq \{v_i\mid i\in [\enn]\} \cup X\cup \overline{X} \cup \{c_j\mid j\in [\emm]\} \cup {F}
  \cup \bigcup_{C_j\in \phi\wedge \lit_i \in C_j}A_{i,j}$, and\\
  \noindent $W \coloneqq \{w_i\mid i\in [\enn]\} \cup Y \cup \overline{Y} \cup \{d_j\mid j\in [\emm]\} \cup {E} \cup \bigcup_{C_j\in \phi\wedge \lit_i \in C_j}B_{i,j}$.\\
\else  \begin{align*}
    U &\coloneqq \{v_i\mid i\in [\enn]\} \cup X\cup \overline{X} \cup \{c_j\mid j\in [\emm]\} \cup {F}
         \cup \bigcup_{C_j\in \phi\wedge \lit_i \in C_j}A_{i,j},\\
    W &\coloneqq \{w_i\mid i\in [\enn]\} \cup Y \cup \overline{Y} \cup \{d_j\mid j\in [\emm]\} \cup {E} \cup \bigcup_{C_j\in \phi\wedge \lit_i \in C_j}B_{i,j}.
  \end{align*}
  \fi
  \iflong  Note that we use the same symbol~$x_i$ for both the variable and the variable-agent to strengthen the connection.
  The meaning will, however, be clear from the context.\fi

  \myparagraph{The preference lists.}
  The preference lists of the agents are shown in \cref{fig:cesm-length=3-nph}.
  Herein, the preferences of the switch-agents of each occurrence of the literal correspond to those given in \cref{lem:switch}.
  Note that all preferences are specified up to defining the agents~$\alpha_{i,j}$ and $\delta_{i,j}$, which we do now.
  Defining them in an appropriate way will connect the two groups of switch-agents that correspond to the same literal as well as literals to clauses.
  For each literal~$\lit_i\in X\cup \overline{X}$, recall that $\oc_1(i)$ and $\oc_2(i)$ are the indices of the clauses which contain~$\lit_i$ with $\oc_1(i)< \oc_2(i)$.
  Define the agents~\myemph{$\alpha_{i,\oc_1(\lit_i)}$}, \myemph{$\delta_{i,\oc_1(\lit_i)}$}, \myemph{$\alpha_{i,\oc_2(\lit_i)}$}, and \myemph{$\delta_{i,\oc_2(\lit_i)}$} as follows:
    \begin{align}\label{def:alpha+delta} %
      \alpha_{i,\oc_1(\lit_i)} \coloneqq \lit_i,  \delta_{i,\oc_1(\lit_i)} \coloneqq b^0_{i,\oc_2(\lit_i)},        \alpha_{i,\oc_2(\lit_i)} \coloneqq a^6_{i,\oc_1(\lit_i)}, \delta_{i,\oc_2(\lit_i)} \coloneqq y(\lit_i).
    \end{align}

  \iflong
  For an illustration, assume that literal~$x_4$ appears in $C_3$ and $C_5$, and literal~$\overline{x}_4$ only appears in $C_2$ and $C_6$ with
  \begin{align*}
    C_2 = (x_2\vee \overline{x}_4 \vee \overline{x}_5),  ~~   C_3 = (x_1\vee {x}_4 \vee x_6), ~~    C_5 = (x_3 \vee {x}_4), ~~~ C_6=(\overline{x}_4 \vee {x}_5). 
  \end{align*}
  \noindent Then, $\alpha_{4,3}=x_4$, $\delta_{4,3}=b^0_{4,5}$, $\alpha_{4,5}=a^6_{4,3}$, $\delta_{4,5}=y_4$,
  $\alpha_{4,2}=\overline{x}_4$,  $\delta_{4,2}=b^0_{4,6}$,
  $\alpha_{4,5}=a_{4,6}^6$, and $\delta_{4,6}=\overline{y}_4$.
  The relevant part of the acceptability graph for the variable-agents, switch-agents, and clause-agents which correspond to literal~$x_4$ and clauses~$C_3,C_5$ are depicted in \cref{fig:cesm-length=3-nph-ex}.
  \begin{figure}[t!]
    \centering
    \begin{tikzpicture}
      \begin{scope}[yshift=-2cm,xshift=-4.5cm]
         \foreach \i / \nn /  \x / \y / \p / \typ in {
           w4/{w_4}/1.1/-1/above/W%
         } {
           \node[agent\typ] at (\x*\xs, -\y*\ys) (\i) {};
           \node[\p = 0pt of \i] {\small $\nn$};
         }
      \end{scope}
      \begin{scope}[yshift=-2cm,xshift=10cm]
         \foreach \i / \nn /  \x / \y / \p / \typ in {
           v4/{v_4}/-1.5/-1/above/U%
         } {
           \node[agent\typ] at (\x*\xs, -\y*\ys) (\i) {};
           \node[\p = 0pt of \i] {\small $\nn$};
         }
      \end{scope}

      \begin{scope}[xshift=-3.55cm,yshift=-4cm]
        \alphaagent{$x_4$}{x4}
        \switchgadgetV{4}{3}{{$x_4$}}{{$y_4$}}{x4}{a66}
        \path[draw] (w4) edge[bend right=40] (x4);

        \begin{scope}[yshift=.4cm]
          \clausegadget{3}{3}{beta}{gamma}{50}{-.3}{4.4}
          \foreach  \i / \nn /  \x  / \y / \p / \typ in {
            e31/{e^1_3}/0/-2/right/W,
            f31/{f^1_3}/4/-2/left/U,%
            e36/{e^6_3}/0/-3.6/right/W,
            f36/{f^6_3}/4/-3.6/left/U%
          }{
            \node[agent\typ] at (\x*\xs, \y*\ys) (\i) {};
            \node[\p = 0pt of \i, inner sep=0pt] {\small $\nn$};
          }
          \foreach \i / \j / \ag in {3/31/0, 3/36/50} {
            \path[draw] (c\i) edge[bend right=\ag] (e\j); 
            \path[draw] (d\i) edge[bend right=-\ag] (f\j); 
          }
        \end{scope}
   
      \end{scope}
      \begin{scope}[xshift=3.55cm,yshift=-4cm]
        \deltaagent{$y_4$}{y4}
        \switchgadgetV{4}{5}{{$x_4$}}{{$y_4$}}{a66}{a666}
        
        \path[draw] (a666) edge (y4);
        \path[draw] (v4) edge[bend right=-40] (y4);

        \begin{scope}[yshift=.4cm]
          \clausegadget{5}{5}{beta}{gamma}{50}{-.3}{4.4}
          \foreach  \i / \nn /  \x  / \y / \p / \typ in {
            e53/{e^3_5}/0/-2/right/W,
            f53/{f^3_5}/4/-2/left/U%
          }{
            \node[agent\typ] at (\x*\xs, \y*\ys) (\i) {};
            \node[\p = 0pt of \i, inner sep=0pt] {\small $\nn$};
          }
          \foreach \i / \j in {5/53} {
            \path[draw] (c\i) edge[bend right=0] (e\j); 
            \path[draw] (d\i) edge[bend left=0] (f\j); 
          }
        \end{scope}
      \end{scope}

    \end{tikzpicture}
    \caption{Relevant part of the acceptability graph for variable~$x_4$ and clauses~$C_3$, and $C_5$ discussed in the proof of \cref{thm:cesm-length=3-nph}.}\label{fig:cesm-length=3-nph-ex}
  \end{figure}
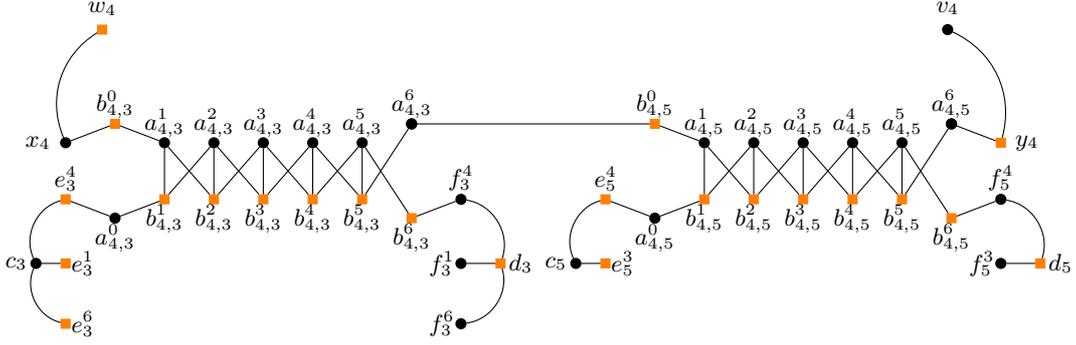
  \fi

  This completes the construction of the instance for \dCESMs[3], which can clearly be done in polynomial-time. 
  Let $\ppp$ denote the constructed instance with $\ppp=(U \uplus W, (\succ_x)_{x\in U\cup W})$.
  It is straight-forward to verify that $\ppp$ is bipartite and contains no ties and that each preference list~$\succ_x$ has length bounded by three.
    Before we give the correctness proof, for each literal~$\lit_i\in X\cup \overline{X}$ and each clause~$C_j$ with $\lit_i\in C_j$ we define the following three matchings:
      \begin{equation}
        \allowdisplaybreaks
      \begin{split}
        {\color{truecolor}\truematch_{i,j}} & \coloneqq \{\{\alpha_{i,j}, b^{0}_{i,j}\}, \{a^{6}_{i,j}, \delta_{i,j}\}\} \cup \{\{a_{i,j}^{z-1},b_{i,j}^{z}\}\mid z\in [6]\},\\
        {\color{falsecolor}\falsematch_{i,j}} &\coloneqq \{\{a^0_{i,j}, e^i_j\}, \{b^6_{i,j}, f^i_{j}\}\}\cup\{\{a^z_{i,j},b_{i,j}^{z-1}\}\mid z\in [6]\},\\
        {\color{doncolor}\donmatch_{i,j}} &\coloneqq \{\{\alpha_{i,j}, b^0_{i,j}\}, \{a^0_{i,j}, e^i_j\}, \{a^6_{i,j}, \delta_{i,j}\}, \{f^i_j, b^6_{i,j}\}, \\
        & \phantom{{}\coloneqq{}\{} \{a^1_{i,j},b^2_{i,j}\}, \{a^2_{i,j},b^1_{i,j}\}, \{a^3_{i,j},b^3_{i,j}\}, \{a^4_{i,j},b^5_{i,j}\},\{a^5_{i,j},b^4_{i,j}\}\}.
      \end{split}\label{eq:defmatchings}
    \end{equation}
    Now we show the correctness, i.e., $\phi$ admits a satisfying assignment if and only if $\ppp$ admits a perfect and \ces{} (resp.\ \es) matching.
    For the ``only if'' direction, assume that $\sigma\colon X \to \{\tr,\fa\}$ is a satisfying assignment for~$\phi$.
    Then, we define a perfect matching~$M$ as follows.
    \begin{compactitem}[--]
    \item For each variable~$x_i\in X$, let $M(\overline{x}_i)\coloneqq w_i$ and $M(v_i)\coloneqq \overline{y}_i$ if $\sigma(x_i)=\tr$; otherwise, let $M({x}_i)\coloneqq w_i$ and $M(v_i)\coloneqq {y}_i$.
    \item For each clause~$C_j\in \phi$, fix an arbitrary literal whose truth value satisfies~$C_j$ and denote the index of this literal as~$\ii(j)$.
    Then, let $M(c_j) \coloneqq e_j^{\ii(j)}$ and  $M(f_j^{\ii(j)}) \coloneqq d_j$.
    \item    
    Further, for each literal~$\lit_i\in X\cup \overline{X}$ and each clause~$C_j$ with $\lit_i\in C_j$, do the following: 
    \begin{compactenum}[(a)]
      \item If $\ii(j)=i$,
      then add to~$M$ all pairs from~$\truematch_{i,j}$.
      \item If $\ii(j)\neq i$ and $\lit_i$ is set true under~$\sigma$ (i.e., $\sigma(x_i)=\tr$ iff.\ $\lit_i=x_i$),
      then add to~$M$ all pairs from~$\donmatch_{i,j}$.
      \item If $\ii(j)\neq i$ and $\lit_i$ is set to false under~$\sigma$ (i.e., $\sigma(x_i)=\tr$ iff.\ $\lit_i=\overline{x}_i$),
      then add to~$M$ all pairs from~$\falsematch_{i,j}$.
    \end{compactenum}
  \end{compactitem}
  One can verify that $M$ is perfect.
  Hence, it remains to show that $M$ is \ces{}.
  Note that this would also imply that $M$ is \es.

  Suppose, for the sake of contradiction, that $M$ admits an \ebcs{}~$\rho$.
  First, observe that for each variable-agent~$z\in X\cup \overline{X} \cup Y \cup \overline{Y}$
  it holds that $M(z)$ either is matched with his most-preferred partner (i.e., either $v_i$ or $w_i$)
  or only envies someone who is matched with his most-preferred partner. 
  Hence, no agent from~$X\cup \overline{X} \cup Y \cup \overline{Y}$ is involved in~$\rho$.
  Analogously, no agent from~$E\cup F$ is involved in~$\rho$.

  Next, we claim the following.
  \begin{claim}\label{cl:not-alpha}
    For each literal~$\lit_i\in X\cup \overline{X}$ and each clause~$C_j$ with $\lit_i\in C_j$,
    it holds that neither~$\alpha_{i,j}$ nor~$\delta_{i,j}$ is involved in~$\rho$.
  \end{claim}
  \iflong
  \begin{proof}  \renewcommand{\qedsymbol}{(of
       \cref{cl:not-alpha})~$\diamond$}
    Suppose, for the sake of contradiction, that there exists some~$\lit_i\in X\cup \overline{X}$ with $\lit_i \in C_j$ such that $\alpha_{i,j}\in \rho$ or $\delta_{i,j}\in \rho$.

    If $\alpha_{i,j}\in \rho$, then by the definition of $\alpha_{i, j}$, it follows that $j= \oc_2(\lit_i)$ as otherwise $\alpha_{i,j}=\lit_i$ which would be a contradiction because we have just shown that no agent from~$X\cup \overline{X}$ is involved in~$\rho$.
    This implies that $\alpha_{i,j}=a^6_{i,\oc_{1}(\lit_i)} \in \rho$.
    We infer from the preferences of $a^{6}_{i,\oc_1(\lit_i)}$ that $M(a^{6}_{i,\oc_1(\lit_i)})=\delta_{i,\oc_1(\lit_i)}$.
    By the definition of~$M$, we have that $\truematch_{i, \oc_1(\lit_i)}\subseteq M$ or $\donmatch_{i,\oc_1(\lit_i)}\subseteq M$.
    From \cref{lem:switch}\eqref{switch:T} and \cref{lem:switch}\eqref{switch:D}~(setting $\alpha=\alpha_{i,\oc_1(\lit_i)}$, $\beta=e^i_{\oc_1(\lit_i)}$, $\gamma=f^i_{\oc_1(\lit_i)}$, and $\delta=\delta_{i,\oc_1(\lit_i)}$) we infer that $\rho$ involves $f^i_{\oc_1(\lit_i)} \in F$ or $\rho$ involves~$\alpha_{i,\oc_1(\lit_i)}$; observe that $\alpha_{i,\oc_1(\lit_i)}=\lit_i \in X$.
    This is a contradiction since we have already proved that no agent from~$X\cup \overline{X} \cup F$ is involved in~$\rho$.

    If $\delta_{i,j}\in \rho$, then by the definition of $\delta_{i,j}$, we have $j=\oc_1(\lit_i)$ as otherwise $\delta_{i,j}=y(\lit_i)$ which would be a contradiction because we have just shown that no agent from~$Y\cup \overline{Y}$ is involved in~$\rho$.
    This implies that $\delta_{i,j}=b^0_{i,\oc_{2}(\lit_i)} \in \rho$. %
    By the preferences of $b^0_{i,\oc_{2}(\lit_i)}$ we infer that $M(b^0_{i,\oc_2(\lit_i)})=\alpha_{i,\oc_2(\lit_i)}$.
    By the definition of~$M$, we have that $\truematch_{i, \oc_2(\lit_i)}\subseteq M$ or $\donmatch_{i,\oc_2(\lit_i)}\subseteq M$.
    By \cref{lem:switch}\eqref{switch:T} and \cref{lem:switch}\eqref{switch:D}~(setting $\alpha=\alpha_{i,\oc_2(\lit_i)}$, $\beta=e^i_{\oc_2(\lit_i)}$, $\gamma=f^i_{\oc_2(\lit_i)}$, and $\delta=\delta_{i,\oc_2(\lit_i)}$),
    we infer that $\rho$ involves $e^i_{\oc_2(\lit_i)} \in E$ or $\delta_{i,\oc_2(\lit_i)}$; observe that $\delta_{i,\oc_2(\lit_i)}=y(\lit_i)$.
    This is a contradiction since we have already proved that no agent from~$Y\cup \overline{Y} \cup E$ is involved in~$\rho$.
  \end{proof}
  \fi
  Now, using the above observations and claim, we continue with the proof.
  We successively prove that no agent is involved in $\rho$, starting with the agents in~$U$.
  \begin{compactitem}[--]
  \item \looseness=-1 If $v_i$ is involved in~$\rho$ for some~$i\in [\enn]$, then he only envies someone who is matched with~$y_i$.
  By the preferences of~$y_i$, %
  this means that $M(y_i)=a_{i,\oc_2(x_i)}^6$ and that $v_i$ envies~$a_{i,\oc_2(x_i)}^6$.
    Hence, $a_{i,\oc_2(x_i)}^6$ is also involved in~$\rho$.
    Moreover,  since $M(a_{i,\oc_2(x_i)}^6)=y_i$, we have $\truematch_{i,\oc_2(x_i)}\subseteq M$ or $\donmatch_{i,\oc_2(x_i)}\subseteq M$.
    By \cref{lem:switch}\eqref{switch:T} and \cref{lem:switch}\eqref{switch:D}~(setting $\alpha=\alpha_{i,\oc_2(x_i)}$, $\beta=e^i_{\oc_2(x_i)}$, $\gamma=f^i_{\oc_2(x_i)}$, and $\delta=\delta_{i,\oc_2(x_i)}$),
    $\rho$ involves an agent from~$\{\alpha_{i,\oc_2(x_i)}, f^i_{\oc_2(x_i)}\}$.
    Since no agent from~$F$ is involved in~$\rho$,
    it follows that $\rho$ involves~$\alpha_{i,\oc_2(x_i)}$, a contradiction to~\cref{cl:not-alpha}.

    \item Analogously, if $c_j\in \rho$ for some~$j\in [\emm]$, then this means that
    $E_j$ contains two agents~$e^i_j$ and $e^t_j$ such that
    $M(c_j)=e^t_{j}$ but $c_j$ prefers $e^i_j$ to~$e^t_j$,
    and $M(e^i_j)\in \rho$.
    Since $M$ is perfect and $c_j$ is not available,
    it follows that $M(e^i_j)=a^0_{i,j}$, implying that $a^0_{i,j}\in \rho$.
    Moreover, by the definition of~$M$ we have that $\falsematch_{i,j}\subseteq M$ or $\donmatch_{i,j}\subseteq M$.
    By Lemmas~\ref{lem:switch}\eqref{switch:F}--\eqref{switch:D}~(setting $\alpha=\alpha_{i,j}$, $\beta=e^i_{j}$, $\gamma=f^i_{j}$, and $\delta=\delta_{i,j}$),
    $\rho$ involves an agent from~$\{\alpha_{i,j}, f^i_{j}\}$, a contradiction since
    no agent from~$F_j$ is involved in~$\rho$ and by \cref{cl:not-alpha} $\alpha_{i,j}$ is not in~$\rho$. 
    \item Analogously, we can obtain a contradiction %
      if $w_i$ with $i\in [\enn]$ is in~$\rho$:
    By the definition of~$M$, if $w_i\in \rho$, then $M(x_i)=b_{i,\oc_1(x_i)}^0$ and $w_i$ envies~$b_{i,\oc_1(x_i)}^0$.
    Hence, $b_{i,\oc_1(x_i)}^0$ is also involved in~$\rho$.
    Moreover, since $M(b_{i,\oc_1(x_i)}^0)=x_i$, it follows that $\truematch_{i,\oc_1(x_i)}\subseteq M$ or $\donmatch_{i,\oc_1(x_i)}\subseteq M$.
    By Lemmas~\ref{lem:switch}\eqref{switch:T} and \eqref{switch:D}~(setting $\alpha=\alpha_{i,\oc_1(x_i)}$, $\beta=e^i_{\oc_1(x_i)}$, $\gamma=f^i_{\oc_1(x_i)}$, and $\delta=\delta_{i,\oc_1(x_i)}$),
    $\rho$ involves an agent from~$\{e^i_{\oc_1(x_i)}, \delta_{i, \oc_1(x_i)}\}$.
    Since no agent from~$E$ is involved in~$\rho$,
    it follows that $\rho$ involves~$\delta_{i,\oc_1(x_i)}$, a contradiction to~\cref{cl:not-alpha}.
    \item Again, analogously, if $d_j\in \rho$ for some~$j\in [\emm]$, then we obtain that
    $\delta_{i, j}$ is involved in~$\rho$, which is a contradiction to~\cref{cl:not-alpha}.
    
    \item Finally, if $\rho$ involves an agent from~$A_{i,j}$ (resp.\ $B_{i,j}$),
      then by \cref{lem:switch}\eqref{switch:T} and \eqref{switch:D}~(setting $\alpha=\alpha_{i,j}$, $\beta=e^i_{j}$, $\gamma=f^i_{j}$, and $\delta=\delta_{i,j}$),
      it follows that $\rho$ involves an agent from~$\{\alpha_{i,j}, f^{i}_{j}\}$ (resp.~$\{\beta_{i,j}, e^{i}_j\}$),
      a contradiction to our observation and to \cref{cl:not-alpha}.      
  \end{compactitem}
  Summarizing, we have showed that $M$ is \ces{} and \es.

  For the ``if'' direction, assume that $M$ is a perfect and \es{} matching for~$\ppp$.
  We show that there is a satisfying assignment for $\phi$.
  Note that this then also implies that, if $M$ is perfect and \ces, then there is a satisfying assignment for $\phi$.
  
  We claim that the selection of the partner of~$w_i$ defines a satisfying truth assignment for~$\phi$.
  More specifically, define a truth assignment~$\sigma\colon X\to \{\tr, \fa\}$ with $\sigma(x_i) = \tr$ if $M(w_i)=\overline{x}_{i}$, and $\sigma(x_i)=\fa$ otherwise.
  We claim that $\sigma$ satisfies $\phi$.
  To this end, consider an arbitrary clause~$C_j$ and the corresponding clause-agent.
  Since $M$ is perfect, it follows that $M(c_j)=e^i_{j}$ for some~$\lit_i\in C_j$.
  Since $e^i_j$ is not available, it also follows that $M(a^0_{i,j})=b^1_{i,j}$.
  By \cref{lem:switch}\eqref{switch:es}~(setting $\alpha=\alpha_{i,j}$, $\beta=e^i_{j}$, $\gamma=f^i_{j}$, and $\delta=\delta_{i,j}$),
  it follows that $\truematch_{i,j}\subseteq M$.
  In particular, $M(\alpha_{i,j})=b^0_{i,j}$ so that $\alpha_{i,j}$ is not available to other agents anymore.

  Now, if we can show that $\lit_i$ is matched to~$b^0_{i,\oc_1(\lit_i)}$,
  then since $M$ is perfect, we have $M(w_i)=\overline{x}_i$ if $\lit_i={x}_i$, and $M(w_i)=\overline{x}_i$ otherwise  By definition, we have $\sigma(x_i)=\tr$ if $\lit_{i}=x_i$ and $\sigma(x_i)=\fa$ otherwise.
  Thus, $C_j$ is satisfied under~$\sigma$, implying that $\sigma$ is a satisfying assignment.  
  It remains to show that $\lit_i$ is matched to~$b^0_{i,\oc_1(\lit_i)}$. %
          We distinguish between two cases; 
  \begin{compactitem}[--]
    \item If $j=\oc_1(\lit_i)$, then by assumption, $\alpha_{i,\oc_1(\lit_i)}$ is matched to $b^0_{i,\oc_1(\lit_i)}$, as required.
    \item If $j=\oc_2(\lit_i)$, then by definition, it holds that $\alpha_{i,j}=a^6_{i,\oc_1(\lit_i)}$
    and $\delta_{i,\oc_1(\lit_i)}=b^0_{i,j}$.
    In other words, $M(a^6_{i,\oc_1(\lit_i)})=\delta_{i,\oc_1(\lit_1)}$.
    By \cref{lem:switch}\eqref{switch:es}~(setting $\alpha=\alpha_{i,\oc_1(\lit_i)}$, $\beta=e^i_{\oc_1(\lit_i)}$, $\gamma=f^i_{\oc_1(\lit_i)}$, and $\delta=\delta_{i,\oc_1(\lit_i)}$),
    it follows that $\truematch_{i,j}\subseteq M$ or $\donmatch_{i,j}\subseteq M$.
    In both cases, it follows that $\alpha_{i,\oc_1(i)}$ is matched to~$b^0_{i,\oc_1(i)}$.
  \end{compactitem}
          \looseness=-1
  \iflong

  We have thus proved that there is a satisfying assignment for $\phi$ if and only if there is a perfect \ces\ (resp.\ \es) matching for $\ppp$.
  \fi
\end{proof}

Next, we show how to complete the preferences of the agents constructed in the proof of \cref{thm:cesm-length=3-nph} to show hardness for complete and strict preferences.

\begin{theorem}\label{thm:cesm-complete-nph}
  \CESMs\ and \CESRs\ are NP-complete even for complete and strict preferences.
\end{theorem}

\begin{proof}
  \iflong
  \CESMs\ is contained in NP by the same argument as in \cref{thm:cesm-length=3-nph}.
  For NP-hardness we adapt the proof of \cref{thm:cesm-length=3-nph}.
  \fi
  Recall that in that proof, for a given \rrsat{} instance~$(X,\phi)$ with $X=\{x_1,x_2,\cdots, x_{\enn}\}$
  and $\phi=\{C_1,C_2,\cdots, C_{\emm}\}$, we construct two disjoint agent sets~$\UU$ and $\WW$ %
  with
  \iflong
  \begin{align*}
    \UU &\coloneqq \{v_i\mid i\in [\enn]\} \cup X\cup \overline{X} \cup \{c_j\mid j\in [\emm]\} \cup {F}
         \cup \bigcup_{C_j\in \phi\wedge \lit_i \in C_j}A_{i,j},\\
    \WW &\coloneqq \{w_i\mid i\in [\enn]\} \cup Y \cup \overline{Y} \cup \{d_j\mid j\in [\emm]\} \cup {E} \cup \bigcup_{C_j\in \phi\wedge \lit_i \in C_j}B_{i,j}, \text{ where }
  \end{align*}
  $\overline{X}\coloneqq \{\overline{x}_i\mid i\in [\enn]\}$, $\overline{Y}\coloneqq \{\overline{y}_i\mid i\in [\enn]\}$,
  $A_{i,j}=\{a^z_{i,j}\mid z\in \{0,1,\ldots,6\}\}$  $B_{i,j}=\{b^z_{i,j}\mid z\in \{0,1,\ldots,6\}\}$,
  $E_j\coloneqq \{e_j^{i}\mid \lit_i\in C_j\}$, 
  $F_j\coloneqq \{f_j^{i}\mid \lit_i\in C_j\}$, and 
  ${E} \coloneqq \bigcup_{C_j\in \phi}E_j$  and~${F} \coloneqq \bigcup_{C_j\in \phi}F_j$.
  \fi
  For each agent~$z\in \UU\cup \WW$ let $\lin_z$ denote the preference list of~$z$ constructed in the proof.
  The basic idea is to extend the preference list~$\lin_z$ by appending to it the remaining agents appropriately. 

  We introduce some more notations.
  Let $\rhd_{\UU}$ and $\rhd_{\WW}$ denote two arbitrary but fixed linear orders of the agents in~$\UU$ and $\WW$, respectively.
  Now, for each subset of agents~$S\subseteq \UU$ (resp.\ $S\subseteq \WW$),
  let \myemph{$[S]_{\rhd}$} denote the fixed order of the agents in~$S$ induced by~$\rhd_{\UU}$ (resp.\ $\rhd_{\WW}$),
  and let \myemph{$S\setminus \lin_z$} denote the subset~$\{t\in S\mid t \notin \lin_z\}$, where $z\in \WW$ (resp.\ $z\in \UU$).
  Finally, for each agent~$z\in \UU$ (resp.\ $z\in \WW$), let \myemph{$\rest_z$} denote the subset of agents which do not appear in~$\lin_z$ or in~$Y\cup \overline{Y}\cup E$ (resp.\ ~$X\cup \overline{X}\cup F$).
  That is, $\rest_z\coloneqq \left(\WW\setminus (Y\cup \overline{Y}\cup F)\right)\setminus \lin_z$
  (resp.\ $\rest_z\coloneqq \left(\UU\setminus (X\cup \overline{Y}\cup F)\right)\setminus \lin_z$).
  
  \noindent Now, we define the preferences of the agents as follows.
  \[
      \forall z\in \UU, ~ z\colon   \lin_z \succ [Y\cup \overline{Y}\cup E\setminus \lin_z]_{\rhd} \succ [\rest_{z}]_{\rhd}, ~\text{ and }
      \forall z\in \WW, ~ z\colon  \lin_z \succ [X\cup \overline{X}\cup F\setminus \lin_z]_{\rhd} \succ [\rest_{z}]_{\rhd}.
    \]
 \iflong For instance, the complete preference list of an agent called~$a^0_{i,j}$ (corresponding to the literal~$\lit_i$ which appears in clause~$C_j$) is
  \begin{align*}
    a^0_{i,j}\colon~~ b^1_{i,j}\succ e^i_j \succ [X\cup \overline{X}\cup F \setminus \{e^i_j\}]_{\rhd} \succ
    [\bigcup_{k=0}^6B^k\cup C\cup V \setminus \{b^1_{i,j}\}]_{\rhd}.
  \end{align*}\fi
  \looseness=-1
  Let $\ppp'$ denote the newly constructed preference profile.
  Clearly, the constructed preferences are complete and strict.
  \iflong
  It remains to show the correctness.
  Assume that $\phi$ admits a satisfying assignment~$\sigma\colon X\to \{\tr, \fa\}$.
  First of all, observe that the following claim holds.
  \begin{claim}\label[claim]{cl:gadget-agents-fancied}
    Besides the agents of~$X\cup \overline{X}\cup F\cup Y\cup \overline{Y}\cup E$ every other agent appears as the most-preferred agent of some other agent.
  \end{claim}
  Using this, we claim the following for each \ces{} matching of~$\ppp'$.
  \fi
  \begin{claim}\label[claim]{cl:not-outside-lin}
    If $M$ is a \ces{} matching for~$\ppp'$,
    then
    \begin{compactenum}[(i)]
      \item\label{rest} for each agent~$z\in \UU\cup \WW$ it holds that $M(z)\notin \rest_z$, and
      \item\label{lin} for each agent~$z\in \UU\cup \WW\setminus (X\cup \overline{X}\cup F \cup Y\cup \overline{Y}\cup E)$ it holds that $M(z)\in \lin_z$.
    \end{compactenum}
  \end{claim}
  \iflong
  \begin{proof}
    \renewcommand{\qedsymbol}{(of
      \cref{cl:not-outside-lin})~$\diamond$}
    Let $M$ be a \ces{} matching of~$\ppp'$.
    By \cref{lem:props-es}, $M$ is maximal.
    Since $G(\ppp')$ is complete it follows that $M$ is perfect.
    For Statement~\eqref{rest}, suppose, for the sake of contradiction, that
    there exists an agent~$z_0\in U\cup W$ such that $M(z_0)\in \rest_{z_0}$.
    This means that $M(z_0)\notin X\cup \overline{X}\cup F\cup Y \cup \overline{Y} \cup E$.
    Now consider the following inductive definition of agents $z_i$ for $i = 1, 2, \ldots$ such that
    \begin{compactenum}[(a)]
    \item $z_i$ prefers $M(z_{i - 1})$ to $M(z_i)$,\label{en:zipref}
    \item $M(z_i) \notin X\cup \overline{X}\cup F\cup Y \cup \overline{Y} \cup E$, and\label{en:mzinotx}
    \item $z_i \notin \{z_0, z_1, \ldots, z_{i - 1}\}$.\label{en:zidistinct}
    \end{compactenum}
    Let $z_{i - 1}$ already be defined.
    Since $M(z_{i - 1}) \notin X\cup \overline{X}\cup F\cup Y \cup \overline{Y} \cup E$, by \cref{cl:gadget-agents-fancied}, there exists an agent who considers $M(z_{i - 1})$ as the most-preferred agent.
    Define $z_i$ to be an arbitrary agent who considers $M(z_{i - 1})$ as the most-preferred agent.
    Clearly, \labelcref{en:zipref}~holds.
    Observe that $z_0$ prefers $M(z_0)$ to $M(z_i)$ since otherwise $(z_0, z_1, \ldots, z_i)$ would form an \ebc.
    Thus, $M(z_i) \in \rest_{z_0}$, which implies \labelcref{en:mzinotx}.
    Finally, for the sake of contradiction assume that \labelcref{en:zidistinct} does not hold, that is, there is an agent $z_j$ with $j < i$ such that $z_j = z_i$.
    By definition, $z_j$ and $z_i$ consider $M(z_{j - 1}) = M(z_{i - 1})$ as the most-preferred agent.
    Since $M$ is a matching, $z_{j - 1} = z_{i - 1}$, a contradiction to the fact that $z_{i - 1} \notin \{z_0, z_1, \ldots, z_{i - 2}\}$.
    Hence, we have found an infinite sequence of pairwise distinct agents, a contradiction, proving Statement~\eqref{rest}.
    
    For Statement~\eqref{lin}, suppose, for the sake of contradiction, that there exists an agent~$z\in U\cup W\setminus (X\cup \overline{X}\cup F \cup Y\cup \overline{Y}\cup E)$ such that $M(z)\notin \lin_{z}$.
    By symmetry, it follows that $z=M(M(z))\notin \lin_{M(z)}$.
    By the first statement, $z\notin \rest_{z}$.
    Hence, $z\in (X\cup \overline{X}\cup F\cup Y \cup \overline{Y} \cup E)\setminus \lin_{z}$, a contradiction.    
    \end{proof}
  \fi
  
  Now, we are ready to show the correctness, i.e., $\phi$ admits a satisfying assignment if and only if $\ppp'$ admits a \ces{} matching.
  
  For the ``only if'' direction, assume that $\phi$ admits a satisfying assignment, say~$\sigma\colon X\to \{\tr,\fa\}$.
  We claim that the \ces{} matching~$M$ for $\ppp$ that we defined in the ``only if'' direction of the proof for \cref{thm:cesm-length=3-nph} is a \ces{} matching for~$\ppp'$.
  Clearly, $M$ is a perfect matching for~$\ppp'$ since~$G(\ppp')$ is a supergraph of~$G(\ppp)$.
  Since each agent~$z\in \UU\cup \WW$ has $M(z)\in \lin_z$,
  for each two agents~$z, z'\in \UU$ (resp.\ $\WW$),
  it holds that $z$ envies $z'$ only if $M(z')\in \lin_z$.
  In other words, if $M$ would admit an \ebc{}~$\rho=(z_0,z_1,\cdots,z_{r-1})$ ($r\ge 2$) for $\ppp'$,
  then for each $i\in \{0,1,\dots,r-1\}$ it must hold that $M(z_{i})\in \lin_{z-1}$ ($z-1$ taken modulo~$r$).
  But then, $\rho$ is also an \ebc{} for~$\ppp$, a contradiction to our ``only if'' part of the proof for \cref{thm:cesm-length=3-nph}.

  For the ``if'' direction, let $M$ be a \ces{} matching for~$\ppp'$.
  Note that in the ``if'' part of the proof of \cref{thm:cesm-length=3-nph} we heavily utilize the properties given in \cref{lem:switch}\eqref{switch:es}.
  Now, to construct a satisfying assignment for~$\phi$ from~$M$, we will prove that the lemma also holds for profile~$\ppp'$.
  \ifshort
  To this end, for each literal~$\lit_i\in X\cup \overline{X}$ and each clause~$C_j$ with $\lit_i\in C_j$,
  recall the three matchings $\truematch_{i,j}$, $\falsematch_{i,j}$, $\donmatch_{i,j}$ and the agents  $\alpha_{i,j}$ and $\delta_{i,j}$ that we have defined in equations~\eqref{eq:defmatchings} and \eqref{def:alpha+delta} (on Page~\pageref{def:alpha+delta}).
  \else
  To this end, for each literal~$\lit_i\in X\cup \overline{X}$ and each clause~$C_j$ with $\lit_i\in C_j$,
  recall the three matchings we have defined in equation~\eqref{eq:defmatchings} (on Page~\pageref{def:alpha+delta}):
  \begin{equation*}
    \begin{split}
      {\color{truecolor}\truematch_{i,j}} & \coloneqq \{\{\alpha_{i,j}, b^{0}_{i,j}\}, \{a^{6}_{i,j}, \delta_{i,j}\}\} \cup \{\{a_{i,j}^{z-1},b_{i,j}^{z}\}\mid z\in [6]\},\\
      {\color{falsecolor}\falsematch_{i,j}} &\coloneqq \{\{a^0_{i,j}, e^i_j\}, \{b^6_{i,j}, f^i_{j}\}\}\cup\{\{a^z_{i,j},b_{i,j}^{z-1}\}\mid z\in [6]\},\\
      {\color{doncolor}\donmatch_{i,j}} &\coloneqq \{\{\alpha_{i,j}, b^0_{i,j}\}, \{a^0_{i,j}, e^i_j\}, \{a^6_{i,j}, \delta_{i,j}\}, \{f^i_j, b^6_{i,j}\}, \\
      & \phantom{{}\coloneqq{}\{} \{a^1_{i,j},b^2_{i,j}\}, \{a^2_{i,j},b^1_{i,j}\}, \{a^3_{i,j},b^3_{i,j}\}, \{a^4_{i,j},b^5_{i,j}\},\{a^5_{i,j},b^4_{i,j}\}\}.
    \end{split}
  \end{equation*}
  \noindent The agents~\myemph{$\alpha_{i,j}$} and \myemph{$\delta_{i,j}$} were defined according to equation~\eqref{def:alpha+delta}; for the sake of completeness, we repeat here:
  \begin{align*} %
     \alpha_{i,\oc_1(\lit_i)} \coloneqq \lit_i,   \delta_{i,\oc_1(\lit_i)} \coloneqq b^0_{i,\oc_2(\lit_i)}, 
     \alpha_{i,\oc_2(\lit_i)} \coloneqq a^6_{i,\oc_1(\lit_i)},   \delta_{i,\oc_2(\lit_i)} \coloneqq y(\lit_i).
  \end{align*}
    \fi

  \begin{claim}\label{cl:complete-cesm}
    Matching~$M$ satisfies for each literal~$\lit_i\in X\cup \overline{X}$ and each clause~$C_j\in \phi$ with $\lit_i\in C_j$, either
    \begin{inparaenum}[(i)]
      \item\label{complete-esm:x-true} $\truematch_{i,j}\subseteq M$, or
      \item\label{complete-esm:x-false} $\falsematch_{i,j}\subseteq M$, or
      \item\label{complete-esm:x-doncare} $\donmatch_{i,j}\subseteq M$.
    \end{inparaenum}
  \end{claim}

  \iflong
  \begin{proof}\renewcommand{\qedsymbol}{(of
      \cref{cl:complete-cesm})~$\diamond$}
    The proof is almost the same as the one given for \cref{lem:switch}\eqref{switch:es}.
    We repeat for the sake of completeness.
  
    Since $M$ is \ces{} for~$\ppp'$,
    by \cref{cl:not-outside-lin}\eqref{lin} it follows that
    \begin{align}\label{eq:only-in-lin}
      \forall \lit_i\in X\cup \overline{X}, \forall C_j\in \phi \text{ with }
      \lit_i\in C_j, \forall z\in \{0,1,\cdots, 6\}\colon
      M(a^z_{i,j})\in \lin_{a^z_{i,j}}.
    \end{align}
    We distinguish between three subcases.

    If $M(a_{i,j}^3) = b_{i,j}^2$, then $M(a_{i,j}^2)\neq b_{i,j}^3$ as otherwise $(b_{i,j}^2,b_{i,j}^3)$ is an \ebp{}.
    Hence, $M(a_{i,j}^2) = b_{i,j}^1$ and $M(b_{i,j}^3)=a_{i,j}^4$ since $b_{i,j}^2$, $a_{i,j}^2$, and $a_{i,j}^3$ are already matched to others.
    By \eqref{eq:only-in-lin}, $M(a_{i,j}^1)=b_{i,j}^0$, $M(a_{i,j}^0) = e^i_{j}$, $M(b_{i,j}^4)=a_{i,j}^5$, $M(b_{i,j}^5)=a_{i,j}^6$, and $M(b_{i,j}^6)=f_j^i$.
    This implies that $M$ contains $\falsematch_{i,j}$.

    If $M(a_{i,j}^3) = b_{i,j}^3$, then $M(a_{i,j}^2)\neq b_{i,j}^2$ and $M(a_{i,j}^4)\neq b_{i,j}^4$ as otherwise $(a_{i,j}^2,a_{i,j}^3)$
    or $(b_{i,j}^3,b_{i,j}^4)$ is an \ebp{}.
    By \eqref{eq:only-in-lin}, $M(a_{i,j}^2) = b_{i,j}^1$ and $M(a_{i,j}^4)=b_{i,j}^5$ since $b_{i,j}^3$ is already matched to~$a_{i,j}^3$.
    Again, by \eqref{eq:only-in-lin}, $M(b_{i,j}^2)=a_{i,j}^1$, $M(b_{i,j}^4) = a_{i,j}^5$, $M(a_{i,j}^0)=e^i_j$, $M(a_{i,j}^6)=\delta_{i,j}$, $M(b_{i,j}^0)=\alpha_{i,j}$, $M(b_{i,j}^6)=f^i_{j}$.
    This implies that $M$ contains $\donmatch_{i,j}$.

    Finally, if $M(a_{i,j}^3) = b_{i,j}^4$, then $M(a_{i,j}^4)\neq b_{i,j}^3$ as otherwise $(a_{i,j}^3,a_{i,j}^4)$ is an \ebp{}.
    Hence, by \eqref{eq:only-in-lin}, $M(a_{i,j}^4) = b_{i,j}^5$ and $M(b_{i,j}^3)=a_{i,j}^2$ since $b_{i,j}^5$ and $a_{i,j}^2$ are the only agents available to~$a_{i,j}^4$ and $b_{i,j}^3$, respectively.
    By the acceptability relations, $M(a_{i,j}^5)=b_{i,j}^6$, $M(a_{i,j}^6) = \delta_{i,j}$,
    $M(b_{i,j}^2)=a_{i,j}^1$,
    $M(b_{i,j}^1)=a_{i,j}^0$,
    and $M(b_{i,j}^0)=\alpha_{i,j}$.
    This implies that $M$ contains $\truematch_{i,j}$.
    Together, this completes the proof for Statement~\eqref{cl:complete-cesm}.
  \end{proof}
  \fi

  Now we show that the function~$\sigma\colon X\to \{\tr,\fa\}$ with $\sigma(x_i) = \tr$ if $M(w_i)=\overline{x}_{i}$, and $\sigma(x_i)=\fa$ otherwise
  is a satisfying truth assignment for~$\phi$.
  Clearly, $\phi$ is a valid truth assignment since by \cref{cl:not-outside-lin}\eqref{lin} every variable agent~$w_i$ is matched to either~$x_i$ or $\overline{x}_i$.
  We claim that $\sigma$ satisfies $\phi$.
  To this end, consider an arbitrary clause~$C_j$ and the corresponding clause-agent~$c_j$.
  By \cref{cl:not-outside-lin}\eqref{lin}, we know that $M(c_j)=e^i_{j}$ for some~$\lit_i\in C_j$.
  Since $e^i_j$ is not available, by \cref{cl:not-outside-lin}\eqref{lin},
  it also follows that $M(a^0_{i,j})=b^1_{i,j}$.
  By \cref{cl:complete-cesm}, it follows that $\truematch_{i,j}\subseteq M$.
  In particular, $M(\alpha_{i,j})=b^0_{i,j}$ so that $\alpha_{i,j}$ is not available to other agents anymore.

  We aim to show that $\alpha_{i,\oc_1(\lit_i)}$ is matched to~$b^0_{i,\oc_1(\lit_i)}$ by~$M$, which implies that $\lit_i$ is not available to~$w_i$ since $\alpha_{i,\oc_1(\lit_i)}=\lit_i$ by the definition of $\alpha_{i,\oc_1(\lit_i)}$.
  We distinguish two cases:
 \iflong \begin{compactitem}[--]
   \item \fi
   If $j=\oc_1(\lit_i)$, then by the definition of $\alpha_{i, j}$, it follows that $\alpha_{i,\oc_1(\lit_i)}$ is matched to $b^0_{i,\oc_1(\lit_i)}$\iflong, as required\fi.
   \iflong \item \fi If $j = \oc_2(\lit_i)$, then by the definition of $\alpha_{i,j}$, we have $\alpha_{i,j}=a^6_{i,\oc_1(\lit_i)}$
      and by the definition of $\delta_{i,\oc_1(\lit_i)}$ we have $\delta_{i,\oc_1(\lit_i)} = b^0_{i,\oc_2(\lit_i)} = b^0_{i,j}$.
      In particular, since $M(\alpha_{i, j}) = b^0_{i, j}$ we have $M(a^6_{i,\oc_1(\lit_i)})=\delta_{i,\oc_1(\lit_1)}$.
      By \cref{cl:complete-cesm},
      it follows that $\truematch_{i,\oc_1(\lit_i)}\subseteq M$ or $\donmatch_{i,\oc_1(\lit_i)}\subseteq M$.
    In both cases, it follows that $\alpha_{i,\oc_1(\lit_i)}$ is matched to~$b^0_{i,\oc_1(\lit_i)}$.
 \iflong \end{compactitem}\fi
  \looseness=-1
  We have just shown that $\lit_i$ is \emph{not} available to~$w_i$.
  Hence, by \cref{cl:not-outside-lin}\eqref{lin}, $M(w_i)=\overline{x}_i$ if $\lit_i={x}_i$, and $M(w_i)=\overline{x}_i$ otherwise.
  By definition, we have that $\sigma(x_i)=\tr$ if $\lit_{i}=x_i$ and $\sigma(x_i)=\fa$ otherwise.
  This implies that $C_j$ is satisfied under~$\sigma$, implying that $\sigma$ is a satisfying assignment.
\end{proof}

\section{Algorithms for Bounded Preferences Length}
\label{sec:d-ces}

\ifshort
We first observe that when bounding the preference length by two (coalitional) \esty\ can be decided in linear time.
\begin{theorem}
  \label{thm:maxdeg2}
  \dESMs[2], \dESRs[2], \dCESMs[2], and \dCESRs[2] can be solved in linear time.
\end{theorem}
\noindent The reason is that for preference lengths at most two the
acceptability graph becomes a disjoint union of paths and cycles.
Since we are looking for perfect matchings, they have to be of even length.
Moreover, since the agents involved in an \ebc\ (together with their partners) form a cycle in the acceptability graph, it suffices to check for each cycle whether one of its two perfect matchings is \ces.
\else

In this section we give algorithms for the profiles with bounded preference length.
In \cref{sec:maxdeg2} we give linear-time algorithms for length at most two, and in \cref{sec:FPT} we give a fixed-parameter algorithm for length at most three.

\subsection{Preference list length at most two}
\label{sec:maxdeg2}
Bounding the length of the preference lists by two allows us to decide whether a \ces{} and perfect matching exists in linear time.
The reason is that for preferences of length at most two, the underlying acceptability graph consists of disjoint paths and cycles.
Together with \cref{lm:coal_means_cycle}, we infer that only agents from a cycle may induce \ebc{s}, and hence can be checked for \cesty{} in linear time. %

\begin{algorithm}[t!]
  \PrintSemicolon

  \KwIn{A preference profile~$\ppp$ with preferences length at most two.}
  
  \KwOut{A \ces{} matching for $\ppp$ or \KwNo if none exists.}

  $M\leftarrow \emptyset$

  \lIf{$G(\ppp)$ contains a connected component of odd size}{%
    \Return \KwNo
  }
  
  \ForEach{connected component~$C$ in $G(\ppp)$}{
    \If{$G(\ppp)[C]$ is a path}{
      Let $E(G(\ppp)[C]) = \{\{x_0,x_1\}, \{x_1,x_2\},\ldots,\{x_{2t-2}, x_{2t-1}\}\}$

      $M\leftarrow M\cup \{\{x_{2i-2}, x_{2i-1}\}\mid i\in [t]\}$
    }

    \If{$G(\ppp)[C]$ is a cycle}{
      Let $E(G(\ppp)[C]) = \{\{x_0,x_1\}, \{x_1,x_2\}, \ldots,\{x_{2t-2}, x_{2t-1}\}, \{x_{2t-1},x_0\}\}$

      $M_1\leftarrow \{\{x_{2i-2},x_{2i-1}\} \mid i\in [t]\}$

      $M_2\leftarrow \{\{x_{2i-1},x_{2i}\} \mid i\in [t-1]\}\cup \{\{x_{2t-1},x_0\}\}$

      \lIf{$M_1$ is \ces{} for $\ppp$ restricted to~$C$}{%
      \label{alg:d=2-check1}  $M\leftarrow M\cup M_1$%
      }
      \lElseIf{$M_1$ is \ces{} for $\ppp$ restricted to~$C$}{%
       \label{alg:d=2-check2}  $M\leftarrow M\cup M_2$%
      }\lElse{\Return \KwNo}
    }
  }

  \Return $M$
  \caption{Algorithm for checking \cesty{} when each preference list has length bounded by two.}\label{alg:d=2}
\end{algorithm}

\begin{theorem}
  \label{thm:maxdeg2}
  \dESMs[2], \dESRs[2], \dCESMs[2], and \dCESRs[2] can be solved in linear time.
\end{theorem}

\begin{proof}
  Let $\ppp$ be a preference profile with preference-list length bounded by two, meaning that $G(\ppp)$ consists of solely paths and cycles.
  The algorithm works as follows and we give pseudo code for \cesty{} in \cref{alg:d=2}.
  For \esty\ the pseudo code is analogous; we only need to exchange \ces{} with \es{} in lines~\ref{alg:d=2-check1}--\ref{alg:d=2-check2}.

  We consider each path and cycle independently and accept if and only if each one admits a perfect (coalitional) \es\ matching.
  Since we are aiming for a perfect matching that is \es{} or \ces{}, by \cref{lm:coal_means_cycle}, it suffices to only consider cycles of even length as paths do not induce any \ebc{s} and cycles of odd length cannot result in a perfect matching.

  For cycles of even length, since there are only two possible perfect matchings,
  it suffices to check whether one of them avoids \ebc{s} or \ebp{s}.
  Since constructing all perfect matchings for a path or a cycle (there are at most two) and checking each of them for \cesty{} or \esty{} can be done in linear time, the described algorithms run in linear time. 
\end{proof}

\fi 

\ifshort
\paragraph{Fixed-parameter algorithm for \dESRs[3].}
\else
\subsection{Fixed-parameter algorithm for \dESRs[3]}
\label{sec:FPT}
\fi
We now turn to preference length at most three%
. %
In \cref{thm:cesm-length=3-nph} we have seen that even this case remains NP-hard, even for bipartite preference profiles.
Moreover, the proof suggests that a main obstacle that one has to deal with when solving \dESMs[3] (and hence \dESRs[3]) are the switch gadgets.
Here we essentially show that they are indeed the \emph{only} obstacles, that is, if there are few of them present in the input, then we can solve the problem efficiently.
We capture the essence of the switch gadgets with the following structure that we call hourglasses.
\begin{definition}\label{def:hourglass}
  Let $\ppp$ be a preference profile and $V_H\subseteq V$ a subset of $2h$~agents with $V_H=\{u_i,w_i\mid 0\le i \le h-1\}$.
  We call the subgraph~$G(\ppp)[V_H]$ induced by~$V_H$ an \myemph{hourglass} of height~$h$ if it satisfies the following:
  \begin{compactitem}[--]
    \item For each $i\in \{0,h-1\}$ the vertex degree of $u_i$ and $w_i$ are both at least two in~$G(\ppp)[V_H]$;
    \item For each $i \in [h-2]$, the vertex degree of $u_i$ and $w_i$ are exactly three in $G(\ppp)[V_H]$;
    \item For each $i \in \{0,1,\dots,h-1\}$ we have $\{u_i,w_i\}\in E(G(\ppp)[V_H])$;
    \item For each $i \in \{0,1,\dots,h-2\}$ we have $\{u_i,w_{i+1}\}, \{u_{i+1},w_i\}\in E(G(\ppp)[V_H])$.
  \end{compactitem}
  We refer to the agents~$u_i$ and $w_i$ from~$V_H$ as \myemph{layer-$i$} agents.
  We call an hourglass~$H$ \myemph{maximal} if no larger agent subset~$V'\supsetneq V(H)$ exists that induces an hourglass.
  
  \looseness=-1
  Given an hourglass~$H$ in~$G(\ppp)$, we call a matching~$M$ for $\ppp$ \myemph{perfect for~$H$} if for each agent~$v\in V(H)$ we have $M(v)\in V(H)\setminus \{v\}$.
  Further, $M$ is \myemph{\es for~$H$} if no two agents from $V(H)$ can form an \ebp.
\end{definition}

Notice that the smallest hourglass has height two and is a four-cycle. 
\iflong
Figure~\ref{fig:maxdg3_hourglass} shows a maximal hourglass of height three which contains two different non-maximal hourglasses of height two.
Let us mention that it is possible that the vertex sets of two distinct maximal hourglasses have nonempty intersection.

\fi
We are ready to show the following fixed-parameter tractability result.

\begin{theorem}\label{thm:fpt-hourglasses}
  An instance of \dESRs[3]\ with $2n$~agents and $\ell$~maximal hourglasses %
  can be solved in $O(6^\ell \cdot n \sqrt{n})$ time. %
\end{theorem}
\looseness=-1

The main ideas are as follows.
The first observation is that a matching for a maximal hourglass can interact with the rest of the graph in only six different ways: The only agents in an hourglass~$H$ of height~$h$ that may have neighbors outside~$H$ are the layer-$0$ and layer\nobreakdash-$(h-1)$ agents; %
let us call them \myemph{connecting agents} of~$H$.
A matching~$M$ may match these agents either to agents inside or outside~$H$.
Requiring $M$ to be perfect means that an even number of the connecting agents has to be matched inside~$H$.
This then gives a bound of at most six different possibilities of the matching~$M$ with respect to whether the connecting agents are matched inside or outside~$H$\iflong~(see \cref{fig:6-categories})\fi.
Let us call this the \myemph{signature} of $M$ with respect to~$H$.
Hence, we may try all $6^\ell$ possible combinations of signatures for all hourglasses and check whether one of them leads to a solution (i.e., an \es{} matching).

\looseness=-1
The second crucial observation is that each \ebp\ of a perfect matching yields a four-cycle and hence, is contained in some maximal hourglass.
Thus, the task of checking whether a combination of signatures leads to a solution decomposes into (a) checking whether each maximal hourglass~$H$ allows for an \es\ matching adhering to the signature we have chosen for~$H$ and (b) checking whether the remaining acceptability graph after deleting all agents that are in hourglasses or matched by the chosen signatures admits a perfect matching.

Task (b) can clearly be done in $O(n\cdot \sqrt{n})$~time by performing any maximum-cardinality matching algorithm (note that the graph~$G(\ppp)$ has $O(n)$~edges).
We then prove that task (a) for all six signatures can be reduced to checking whether a given hourglass admits a perfect and \es\ matching.
This, in turn, we show to be linear-time solvable by giving a dynamic program that fills a table, maintaining some limited but crucial facts about the structure of partial matchings for the hourglass.

\iflong

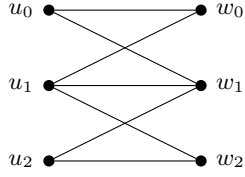
\begin{figure}[t!]
  \centering
  \begin{tikzpicture}
    \def \xs {.5}
    \foreach \i / \nn /  \x / \y / \p / \typ in {
      u0/{u_0}/0/0/left/U, w0/{w_0}/2/0/right/U,
      u1/{u_1}/0/-2/left/U, w1/{w_1}/2/-2/right/U,
      u2/{u_2}/0/-4/left/U, w2/{w_2}/2/-4/right/U%
    } {
      \node[agent\typ] (\i) at (\x, \y*\xs) {};
      \node[\p = 0pt of \i] {\small $\nn$};
    }

    \foreach \i in {0,1,2} {
      \path[draw] (u\i) edge (w\i);
    }
    \foreach \i / \j in {0/1,1/2} {
      \path[draw] (u\i) edge (w\j);
      \path[draw] (u\j) edge (w\i);
    }
  \end{tikzpicture}
  \caption{An hourglass of height three.}
  \label{fig:maxdg3_hourglass}
\end{figure}

In the remainder of this section we prove \cref{thm:fpt-hourglasses}.
To this end, let $\ppp$ denote an input instance of \dESRs[3]. %
We observe that the hourglasses are the only parts where an \ebp\ can form:
\begin{observation}
  \label{sc:md3_ebphg}
  For each \ebp{}~$(x,y)$ of a perfect matching~$M$ of~$\ppp$ there exists an hourglass in~$G(\ppp)$ which contains both $x,y$ and their partners under~$M$. %
\end{observation}
\begin{proof}
  Let $M$ be an arbitrary perfect matching of~$\ppp$ which admits an \ebp{}, say~$\{x,y\}$.
  Then, it must hold that $\{x,M(x)\}$, $\{x,M(y)\}$, $\{y,M(y)\}$, $\{y,M(x)\}\in E(G(\ppp))$.
  This implies that $\{x,y,M(x),M(y)\}$ induces an hourglass of height two.
\end{proof}

\subsubsection{Perfect and exchange-stable matchings for a maximal hourglass}

We can check in linear time whether a perfect and \es{} matching exists for a given hourglass.

\begin{lemma}\label{lem:dp-d=3=hg}
  Let $H$ be a height-$h$ hourglass with agents~$V(H)=\{u_i,w_i\mid i\in \{0,1,\dots,h-1\}\}$. %
  In $O(h)$~time, one can decide whether there exists a perfect and \es{} matching for~$H$. %
\end{lemma}
\begin{proof}
  Clearly, if $h\le 4$, we can check for existence of a perfect and \es{} matching in constant time.
  Otherwise, we distinguish between three cases.

  \noindent \textbf{Case 1: $\{u_0,w_{h-1}\}, \{u_{h-1},w_{0}\}\in E(H)$.}
  Since the maximum degree is three and by the definition of hourglass, this means that $H$ is bipartite.
  We claim that the matching~$M$ with $M\coloneqq \{u_{i-1},w_{i}\mid i\in [h]\}$ (indices taken modulo~$h$) is perfect and \es{} for~$H$.
  Clearly, $M$ is perfect for~$H$.
  Since $h\ge 5$, any alternating cycle wrt.~$M$ has length at least six.
  By \cref{lm:coal_means_cycle}, an \ebc\ implies that there is a cycle of length at most four and thus we obtain that $M$ is \es{} for~$H$.
  Hence, in this case, we immediately return \KwYes by returning~$M$.

  \noindent \textbf{Case 2: $\{u_0,u_{h-1}\}, \{w_0,w_{h-1}\}\in E(H)$.}
  By the definition of hourglasses, there are two possible types of perfect matchings~$M$: Either (a) both $\{u_0, u_{h-1}\}$ and $\{w_{0}, w_{h-1}\}$ are in $M$ or (b) neither is.
  Consider a matching $M$ of type~(a).
  Observe that no agent $u_i$, $1 < i < h - 1$, envies $u_0$ or $u_{h - 1}$ and analogous for agents $w_i$, $1 < i < h - 1$.
  If $M$ is not \es\ then either (i) $(u_0, w_0)$ or $(u_{h - 1}, w_{h - 1})$ form an \ebp\ or (ii) no \ebp\ involves an agent from $\{u_0, u_{h - 1}, w_0, w_{h - 1}\}$.
  This is because, since $h \geq 5$, there is no alternating cycle of length at most~4 involving exactly one agent from $\{u_0, u_{h - 1}, w_0, w_{h - 1}\}$. 
  Hence, there is a perfect \es\ matching of type~(a) for $H$ if and only neither $(u_0, w_0)$ nor $(u_{h - 1}, w_{h - 1})$ are an \ebp\ and the sub-hourglass~$H'$ with $H'=H[\{u_i,w_i\mid i\in [h-2]\}]$ has a perfect \es\ matching.
  Notice that in $H'$ the layer-$0$ agents both have degree two.
  Thus we may use the algorithm for Case~3 below.

  Analogously, there is perfect \es\ matching $M$ of type~(b) if and only if the hourglass obtained from $H$ by removing both $\{u_0, u_{h-1}\}$ and $\{w_{0}, w_{h-1}\}$ has a perfect \es\ matching.
  Hence, again we may use the algorithm for Case~3 below.
  
  \noindent \textbf{Case 3: $d_{H}(u_0)=2$ or $d_H(w_0)=2$.}
  This means that all perfect and \es{} matchings for~$H$ must match~$u_0$ with $w_0$ or match~$u_0$ and $w_0$ with the agents from layer~$1$.
  In either case, such a matching must match the first layer of unmatched agents within this layer or to agents in a larger layer.
  This pattern will continue until the last but one layer.
  We use the following notions to describe the pattern formally.
  For each $i\in \{0,\ldots, h-1\}$, we say that layer~$i$ is \myemph{horizontally matched} in~$M$ if $\{u_i,w_i\}\in M$.
  For each $i\in [h-1]$,  we say that layers $i-1$ and $i$ are \myemph{cross-matched} in~$M$ if $\{u_{i-1},w_{i}\}, \{u_{i}, w_{i-1}\} \in M$.
  Then, by induction, we observe that for each~$i\in \{0,1,\ldots, h-1\}$,
  layer-$i$ is either horizontally matched or cross-matched.
  To check which match yields a correct answer, we can use dynamic programming.

  The dynamic program aims to fill two tables~$\horitable, \crosstable$, each of size~$O(h)$, which specify for each layer whether there exists a perfect matching that ends with a cross-match or a horizontal match.
  Formally,
  $\horitable[i] = \KwT$ if there exists a perfect and \es{} matching for~$H[\{u_z,w_z\mid 0\le z\le i\}]$ where layer~$i$ is horizontally matched,
  and $\horitable[i] = \KwF$ otherwise.
  Analogously, 
  $\crosstable[i] = \KwT$ if there exists a perfect and \es{} matching for~$H[\{u_z,w_z\mid 0\le z\le i\}]$ where layers~$i-1$ and $i$ are cross-matched,
  and $\crosstable[i] = \KwF$ otherwise.
  Clearly, by definition, there exists a perfect and \es{} matching for the entire hourglass~$H$ that matches the agents from~$H$ among themselves if and only if $\horitable[h-1]$ or $\crosstable[h-1]$ is \KwT.
  Hence, to show our desired statement it suffices to show that we can compute $\horitable[h-1]$ and $\crosstable[h-1]$ correctly in $O(h)$~time.

  Next, we describe how to fill the tables.
  First, we fill the table for layer $i=0$, i.e., $\horitable[0]$ and $\crosstable[0]$.
  The only possibility for a perfect and exchange-stable matching for layer~$0$ is a horizontal match, i.e., $\crosstable[0]\coloneqq \KwF$ and $\horitable[0]\coloneqq \KwT$.
  For $i=1$, we have two possible perfect matchings such that agents from layer up to~$1$ are matched among themselves, and thus can determine $\horitable[1]$ and $\crosstable[1]$ in constant time.

  Now, to compute the remaining entries of tables~$\horitable$ and $\crosstable$, we claim that
  for all $i\in \{2,\dots,h-1\}$, the following holds: %
  \begin{align}
    \horitable[i] & \coloneqq \crosstable[i-1] \vee \big(\horitable[i-1] \wedge \tau_h^{i-1, i}\big), \label{eq:hourg2}\\
    \crosstable[i] & \coloneqq \big(\crosstable[i-2] \vee \horitable[i-2]\big) \wedge \tau_c^{i-1, i}. \label{eq:hourg1}
  \end{align}
  Herein, $\tau_c^{i - 1, i}$ is set $\KwT$ if and only if cross-matching layers $i-1$ and $i$ is \es for~$H[\{u_{i-1}, w_{i-1},u_{i},w_i\}]$.
  Analogously, $\tau_h^{i-1, i}$ is set $\KwT$ if and only if horizontally matching layers $i-1$ and $i$ respectively is \es{} for~$H[\{u_{i-1}, w_{i-1},u_{i},w_i\}]$.

  It remains to prove that the equations~\eqref{eq:hourg2}--\eqref{eq:hourg1} are correct.

  We first prove equation~\ref{eq:hourg2}.
  ``$\horitable[i] \Rightarrow \crosstable[i-1] \vee (\horitable[i-1] \wedge \tau_h^{i-1, i})$'':
  Let $M$ be a matching witnessing that $\horitable[i] = \KwT$.
  If $M$ cross-matches layers $i - 2$ and $i - 1$, then $\crosstable[i-1]=\KwT$.
  Otherwise, $M$ matches layer~$i - 1$ horizontally.
  Thus, $\horitable[i-1] \wedge \tau_h^{i-1, i}=\KwT$.

  ``$\horitable[i] \Leftarrow \crosstable[i-1] \vee (\horitable[i-1] \wedge \tau_h^{i-1, i})$'':
  If $\crosstable[i-1]=\KwT$, then there is a perfect and \es\ matching $M$ for layers 0 to $i - 1$ that cross-matches layers~$i - 2$ and $i - 1$.
  Extending $M$ to layer~$i$ by matching layer~$i$ horizontally yields a perfect and \es\ matching for layers up to~$i$ because of the following:
  \begin{compactitem}[--]
    \item $u_{i}$ (resp.\ $w_i$) does not envy~$u_{i-1}$ (resp.\ $w_{i-1}$).
    \item $u_{i-2}$ (resp.\ $w_{i-2}$) does not envy~$u_i$ (resp.\ $w_i$).
  \end{compactitem}
  If $\crosstable[i-1]=\KwF$, $\horitable[i-1] \wedge \tau_h^{i-1, i}=\KwT$.
  That is, there is a perfect matching~$M$ for layers $0$ to~$i - 1$ which matches layer~$i - 1$ horizontally and, since $\tau_h^{i-1, i}=\KwT$, extending $M$ to layer~$i$ by matching layer~$i$ horizontally does not incur an \ebp.

  Now we show equation~\eqref{eq:hourg1}.
  ``$\crosstable[i]\Rightarrow (\crosstable[i-2] \vee \horitable[i-2]) \wedge \tau_c^{i-1, i}$'':
  Consider a matching~$M$ for layers $0$ to $i$ witnessing $\crosstable[i] = \KwT$.
  Since $M$ is \es\ and layers~$i$ and $i -1$ are cross-matched by~$M$,
  we have $\tau_c^{i-1, i} = \KwT$.
  Clearly, either layer~$i - 2$ is horizontally matched or layers~$i -3$ and $i - 2$ are cross-matched by $M$
  and hence, $\crosstable[i-2]=\KwT$ or $\horitable[i-2]=\KwT$.

  ``$\crosstable[i] \Leftarrow (\crosstable[i-2] \vee \horitable[i-2]) \wedge \tau_c^{i-1, i}$'':
  Since $\crosstable[i-2]\vee \horitable[i-2]=\KwT$,
  there is a perfect and \es\ matching $M$ for layers up to $i - 2$.
  Since $\tau_c^{i-1, i}$ is true, cross-matching layers $i -1$ and $i$ is \es{} for $H[\{u_{i-1}, w_{i-1}, u_i, w_i\}]$.
  Observe that adding to $M$ the two pairs~$\{u_{i-1},w_{i}\}$ and $\{u_i,w_{i-1}\}$ yields an \es\ matching for layers $0$ to $i$ because neither $u_{i - 2}$ nor $w_{i - 2}$ envy $u_{i - 1}$ or $w_{i - 1}$, respectively, and neither $u_i$ nor $w_i$ envies $u_{i-2}$ or $w_{i-2}$, respectively.

  Finally, it is straight-forward to see that the table entries can be computed in $O(h)$~time by using the just proved equations~\eqref{eq:hourg2}--\eqref{eq:hourg1}.
\end{proof}

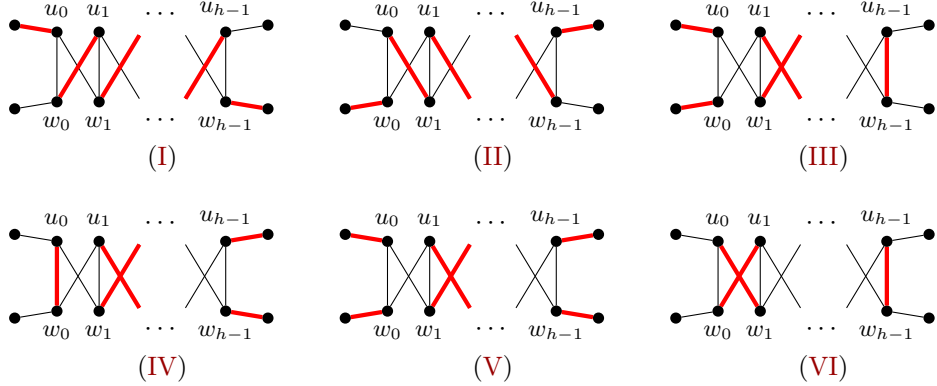
\begin{figure}[t!]
  \begin{center}
  \begin{tikzpicture}[scale=0.925]
    \def \xs {-.3}
    \def \ys {-.5}
   
     \begin{scope}[]
      \sixcat
      \foreach \i / \j in {u00/u0,w4/w40,u1/w0,w1/u2,w3/u4} {
        \draw[red,ultra thick] (\i) edge[] (\j);
      }
      \path (w2) -- node [below= 3ex] {\eqref{v0+wh-1}} (w3);
    \end{scope}

    \begin{scope}[xshift=4.7cm]
      \sixcat
      \foreach \i / \j in {u40/u4,w00/w0,w1/u0,u1/w2,u3/w4} {
        \draw[red,ultra thick] (\i) edge[] (\j);
      }
      \path (w2) -- node [below= 3ex] {\eqref{vh-1+w0}} (w3);
    \end{scope}

    \begin{scope}[xshift=9.4cm]
      \sixcat
      \foreach \i / \j in {u00/u0,w0/w00,u1/w2,u2/w1,u4/w4} {
        \draw[red,ultra thick] (\i) edge[] (\j);
      }
      \path (w2) -- node [below= 3ex] {\eqref{v0+w0}} (w3);
    \end{scope}

    \begin{scope}[yshift=-3cm]
      \sixcat
      \foreach \i / \j in {u40/u4,w40/w4,u0/w0,u1/w2,u2/w1} {
        \draw[red,ultra thick] (\i) edge[] (\j);
      }
      \path (w2) -- node [below= 3ex] {\eqref{vh-1+wh-1}} (w3);
    \end{scope}
    \begin{scope}[yshift=-3cm, xshift=4.7cm]
      \sixcat
      \foreach \i / \j in {u40/u4,w40/w4,u0/u00,w0/w00,u1/w2,u2/w1} {
        \draw[red,ultra thick] (\i) edge[] (\j);
      }
      \path (w2) -- node [below= 3ex] {\eqref{all-out}} (w3);
    \end{scope}
    \begin{scope}[yshift=-3cm, xshift=9.4cm]
      \sixcat
     \foreach \i / \j in {u0/w1,w0/u1,w4/u4} {
       \draw[red,ultra thick] (\i) edge[] (\j);
     }

     \path (w2) -- node [below= 3ex] {\eqref{all-in}} (w3);
   \end{scope}

  \end{tikzpicture}
\end{center}
\caption{Six categories of perfect matchings for an hourglass (see \cref{lem:6_ways_hourglass}). Notice that the edges which are incident to some vertices without labels may not belong to the hourglass. 
  }\label{fig:6-categories}
\end{figure}

\subsubsection{Overview and solution structure}
We now give an overview of how the FPT algorithm works.
Exchange-blocking pairs form only inside hourglasses (Observation~\ref{sc:md3_ebphg}).
Therefore, any perfect matching which is \es{} for all hourglasses is \es{}.
The bound on the preference list length implies that a maximal hourglass $H$ of height $h$ has at most four agents ($u_0$, $w_0$, $u_{h-1}$, and $w_{h-1}$) that might have a partner outside of $H$.
Each of these four vertices can either be matched to a partner inside or outside the hourglass.
\Cref{lem:6_ways_hourglass} below shows that there are only six different ways a perfect matching can assign partners outside the hourglass to these four vertices.
Note that these six possibilities also include the situation that a matching might not assign partners outside the hourglass to these four vertices.
We check each of these possibilities to see if they can be part of an exchange-stable matching on the hourglass.
Finally, we consider all combinations of possible matchings for different hourglasses.
If one of these combination is an \es{} matching for all hourglasses and can be extended to form a perfect matching for the rest of the graph, then we return YES. Otherwise, we return NO.

We now observe that there are only six categories for a perfect matching to match the agents of an hourglass.
We use the following terminology.
Let $H$ be a hourglass in $G(\ppp)$ with~$V(H)=\{u_i,w_i\mid 0\le i \le h-1\}$, and let $M$ be a matching for $\ppp$.
We say that an agent $w_i \in V(H)$ (resp.\ $u_i \in V(H)$) is \myemph{matched straight} if (i) $1 \le i \le h - 2$ and $M(w_i) \in \{u_{i - 1}, u_i, u_{i + 1}\}$, or (ii) $i = 0$ and $M(w_i) \in \{u_i, u_{i + 1}\}$, or (iii) $i = h - 1$ and $M(w_i) \in \{u_{i - 1}, u_{i}\}$ (resp.\ $i > 0$ and $M(u_i) \in \{w_{i - 1}, w_i, w_{i + 1}\}$, or $i = 0$ and $M(u_i) \in \{w_i, w_{i + 1}\}$, or $i = h - 1$ and $M(w_i) \in \{u_{i - 1}, u_{i}\}$).
Otherwise we say that $w_i \in V(H)$ (resp.\ $u_i \in V(H)$) is \myemph{matched askew}.
Note that if an agent in $V(H)$ is matched askew, then it is either $u_0$, $u_{h - 1}$, $w_0$, or $w_{h - 1}$ and the partner is either outside of $V(H)$ or on the other side of the hourglass.
Furthermore, if an agent is matched askew, then its partner is uniquely determined.

\begin{lemma}\label{lem:6_ways_hourglass}
  Let $H$ be a height-$h$ hourglass with~$V(H)=\{u_i,w_i\mid 0\le i \le h-1\}$.
  Then, every perfect matching~$M$ for~$\ppp$ falls into exactly one of the following six categories (see \cref{fig:6-categories}):
  \begin{compactenum}[(I)]
  \item\label{v0+wh-1}
    Vertices $u_0$ and $w_{h - 1}$ are matched askew and all other vertices of $H$ are matched straight.
  \item\label{vh-1+w0}
    Vertices $w_0$ and $u_{h - 1}$ are matched askew and all other vertices of $H$ are matched straight.
  \item\label{v0+w0}
    Vertices $u_0$ and $w_0$ are matched askew and all other vertices of $H$ are matched straight.
  \item\label{vh-1+wh-1}
    Vertices $u_{h - 1}$ and $w_{h - 1}$ are matched askew and all other vertices of $H$ are matched straight.
  \item\label{all-out}
    Vertices $u_0$, $w_0$, $u_{h - 1}$, and $w_{h - 1}$ are matched askew and all other vertices of $H$ are matched straight.
  \item\label{all-in}
    All vertices in $H$ are matched straight.
  \end{compactenum}
\end{lemma}
Below we also say that $M$ \emph{belongs to category $C \in \{\eqref{v0+wh-1}, \ldots, \eqref{all-in}\}$} with respect to $H$ if it it satisfies point $C$ above.

\begin{proof}[\cref{lem:6_ways_hourglass}]
  Consider a specific subgraph~$H'\subseteq H$ in which the agents from layer-$0$ and layer-$h-1$
  have degree two and the remaining layers has degree three.
  Formally, $V(H')=V(H)$ and $E(H')=\{\{u_i,w_i\}\mid 0\le i\le h-1\}\cup \{\{u_i,w_{i+1}\}, \{u_{i+1}, w_{i}\}\mid 0\le i \le h-2\}$.
  Notice that $H'$ is a bipartite graph on two equal-size subsets.
  Hence, under~$M$ an even number,~$\nu$, of agents from~$V(H)$ are matched with edges not contained in~$H'$.
  Note that each such agent is matched askew (and each agent that is matched askew is counted in $\nu$).
  Furthermore, each such agent is either $u_0$, $w_0$, $u_{h - 1}$, or $w_{h - 1}$, hence $\nu \in \{0, 2, 4\}$.

  If $\nu = 0$, then $M$ corresponds to category~\eqref{all-in}.

  If $\nu = 4$, then $M$ corresponds to category~\eqref{all-out}.

  If $\nu = 2$, observe that it is not possible that both $u_0$ and $u_{h - 1}$ are matched with edges outside of~$H'$ as otherwise it would not be possible to match all remaining agents with edges inside $H'$.
  Similarly, it is not possible that both $w_0$ and $w_{h - 1}$ are matched with edges outside of~$H'$.
  Hence, the agents matched outside of $H'$ are either $u_0$ and $w_{h - 1}$ (category~\eqref{v0+wh-1}), $u_{h - 1}$ and $w_0$ (category~\eqref{vh-1+w0}), $u_0$ and $w_{0}$ (category~\eqref{v0+w0}), or $u_{h - 1}$ and $w_{h - 1}$ (category~\eqref{vh-1+wh-1}).
\end{proof}

\subsubsection{Checking for eligible matchings for hourglasses}
Next, we show that, given a height-$h$ hourglass~$H$ and a category~$C \in \{\eqref{v0+wh-1}, \ldots, \eqref{all-in}\}$, we can check in~$O(h)$~time whether there exists an \es{} matching for~$H$ which belongs to category~$i$ (see \cref{lem:6_ways_hourglass} and \cref{fig:6-categories}).

\begin{lemma}
  \label{lem:md3_linearTime}
  Let $H$ be a maximal hourglass in $G(\ppp)$ of height $h$ with $h \geq 5$ and let $C \in \{\eqref{v0+wh-1}, \ldots, \eqref{all-in}\}$.
  In time $O(h)$ it is possible to decide whether there is a matching for $\ppp$ that matches all agents in $H$, is \es\ for $H$ (i.e.\ does not contain an \ebp\ in $V(H)$), and belongs to category~$C$. 
\end{lemma}
\begin{proof}
  We consider each category individually.
  Consider first the case $C = \eqref{v0+wh-1}$.
  Let $M$ be a matching that matches all agents in~$H$.
  Observe that $M(w_0) = u_1$.
  Hence, $M(w_1) = u_2$, $M(w_2) = u_3$, and so forth.
  Since $M$ is perfect, by \cref{lm:coal_means_cycle}, each \ebp{} induces a length-$4$ alternating cycle wrt.~$M$.
  However, since  $h\ge 5$, every alternating cycle wrt.~$M$ containing an edge~$\{x,M(x)\}$ with $x\in V(H)$ has length at least six.
  So no two agents from $V(H)$ form an \ebp, that is, $M$ is \es\ for~$H$.
  Thus, there is always a matching that matches all agents in $H$ and is \es\ for~$H$.

  The case $C = \eqref{vh-1+w0}$ is analogous to $C = \eqref{v0+wh-1}$ and omitted.

  The case $C = \eqref{all-in}$ follows from \cref{lem:dp-d=3=hg}.

  For $C = \eqref{v0+w0}, \eqref{vh-1+wh-1}, \eqref{all-out}$, we show that checking for an \es{} matching in category~$C$ can be reduced to checking whether there exists a perfect and \es\ matching (i.e.\ in category~\eqref{all-in}) for the smaller hourglasses~$H'$ obtained by ignoring the agents from layer-$0$ (category~\eqref{v0+w0}), layer-$(h-1)$ (category~\eqref{vh-1+wh-1}), or both (category~\eqref{all-out}).
  The statement then follows from \cref{lem:dp-d=3=hg}.

  Consider $C = \eqref{v0+w0}$.
  Define the induced subgraph~$H'\coloneqq H[\{u_i,w_i\mid i\in [h-1]\}]$.
  We show that there is an \es\ matching that matches all agents of $H$ and belongs to category~\eqref{v0+w0} if and only if there is a perfect \es\ matching for $H'$.

  For the forward direction, assume that $M$ is an \es{} matching for~$H$ and falls in category~\eqref{v0+w0}. 
  Let $M'$ be the matching obtained from~$M$ by removing the pairs matching the agents from~$\{v_0,w_0\}$.
  By the definition of category~\eqref{v0+w0}, $M'$ is perfect for~$H'$.
  Since $M'\subseteq M$ and $M$ is \es{} for~$H$, matching~$M'$ is also \es\ for $H'$, as required.

  For the backward direction, assume that $M'$ is a perfect and \es{} matching for~$H'$.
  Without loss of generality, assume that both~$u_0$ and $w_0$ have degree three in~$G(\ppp)$ as otherwise no \es{} matching in category~\eqref{v0+w0} exists.
  Let $x$ and $y$ denote the neighbors of $u_0$ and $w_0$, respectively, outside of~$H$.
  Notice that such agents must exist and $x\neq y$ as otherwise no matching in category~\eqref{v0+w0} exists.
  Now, we claim that matching~$M$ derived from~$M'$ by defining $M\coloneqq M'\cup \{\{u_0,x\}, \{w_0,y\}\}$ is \es{} for~$H$.
  Suppose, for the sake of contradiction, that $M$ is not \es{} for~$H$, and let $(a,b)$ denote an \ebp{} with $a,b\in V(H)$.
  Clearly, at least one of $\{a,b\}$, say~$a$, is from layer-$0$ of $H$ as otherwise $(a,b)$ would also be exchange-blocking~$H'$.
  By symmetry, suppose that $a=u_0$ without loss of generality.
  By \cref{sc:md3_ebphg}, there must be an hourglass which contains~$(u_0,b)$, a contradiction to~$H$ being maximal.

  The case $C = \eqref{vh-1+wh-1}$ is analogous to $C = \eqref{v0+w0}$ and omitted.

  Consider $C = \eqref{all-out}$.
  If all edges incident with the agents $u_0$, $u_{h - 1}$, $w_0$, and $w_{h - 1}$ are contained in $V(H)$, then $H$ is a connected component in $G(\ppp)$ and thus, checking for an \es\ matching that matches all agents of $H$ is equivalent to checking for a perfect \es\ matching for $H$ and can be done in the required time using \cref{lem:dp-d=3=hg}.
  If not all edges incident with the agents $u_0$, $u_{h - 1}$, $w_0$, and $w_{h - 1}$ are contained in $V(H)$, then observe that if two agents $u_0$, $u_{h - 1}$, $w_0$, and $w_{h - 1}$ share a neighbor (in $G(\ppp)$) outside of $H$, then there is no matching that matches all agents of $H$ and belongs to category~\eqref{all-out}.
  We thus from now on assume that no two agents $u_0$, $u_{h - 1}$, $w_0$, and $w_{h - 1}$ share a neighbor outside of~$H$.
  Define the induced subgraph $H'\coloneqq H[\{u_i,w_i\mid i\in [h-2]\}]$.
  We claim that there is \es\ matching that matches all agents of $H$ and belongs to category~\eqref{v0+w0} if and only if there is a perfect \es\ matching for $H'$.
  If this is true, then the proof is finished in this case by using the algorithm for \cref{lem:dp-d=3=hg}.

  The forward direction is analogous to the one for $C = \eqref{v0+w0}$ and omitted.

  For the backward direction, assume that $M'$ is a perfect and \es{} matching for~$H'$.
  Let $M$ be the matching that is obtained from $M'$ by matching each of the agents $u_0$, $u_{h - 1}$, $w_0$, and $w_{h - 1}$ such that they are matched askew (note that there is a unique way to do so).
  Clearly, $M$ matches all agents of $H$ and belongs to category~\eqref{all-out}.
  We claim that $M$ is \es\ for~$H$.
  If none of the agents $u_0$, $u_{h - 1}$, $w_0$, and $w_{h - 1}$ are matched inside $H$ then $M$ is \es\ for $H$ by the same argument as for $C = \eqref{v0+w0}$.
  Hence, assume that two of these agents are matched inside~$H$.
  To get a contradiction, assume that $(a, b)$ is an \ebp\ in~$H$.
  Since $M'$ is exchange-stable for $H'$, at least one agent, say $a$, is in layer~0 or layer~$h - 1$ of $H$.
  Note that $a$ is not matched with a partner $x$ outside of $H$, since no agent in $H$ finds $x$ acceptable (recall that no two agents $u_0$, $u_{h - 1}$, $w_0$, and $w_{h - 1}$ share a neighbor outside of~$H$).
  Hence, $M(a) \in V(H)$ and one of $a, M(a)$ is in layer~0 and the other in layer~$h - 1$.
  However, since $h \geq 5$, no neighbor of a neighbor $a$ finds $M(a)$ acceptable except for~$a$ itself.
  This is a contradiction to $(a, b)$ being an \ebp.
  Thus, $M$ is \es\ for $H$, as required.

  Consider $C = \eqref{all-in}$.
  To check for an hourglass in this category, we may employ the algorithm from \cref{lem:dp-d=3=hg} for the hourglass obtained from $H$ by removing all edges with one endpoint in $\{u_0, w_0\}$ and one endpoint in $\{u_{h - 1}, w_{h - 1}\}$.
\end{proof}

\subsubsection{Overlapping hourglasses}
The above results allow us to deal with hourglasses that do not overlap with any other hourglass.
However, it is not always the case that hourglasses do not overlap.
We now show that if two hourglasses overlap, then their size is bounded.
This then allows us to specify the interaction of sets of overlapping hourglasses with the rest of the graph in a similar way as \cref{lem:6_ways_hourglass}.

We first observe the interaction between two maximal hourglasses.
\begin{lemma}\label{lem:HG-intersection}
  Let $H_1$ and $H_2$ be two different maximal hourglasses of profile~$\ppp$ such that $H_1$ has height~$h_1$
  and $H_2$ has height~$h_2$.
  Then, the following holds.
  \begin{compactenum}[(i)]
    \item\label{intersect>=4} If $h_1\ge 4$, then for each two agents~$u$ and $w$ in the same layer of $H_1$, it holds that $u\in V(H_2)$ if and only if $w\in V(H_2)$.
    \item\label{intersect=23} If $V(H_1)\cap V(H_2)\neq \emptyset$, then $h_1=h_1=2$ or $h_1= h_2= 3$.
    \item\label{intersect=2} If $V(H_1)\cap V(H_2)\neq \emptyset$ and $h_1=2$, then $h_2=2$ and $|V(H_1)\cap V(H_2)|=3$.
    \item\label{intersect=3} If $V(H_1)\cap V(H_2)\neq \emptyset$ and $h_1=3$, then $h_2=3$ and $|V(H_1)\cap V(H_2)|=5$.
  \end{compactenum}
\end{lemma}

\begin{proof}
  Let $\ppp$ be a preference profile and $H_1$ and $H_2$ two maximal hourglasses in~$G(\ppp)$ with heights~$h_1$ and $h_2$, respectively.
  Let $V(H_1)=\{u_i,w_i\mid 0\le i \le h_1-1\}$.

  For Statement~\eqref{intersect>=4}, assume that $h_1\ge 4$.
  Let $u_i$ and $w_i$ be two agents which are in layer~$i$ in~$H_1$, $i\in \{0,1,\dots,h_1-1\}$. 
  By symmetry, if suffices to show that if $u_i\notin V(H_2)$ then $w_i\notin V(H_2)$.
  Suppose, for the sake of contradiction, that $u_i \notin V(H_2)$ but $w_i\in V(H_2)$.
  By the definition of hourglasses, it follows that $w_i$ is in either the first or the last layer of~$H_2$ (since $w$ has degree two in~$H_2$).
  Without loss of generality, assume that $w_i$ is in layer~$0$ in~$H_2$.
  We distinguish between two cases.

  Case 1: $i=0$ or $i=h_1-1$. We only consider the case when $i=0$ since the other case is symmetric (by reversing the indices).
  Since $i=0$, meaning that $w_i=w_0$ is in layer-$0$ in~$H_2$, we have that $w_0$ has three neighbors~$u_0$, $u_1$, and $x$ for some agent~$x\in V(G(\ppp))\setminus \{u_j, w_j \mid 0\le j \le 2\}$ (note that $h\ge 4$) such that $u_1, x\in V(H_2)$ (recall that $u_0\notin V(H_2)$). 
  Moreover, since $w_0$ is in layer-$0$ in~$H_2$, it follows that $w_0$ is with either $u_1$ or $x$ in the same layer in~$H_2$.
  
  If $w_0$ and $u_1$ are in the same layer in~$H_2$, then by the maximality of~$H_2$,
  it follows that $w_1\in V(H_2)$ (as otherwise we could enlarge~$H_2$ by adding to it $u_0$ and $w_1$).
  Notice that $w_1$ and $u_0$ are adjacent in $G(\ppp)$ but $u_0\notin V(H_2)$.
  Since $h_1\ge 4$, by the definition of hourglasses, this means that the third neighbor of~$w_1$, namely $u_2$, is also in~$H_2$
  such that $w_1$ and $u_2$ are in the last layer in~$H_2$ (recall that $w_0$ and $u_1$ are in layer-$0$ in $H_2$,
  and $w_1$ and $u_1$ are adjacent).
  This implies that $H_2$ has two layers such that $\{w_0,u_2\}\in E(H_2)$ (note that $\{u_1,u_2\}\notin E(G(\ppp))$ since $u_1$ is adjacent to~$w_0,w_1,w_2$ already).
  However, this also means that $x=u_2$ such that $V(H_2)=\{w_0,u_1,w_1,u_2\}\subsetneq V(H_1)$, a contradiction to $H_2$ being maximal.

  If $w_0$ and $x$ are in the same layer in $H_2$, then $u_1$ is with either $w_1$ or $w_2$ in the same layer in $H_2$.
  If $u_1$ and $w_1$ are in the same layer in~$H_2$, then since $u_0\notin V(H_2)$, this further implies that $w_1$ and hence $u_1$ are in the last layer in~$H_2$.
  Hence, $h_2=2$ and $\{w_1,x\}\in E(H_2)$.
  This implies that $x=u_2$ since $w_1$ is adjacent to $u_0$, $u_1$, and $u_2$.
  In other words, $\{w_0,u_2,u_1,w_1\}V(H_2)\subsetneq V(H_1)$, a contradiction to $H_2$ being maximal.

  If $u_1$ and $w_2$ are in the same layer in~$H_2$, then $u_1$ and $w_2$ are not in the last layer in $H_2$ as otherwise we could add $\{w_1,u_2\}$ to $H_2$ to extend the height by one, a contradiction.
  Since $u_1$ and $w_2$ are not in the last layer in~$H_2$, this means that $u_1$ has degree three in~$H_2$.
  Moreover, $u_1$ and $w_2$ are both in layer-$1$ in~$H_2$; recall that $\{u_1,w_0\}\in E(H_2)$ and $w_0$ is in layer-$0$ in~$H_2$.
  This implies that $\{w_0,x\}\in E(H_2)$ such that $x\in \{u_2,u_3\}$.
  Since $h_1\ge 4$ we infer that $x\neq u_2$ and hence $x=u_3$.
  Again, since $u_1$ and $w_2$ are in layer-$1$ but not the last layer in~$H_2$, we also infer that $u_2$ and $w_1$ are in layer-$2$ in~$H_2$.
  Since $u_0\notin V(H_2)$, it follows that $w_1$ is in the last layer in~$H_2$ and hence $h_2=3$.
  This means that $V(H_2)=\{u_3, w_0, u_1, w_2, u_2, w_1\}\subsetneq V(H_1)$, a contradiction to $H_2$ being maximal.  

  Case 2: $i\in [h_1-2]$. This means that $w_i$ is adjacent to~$u_{i-1}, u_{i}$, and $u_{i+1}$ in $H_1$ such that $u_{i-1}$
  and $u_{i+1}$ are both in~$V(H_2)$.
  Furthermore, $w_i$ is with either $u_{i-1}$ or $u_{i+1}$ in layer-$0$ or the last layer in $H_2$.
  We only consider the case that that $w_i$ and $u_{i-1}$ are in layer-$0$ in~$H_2$ as the other three cases are symmetric and analogous. %
  
  If $w_i$ and $u_{i-1}$ are in layer-$0$ in~$H_2$, then $w_{i-1}$ must also be in~$V(H_2)$ as otherwise we could enlarge $H_2$ by one more layer by adding to it $u_{i}$ and $w_{i-1}$, a contradiction.
  Now, since $w_{i-1}\in V(H_2)$ but $v_i\notin V(H_2)$, it follows that $w_{i-1}$ is in the last layer in~$H_2$.
  Similarly as above, we infer that $h_2=2$ and hence, $V(H_2)=\{u_{i-1}, w_{i}, u_{i+1},w_{i-1}\}\subsetneq V(H_1)$, a contradiction to $H_2$ being maximal.

  Overall, we have shown that $u_i$ and $w_i$ are either both in $H_2$ or both not in~$H_2$.

  For Statement~\eqref{intersect=23} assume that $V(H_1)\cap V(H_2)\neq \emptyset$.
  We first show that $h_1,h_2\le 3$.
  To achieve this, it suffices to show that $h_1 \le 3$ since $H_1$ and $H_2$ are symmetric.
  Suppose, for the sake of contradiction, that $h_1 \ge 4$.
  Since $H_1$ and $H_2$ are two different hourglasses, by Statement~\eqref{intersect>=4},
  there are two layers $i$ and $j$ such that
  $u_i$ and $w_i$ are in $V(H_2)$ while $u_j$ and $w_j$ are not.
  Without loss of generality, assume that $i < j$ (the case that $j<i$ can be shown by reversing the indices).
  Then, there must also be a layer~$i$ such that $u_i$ and $w_i$ are in~$V(H_2)$ while $u_{i+1}$ and $w_{i+1}$ are not.
  This means that $u_i$ and $w_i$ are in layer-$0$ or layer-$(h-1)$ in $H_2$.
  If $u_i$ and $w_i$ are in different layers in $H_2$, then it must hold that $h_2=2$ since both $u_i$ and $w_i$ have degree two in $H_2$.
  However, we could enlarge the height of~$H_2$ by one by adding to it the two agents~$u_{i+1}$ and $w_{i+1}$, a contradiction.
  Similarly, if $u_i$ and $w_i$ are in the same layer in~$H_2$, then we could as well enlarge the height of~$H_2$ by one by adding to it the two agents~$u_{i+1}$ and $w_{i+1}$, a contradiction.

  Now, to show the statement, it remains to show that $h_1=h_2$.
  By our above reasoning, suppose, for the sake of contradiction, that $h_1=2$ while $h_2=3$.
  Then, by the maximality of $H_1$ and by the definition of hourglasses, either $2$ or $3$ agents from~$V(H_1)$ are contained in $V(H_2)$.
  If $|V(H_1)\cap V(H_2)|=2$, then without loss of generality, assume that $V(H_1)\cap V(H_2) =\{u_0,w_0\}$.
  Then, this also means that $u_0$ (resp.\ $w_0$) is in the first or the last layer in~$H_2$.
  It cannot be the case that $u_0$ and $w_0$ are in the same layer in~$H_2$ as otherwise we could enlarge $H_2$ by adding to it the pair~$\{u_1,w_1\}$, a contradiction to $H_2$ being maximal.
  If $u_0$ and $w_0$ are in different layers, then $H_2$ has only two layers, a contradiction too.

  Now, it remains to consider the case with $|V(H_1)\cap V(H_2)|=3$.
  Without loss of generality, assume that $V(H_1)\cap V(H_2)=\{u_0,w_0,u_1\}$.
  This means that $u_0$ (resp.\ $u_1$) is in the first or the last layer in~$H_2$ since they both have degree two in~$H_2$.
  Agents~$u_0$ and $u_1$ cannot be in the same layer in~$H_2$ as otherwise $\{u_0,u_1\}\in E(H_2)$ such that $u_0$ or $u_1$ would have degree four in~$G(\ppp)$ (since $u_0$ or $u_1$ would need to be adjacent to the agent which is in the same layer as~$w_0$).
  Hence, without loss of generality, assume that $u_0$ and $u_1$ are in layer-$0$ and layer-$2$ in~$H_2$.  
  Moreover, $w_0$ must be in layer-$1$ in~$H_2$ as otherwise either $u_0$ or $w_0$ would have degree at least four in~$G(\ppp )$, a contradiction.
  This means that there are two other agents~$x,y\notin V(H_1)$ such that $u_0$ and $x$ are layer-$0$ while $w_0$ and $y$ are in layer-$2$ in~$H_2$, respectively.
  However, this implies that we could enlarge $H_1$ by adding to it agents~$x$ and $y$, a contradiction.  
  This completes the proof for Statement~\eqref{intersect=23}.

  For Statement~\eqref{intersect=2}, assume that $V(H_1)\cap V(H_2)\neq \emptyset$ and $V(H_1)=\{u_0,w_0,u_1,w_1\}$.
  By Statement~\eqref{intersect=23}, we know that $h_2=2$ such that $|V(H_1)\cap V(H_2)| \in \{2,3\}$.
  Using a reasoning similar to the one given for Statement~\eqref{intersect=23} we can infer that $|V(H_1)\cap V(H_2)|=3$.

  Finally, for Statement~\eqref{intersect=3}, assume that $V(H_1)\cap V(H_2)\neq \emptyset$ and $V(H_1)=\{u_0,w_0,u_1,w_1\,u_2,w_2\}$.
  By Statement~\eqref{intersect=23}, we know that $h_2=3$. %

  Since $H_1$ and $H_2$ are two different maximal hourglasses, let $u_i$, $0\le i\le 2$, be an agent which does not appear in~$V(H_2)$.
  Notice that we can assume that $i\in \{0,1\}$ as the case with $i=2$ is symmetrical to the case~$i=0$.
  We distinguish between four subcases.
  
  If $i=0$ and $w_0\in V(H_2)$, then there exists another neighboring agent~$x\in V(H_2)\setminus \{u_1,u_0\}$ such that $\{x,w_0\}\in E(H_2)$, $u_{1}\in V(H_2)$, and $w_0$ is in layer-$0$ or layer-$2$ in $H_2$.
  This also means that $w_0$ is with either $x$ or $u_1$ in the same layer in~$H_2$.
  If $w_0$ and $x$ are in the same layer in~$H_2$, then $u_1$ is in layer-$1$ in~$H_2$ so that $w_1$ and $w_2$ are also in~$V(H_2)$.
  Since $u_0\notin V(H_2)$ it follows that $w_1$ is in layer-$0$ or layer-$2$ in~$H_2$. %
  Hence, $w_2$ and $u_1$ are in layer-$2$ in~$H_2$.
  Moreover, $u_2\in V(H_2)$ since $w_1$ is only adjacent to~$u_0,u_1$, and $u_2$ but $u_0\notin V(H_2)$.
  Altogether, we infer that $V(H_2)=\{u_1,u_2,w_0,w_1,w_2,x\}$, meaning that $|V(H_2)\cap V(H_1)|=5$.

  If $i=0$ and $w_0\notin V(H_2)$, then either both $u_1$ and $w_1$ are in $V(H_2)$ or both are not.
  If both $u_1$ and $w_1$ are in $V(H_2)$, then they must be in the same first or last layer as otherwise $h_2=2$.
  But then, we could enlarge $H_2$ by adding to it $\{u_0,w_0\}$, a contradiction.
  If both $u_1$ and $w_1$ are not in~$V(H_2)$, then analogously, both $u_2$ and $w_2$ are in $V(H_2)$ or both are not.
  In either case, we could obtain the contradiction that we could enlarge $H_2$.

  If $i=1$ and $w_1\in V(H_2)$, then it must hold that $u_0,u_2\in V(H_2)$ such that $w_1$ is with either~$u_0$ or $u_2$ in the same first or last layer in~$H_2$.
  Without loss of generality, assume that $w_1$ and $u_0$ are in the same layer in~$H_2$ (the other case is symmetric).
  Then, $u_2$ must be in layer-$1$ in~$H_2$, meaing that $u_2$ has degree three in~$H_2$.
  This implies that there exists an agent~$x\in V(H_2)\setminus \{w_2\}$ such that $\{x,u_2\}\in E(H_2)$.
  Furthermore, we also infer that $w_2\in V(H_2)$.
  Since $u_1\notin V(H_2)$, it follows that $w_2$ in the first layer or the last layer, and hence not with $u_2$ in layer-$1$ in $H_2$.
  Thus, there exists an agent~$y\in V(H_2)\setminus \{u_2\}$ such that $y$ and $w_2$ are in the same layer in~$H_1$.
  Now, if $\{x,y\}\cap V(H_1)=\emptyset$, then we could enlarge $H_1$ by adding to it~$\{x,y\}$ since $\{x, u_2\}, \{y,w_2\}$ are two edges in~$E(G(\ppp))$, a contradiction.
  Hence, $|\{x,y\}\cap V(H_1)|\ge 1$.
  Since $\{u_0,w_1,u_2,w_2\}\subseteq V(H_2)$ and $u_1\notin V(H_2)$.
  We obtain that exactly one of~$\{x,y\}$ is in~$V(H_1)$, i.e., $|V(H_1)\cap V(H_2)|=5$, as required.

  If $i=1$ and $w_1\notin V(H_2)$, then either both $u_0$ and $w_0$ are in $V(H_2)$ or both are not.
  If both $u_0$ and $w_0$ are in $V(H_2)$, then they must be in the same first or last layer as otherwise $h_2=2$.
  But then, we could enlarge $H_2$ by adding to it $\{u_1,w_1\}$, a contradiction.
  If both $u_0$ and $w_0$ are not in~$V(H_2)$, then analogously, both $u_2$ and $w_2$ are in $V(H_2)$ or both are not.
  In either case, we could obtain the contradiction that we could enlarge $H_2$.
\end{proof}

By \cref{lem:HG-intersection} we know that hourglasses of height at least four do not overlap with other hourglasses.
We now show that the number of possible matchings for sets of overlapping hourglasses of height two or three is a small constant.

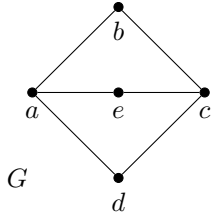
\begin{figure}[t]
  \centering
  \begin{tikzpicture}
    \node[agent,label=below:$e$] (a) at (0,0) {};
    \node[agent,right=of a,label=below:$c$] (b) {}
    edge (a);
    \node[agent,above=of a,label=below:$b$] (c) {}
    edge (b);
    \node[agent,left=of a,label=below:$a$] (d) {}
    edge (a)
    edge (c);
    \node[agent,below=of a,label=below:$d$] (e) {}
    edge (b)
    edge (d);

    \node[left=of e] {$G$};

  \end{tikzpicture}
  \caption{Possible subgraph induced by overlapping maximal hourglasses of height two.}
  \label{fig:overlapping-hourglass-height-two}
\end{figure}

\begin{lemma}\label{lem:overlapping-hourglass-height-two}
  Let $C$ be a collection of at least two maximal hourglasses of height two in $G(\ppp)$ such that the union of their vertex sets induces a connected subgraph of $G(\ppp)$ and no maximal hourglass of height two in $G(\ppp)$ can be added to $C$ maintaining this connectedness property.
  Then, there are at most three matchings that match all vertices in hourglasses in $C$.
\end{lemma}
\begin{proof}
  We observe that each subgraph induced by the vertex sets of hourglasses in $C$ is isomorphic to a supergraph of the graph $G$ shown in \cref{fig:overlapping-hourglass-height-two}.
  Let $C$ contain two hourglasses $H_1$ and $H_2$.
  Let $H_1$ contain the cycle $(a, b, c, d)$.
  Two hourglasses of height two overlap in exactly three vertices by \cref{lem:HG-intersection}\eqref{intersect=2}.
  Without loss of generality, assume $V(H_1) \cap V(H_2) = \{a, b, c\}$.
  Since $G(\ppp)$ has maximum degree~3 and $H_2$ contains a cycle with four vertices, $d$ has two neighbors in $V(H_1) \cap V(H_2)$ that have a common neighbor in $V(H_1) \cap V(H_2)$.
  In other words, $d$'s neighbors are $a$ and $c$.
  Thus, the subgraph induced by $V(H_1) \cup V(H_2)$ is isomorphic to a supergraph of~$G$.

  Now consider another hourglass, $H_3$ such that $V(H_3) \cap V(H_1) \neq \emptyset$.
  By the above, the subgraph induced by $V(H_1) \cup V(H_2)$ is isomorphic to a supergraph of~$G$.
  By symmetry, assume without loss of generality that $V(H_1) = \{a, b, c, d\}$ and $V(H_2) = \{a, b, c, e\}$.
  Since hourglasses overlap in exactly three vertices, \cref{lem:HG-intersection}\eqref{intersect=2} we have $V(H_2) \cap V(H_1) \neq \emptyset$.
  We claim that $V(H_3) \subset V(H_1) \cup V(H_2)$.
  For the sake of a contradiction, assume that $f \in V(H_3) \setminus (V(H_1) \cup V(H_2))$.
  Since $f$ is part of a cycle in $H_3$, it has two neighbors among $V(H_1) \cup V(H_2)$.
  Since $a$ and $c$ have degree three, these two neighbors are among $b, e, d$.
  By symmetry assume $b, d \in V(H_3)$ without loss of generality.
  Notice that $e \notin V(H_3)$ as it would have to be a common neighbor of $b$ and $d$ and hence would have to have degree four.
  Thus, exactly one of $a, c \in V(H_3)$, a contradiction to the fact that $|V(H_1) \cap V(H_3)| = |V(H_2) \cap V(H_3)| = 3$.
  Thus, $V(H_3) \subset V(H_1) \cup V(H_2)$ and thus if $C$ contains at least three hourglasses, its induced subgraph is isomorphic to a supergraph of $G$, as claimed.

  Consider a matching that matches all agents in $C$.
  Since $C$ induces a subgraph isomorphic to $G$, matching $M$ matches $a$ and $c$, and these two have degree three in $G$, two of $b, e, d$ are matched to $a$ and $c$, respectively, and the remaining agent is matched to a uniquely determined agent outside of $G$.
  There are hence at most three possible matchings that match all agents in the hourglasses in $C$.
\end{proof}

Now we turn to hourglasses of height three.

\begin{figure}[t]
  \centering
  \begin{tikzpicture}
    \node[agent,label=below:$a$] (a) at (0,0) {};
    \node[agent,label=below:$b$] (b) at (2,0) {}
    edge (a);
    \node[agent,label=below:$c$] (c) at (4,0) {}
    edge (b);
    \node[agent,label=above:$d$] (d) at (2,0.75) {}
    edge (b);
    \node[agent,label=below:$e$] (e) at (1,1) {}
    edge (a)
    edge (d);
    \node[agent,label=below:$f$] (f) at (3,1) {}
    edge (c)
    edge (d);
    \node[agent,label=below:$g$] (g) at (2,2) {}
    edge (e)
    edge (f);

    \node[left=of e] {$G$};

  \end{tikzpicture}
  \caption{Possible subgraph induced by overlapping maximal hourglasses of height three.}
  \label{fig:overlapping-hourglass-height-three}
\end{figure}
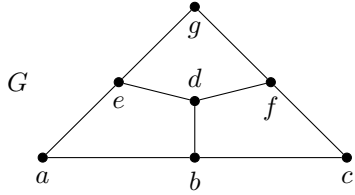

\begin{lemma}\label{lem:overlapping-hourglass-height-three}
  Let $C$ be a collection of at least two maximal hourglasses of height three in $G(\ppp)$ such that the union of their vertex sets induces a connected subgraph of $G(\ppp)$ and no maximal hourglass of height three in $G(\ppp)$ can be added to $C$ maintaining this connectedness property.
  Then, there is a constant number of matchings that match all vertices in hourglasses in $C$.
  Moreover, there are at most three different sets $S$ of vertices that are not contained in hourglasses in $C$ and such that there exists a matching $M$ that matches all vertices in hourglasses in $C$ and such that $S$ is exactly the set of vertices that are not contained in hourglasses in $C$ and matched to vertices in hourglasses in $C$ by~$M$.
\end{lemma}
\begin{proof}
  We observe that each subgraph induced by the vertex sets of hourglasses in $C$ is isomorphic to a supergraph of the graph $G$ shown in \cref{fig:overlapping-hourglass-height-three}.
  Let $C$ contain two hourglasses $H_1$ and $H_2$.
  Let $V(H_1) = \{a, b, c, d, e, f\}$ and let the edges mandated by the definition of hourglasses be as shown in \cref{fig:overlapping-hourglass-height-three}.
  Let thus $a, e$ be in layer~0 and $f, c$ in layer~2 of $H_1$.
  By \cref{lem:HG-intersection}\eqref{intersect=3} we have $|V(H_1) \cap V(H_2)| = 5$, that is, there is exactly one vertex $g \in V(H_2) \setminus V(H_1)$.
  Since the minimum degree is two in $H_2$, vertex $g$ has at least two neighbors in $V(H_1)$.
  Moreover, these vertices must have degree two in $V(H_1)$, that is, $g$ is neighbors with two of $a, e, f, c$.
  Since $g$ is part of a cycle with four vertices in $H_2$, it has two neighbors among these four vertices that have a common neighbor.
  Moreover, that common neighbor must be in $V(H_1)$ because $|V(H_1) \cap V(H_2)| = 5$.
  That is, $g$ is neighbors with either $a, c$ or $e, f$, say $e, f$.
  Thus, $V(H_1) \cup V(H_2)$ induces a subgraph that is isomorphic to a supergraph of $G$.
  Now observe that because $|V(H_1) \cap V(H_2)| = 5$ and maximum degree three we have either $a, b \in V(H_2)$ or $b, c \in V(H_2)$.
  By symmetry, assume without loss of generality that $a, b \in V(H_2)$, that is $V(H_2) = \{a, b, e, d, f, g\}$.

  Now let $H_3$ be a maximal hourglass of height three such that $V(H_3) \cap V(H_1) \neq \emptyset$.
  By \cref{lem:HG-intersection}\eqref{intersect=3} we have $|V(H_3) \cap V(H_1)| = 5$ and thus $V(H_3) \cap V(H_2) \neq \emptyset$, meaning that $|V(H_3) \cap V(H_2)| = 5$.
  We claim that $V(H_3) \subseteq V(H_2) \cup V(H_1)$.
  For a contradiction assume that there exists $h \in V(H_3) \setminus (V(H_2) \cup V(H_1))$.
  By the same argument as for $g$ above and since $e, f$ have degree three already, we get that $h$ is neighbors with $a$ and $c$ and that $a, b, c \in V(H_3)$.
  Since $|V(H_3) \cap V(H_1)| = 5$ exactly two of $e, d, f$ are in $V(H_3)$.
  This is a contradiction to $|V(H_3) \cap V(H_2)| = 5$.
  Thus, indeed $V(H_3) \subseteq V(H_2) \cup V(H_1)$.
  It follows that each subgraph induced by the vertex sets of hourglasses in $C$ is isomorphic to a supergraph of the graph $G$ shown in \cref{fig:overlapping-hourglass-height-three}.

  Now let $M$ be a matching that matches all vertices of hourglasses in $C$.
  It is clear that, because there is a constant number of vertices in $C$ and by maximum degree three, there is a constant number of matchings matching vertices in $C$.
  We claim that exactly one of $a, c, g$ is matched to vertices not in hourglasses in $C$ and all other vertices of $C$ are matched among themselves.
  Note that this implies the statement of the lemma because each of $a, c, g$ has two neighbors in $C$ and so the partner outside (if any) is uniquely determined.
  Indeed, since $d$ has degree three, it is matched to either $e$, $f$, or $b$.
  The remaining two are matched to $a, b$, or $c$.
  Thus the statement holds.
\end{proof}

\subsubsection{Computing hourglasses}
Our FPT algorithm starts by computing the collection of maximal hourglasses.
We now gather the crucial tool to compute this collection.

\SetKwFunction{Fhourglass}{MaxHG}
\begin{algorithm}[t!]
  \PrintSemicolon
  \KwIn{A preference profile~$\ppp$ with agent set~$V$ and with preferences length at most three, and an edge $\{u, w\}$ in $G(\ppp)$.}
  
  \KwOut{A maximal hourglass in~$G(\ppp)$ that contains $\{u, w\}$ or \KwNo if none exists.}
  \SetKwProg{Fn}{}{:}{}

  \SetKwFunction{FMain}{Main}

  \Mycomment{Recursively extend an hourglass~$G'[U\cup W]$ to one more layer}
  \Fn{\Fhourglass{$u$, $w$, $U$, $W$, $G'$}}{
    \If{$\exists x\in N_{G'}(u)\setminus \{w\}$ \KwAnd $y\in N_{G'}(w)\setminus \{u\}$ \KwST $\{x,y\}\in E(G')$}{

      $U\leftarrow U\cup \{y\}$

      $W\leftarrow W\cup \{x\}$

      $G'-x-y$

      \Fhourglass{$x$, $y$, $U$, $W$, $G'$}
    }
  }

  \Fn{\FMain{$u$, $w$}}{
    $G_1 \leftarrow G$;

    $U_1 \leftarrow \emptyset$;
    $W_1 \leftarrow \emptyset$;

    \Mycomment{Try to find a four-cycle and build the first half of a maximal hourglass}
    \label{alg:mh:first-cycle}\If{$\exists a\in N_{G_1}(u)\setminus \{w\}$ \KwAnd $b\in N_{G_1}(w)\setminus \{u\}$ \KwST $\{a,b\}\in E(G')$}{
      $U_1\leftarrow U_1\cup \{u,b\}$

      $W_1\leftarrow W_1\cup \{w,a\}$

      $G_1-\{u,a\}-\{w,b\}$

      \Fhourglass{$a$, $b$, $U_1$, $W_1$, $G_1$}
    }

    \Mycomment{Try to extend to build the second half of the maximal hourglass}

    \label{alg:mh:second-cycle}\If{$\exists a\in N_{G_1}(u)\setminus \{w\}$ \KwAnd $b\in N_{G_1}(w)\setminus \{u\}$ \KwST $\{a,b\}\in E(G')$}{
      
      $U_1\leftarrow U_1\cup \{b\}$

      $W_1\leftarrow W_1\cup \{a\}$

      \Fhourglass{$a$, $b$, $U_1$, $W_1$, $G_1$}
    }

    \lIf{$|U_1| \ge 2$}{%
      \Return $(U_1,W_1)$%
    }
  \Return \KwNo
}
\caption{Algorithm to find a maximal hourglass}\label{alg:maximal-hg}
\end{algorithm}

\begin{lemma}\label{lem:max-hourglass-computed-lintime}
  There is an algorithm~$\mathcal{A}$ such that for each maximal hourglass $H$ in $G(\ppp)$ there is an edge $\{u, w\}$ such that on input of $u, w$ algorithm $\mathcal{A}$ computes $H$ if the height of $H$ is at least three, or computes a set of hourglasses containing $H$ if the height of $H$ is two.
  Algorithm $\mathcal{A}$ runs in $O(1)$ time if there is no maximal hourglass containing $u, w$ and in $O(h)$ time if there is such an hourglass, where $h$ is the hourglass' height.
\end{lemma}
\begin{proof}
  Algorithm $\mathcal{A}$ works as follows.
  First, it computes all maximal hourglasses of height two containing the input edge $\{u, w\}$.
  Observe that this can be done in constant time by enumerating all cycles on four vertices containing $\{u, w\}$, then checking whether they are contained in an hourglass of size three that contains $\{u, w\}$, and if not, reporting them as output.
  (All hourglasses of height three that contain $\{u, w\}$ can be enumerated in constant time since the maximum degree is three.)
  If no hourglasses are found in this step, it runs \cref{alg:maximal-hg} with input~$u, w$.
  
  Let $V(H) = \{u_i, w_i \mid i \in \{0, 1, \ldots, h - 1\}\}$.
  If the height of $H$ is two, then it is clear that $\mathcal{A}$ finds $H$ on input of one of its edges.
  If the height of $H$ is at least three, then we claim that \cref{alg:maximal-hg} finds $H$ on input of $u_1, w_1$.
  Indeed, in this case $a, b$ in \cref{alg:mh:first-cycle} is either $u_0, w_0$ or $u_2, w_2$ and $a, b$ in \cref{alg:mh:second-cycle} is the other pair.
  By a similar argument, \Fhourglass\ will compute the remaining vertices of $H$ in the two calls (note that in each call, the next pair $u_i, w_i$ to add is unique).

  For the running time it only remains to observe that \cref{alg:maximal-hg} runs in $O(h)$ time.
  This follows from the fact that checking for \cref{alg:mh:first-cycle}, \cref{alg:mh:second-cycle}, and the if-condition in the calls of \Fhourglass\ can be done in constant time since the maximum degree is three.
\end{proof}

\subsubsection{FPT algorithm}
We now describe an algorithm which solves \dESMs[3] in $f(6^{\ell}\cdot n\cdot \sqrt{n})$ time, where $\ell$ denotes the number of maximal hourglasses.
We are given an instance~$\ppp$ of \dESRs[3].
We first find a collection of all maximal hourglasses in~$G(\ppp)$ as follows.
We maintain for each agent $v$ a list $L(v)$ of hourglasses of height at most three that contain $v$, we initialize these as empty lists.  
Next, for each edge $\{u, w\}$ in $G(\ppp)$ we run \cref{alg:maximal-hg} with input $u, w$.
If the result is an hourglass of height at least four, we remove its agents from the graph and continue with the next edge.
If the result is an hourglass of height at most three, we save a pointer to this hourglass in $L(v)$ for each vertex~$v$ in the hourglass.
After we have done this for each edge, we have a collection of maximal hourglasses $H_1,\ldots, H_{\ell'}$ by \cref{lem:max-hourglass-computed-lintime}.
Since no two hourglasses of height at least four overlap by \cref{lem:HG-intersection}, each hourglass in $G(\ppp)$ occurs among the hourglasses we have computed, but some may occur multiple times.
To remove duplicates we proceed as follows.
First, since we have removed the agents of hourglasses of height at least four, no such hourglass occurs twice.
We iterate over all agents in $V$ and for each $v \in V$ we check all pairs of hourglasses $L(v)$ and if they contain the same vertices, we keep only the first one.

Let us now tally the running time of computing the collection of hourglasses.
When iterating over the edges of $G(\ppp)$, we can compute a maximal hourglass containing the edge in question in $O(h)$ time where $h$ is the height of the hourglass by \cref{lem:max-hourglass-computed-lintime} (or in $O(1)$ time determine that there is no hourglass containing the edge).
Since the vertices of hourglasses of height at least four are removed, the time spent in loop iterations for such hourglasses is $O(n)$.
Since the other hourglasses have constant size, they are discovered a constant number of times, each of which takes constant time, and hence the time spent in loop iterations for hourglasses of size at most three is $O(n)$ as well.
The deduplication step takes $O(n)$ time because the length of $L(v)$ is constant for each vertex $v$: Each hourglass occurring in $L(v)$ is contained in the set $N_6(v)$ of vertices that have distance at most 6 from $v$.
Since the maximum degree is constant, $|N_6(v)|$ is constant, and the number of edges in $N_6(v)$ is constant.
Hence, each hourglass in $L(v)$ is discovered at most a constant number of times.
Thus the deduplication step takes constant time for each vertex and $O(n)$ time overall.
After the deduplication step, the number of maximal hourglasses we have found is~$\ell$.

We now compute an auxiliary graph that contains all maximal hourglasses of height two or three and an edge between two hourglasses if they overlap.
Note that this auxiliary graph can be computed in $O(n)$ time by using the lists~$L(v)$.
We group the hourglasses into collections $C_1, \ldots, C_{\ell'}$ given by the connected components of the auxiliary graph.
Note that, by \cref{lem:HG-intersection} each collection $C_i$ contains either only hourglasses of height two or only of height three.

Next, we iterate over all possible sets of the following combinations:
\begin{enumerate}
\item Each combination of a collection $C_i$ of hourglasses of height two and one of the three possible matchings that match all vertices in $C_i$ (see \cref{lem:overlapping-hourglass-height-two}) if that matching is \es\ for $C_i$.
\item Each combination of a collection $C_i$ of hourglasses of height three and one of the three possible vertex sets $S$ that are matched to vertices in $C_i$ by a matching $M$ that matches all vertices of $C_i$ (see \cref{lem:overlapping-hourglass-height-three}) if that matching is \es\ for $C_i$. (Note that this can be checked in constant time because there is a constant number of matchings that match all vertices of $C_i$ by \cref{lem:overlapping-hourglass-height-three}.)
\item Each combination of an hourglass $H_i$ of height at least four and the six categories described in~\cref{lem:6_ways_hourglass} if there is a matching of that category that matches all vertices of $H_i$ and is \es\ for $H_i$. (Whether such a matching exists can be checked in overall $O(n)$ time by \cref{lem:md3_linearTime}.)
\end{enumerate}
For each combination, we check if any agent outside an hourglass is assigned more than one partner (note that the partners are in each combination uniquely determined).
If this is the case then we go to the next combination.
Otherwise, we delete all maximal hourglasses and all agents already matched by this combination and compute a perfect matching on the remaining instance.
If such a perfect matching exists then return the corresponding perfect matching, otherwise try the next combination.
Should no combination result in a perfect matching, return \KwNo.

Let us investigate the running time of this algorithm.
By \cref{lem:overlapping-hourglass-height-two,lem:overlapping-hourglass-height-three,lem:6_ways_hourglass} there are in total $6^{\ell}$ possible combinations, so checking the hourglass-category combinations takes overall $O(6^\ell \cdot n)$ time.
Using a maximum-cardinality matching algorithm (see e.g., Hopcroft and Karp~\cite{HopcroftKarp1973}, running time proven by Micali and Vazirani~\cite{MicaliVazirani1980}) we check for a perfect matching in $O(n\cdot \sqrt{n})$ time (recall that the acceptability graph has~$O(n)$ edges).
In total this gives us a running time of $O(6^\ell \cdot(n + n\cdot \sqrt{n})) = O(6^\ell \cdot(n + n\cdot \sqrt{n}))$ which, if correct, shows that the problem is fixed-parameter tractable with respect to the number of maximal hourglasses.

It remains to show that the algorithm is correct.
If $\ppp$ admits a perfect and \es{} matching~$M$, then~$M$ is \es{} for each maximal hourglass~$H_i$, $i\in [\ell]$, such that each agent in~$V(H_i)$ is matched and analogously for each collection $C_i$.
By \cref{lem:HG-intersection} two hourglasses overlap only if both have height two or both have height three.
Thus, matching $M$ decomposes into disjoint matchings that match all agents in (1) each hourglass of height at least four, (2) each collection $C_i$ of hourglasses of height two, (3) each collection $C_i$ of hourglasses of height three, and (4) remaining agents that are not matched by the matchings corresponding to (1), (2), or (3).
By \cref{lem:6_ways_hourglass}, for each maximal hourglass~$H_i$ of height at least four, $M$ corresponds to one of the (up to) six possible categories.
And by \cref{lem:md3_linearTime} we will correctly obtain a matching of the right category in one of the combinations.
By \cref{lem:overlapping-hourglass-height-two} we correctly determine a matching for collections $C_i$ hourglasses of height two in one of the combinations..
By \cref{lem:overlapping-hourglass-height-three} we correctly determine a matching for collections $C_i$ of hourglasses of height three in one of the combinations..
Finally, the Hopcroft-Karp algorithm will find a perfect matching for the remaining agents.
Thus, in one of the combinations we have obtained a perfect matching.
By \cref{sc:md3_ebphg} and since we have checked for \esty\ by direct checks for collections $C_i$ and via \cref{lem:md3_linearTime} for hourglasses of height at least four, the found perfect matching is \es{} for~$\ppp$.

Hence, we have proved \cref{thm:fpt-hourglasses}.

\fi

\section{Paths to Exchange-Stability} \label{sec:local}
We now study the parameterized complexity of \LESMs{} with respect to the number of swaps.
Observe that it is straightforward to decide an instance of \LESMs\ with $2n$ agents in $O((2n)^{2k + 2})$ time by trying $k$ times all of the $O(n^2)$ possibilities for the next swap and then checking whether the resulting matching is \es.
The next theorem shows that the dependency of the exponent in the running time cannot be removed unless
\iflong
the unlikely collapse FPT${}={}$W[1] occurs.
\fi

\begin{theorem}
  \label{thm:swps}
  \LESM is W[1]-hard with respect to the number~$k$ of swaps.
\end{theorem}

\iflong \begin{proof} 
  \else \begin{proof}[Sketch]\fi
We provide a parameterized reduction from the W[1]-complete \IS{} problem, parameterized by the size of the independent set~\cite{CyFoKoLoMaPiPiSa2015}.
Therein, we are given a graph~$H$ and an integer~$h$ and want to decide whether there is an $h$-vertex \emph{independent} set, i.e., a subset of $h$ pairwise nonadjacent vertices.

\looseness=-1
Let $I=(H, h)$ be an instance of \IS\ with vertex set $V(H)=\{v_1,v_2,\ldots,v_n\}$ and edge set $E(H)$. %
We construct an instance~$I'=(\ppp, M_0, 2h)$ of \LESMs\ where $\ppp$ has two disjoint agent sets~$U$ and $W$, each of size~$2n+h$.
\iflong

\fi
Both~$U$ and $W$ consist of $h$ \myemph{selector-agents} and $2n$ \myemph{vertex-agents} with preferences which encode the adjacency of the vertices in~$V(H)$. 
More precisely, for each~$j\in [h]$, we create two selector-agents, called~$s_j$ and $t_j$,
and add them to~$U$ and $W$, respectively.
For each~$i\in [n]$, we create four vertex-agents, called~$x_i,u_i, y_i,w_i$,
 add $x_i$ and $u_i$ to~$U$,
 and add $y_i$ and $w_i$ to~$W$.
 Altogether, we have $U=\{s_j\mid j\in [h]\}\cup \{u_i,x_i\mid i \in [n]\}$
 and $W=\{t_j\mid j\in [h] \}\cup \{w_i,y_i \mid i\in [n]\}$.

\iflong
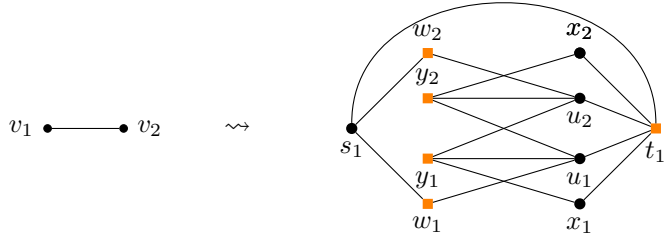
\begin{figure}[t!]
  \centering
  \begin{tikzpicture}		
    \node[draw, circle, black, fill=black, inner sep=1pt] (v1) at (-3,0) {};		
    \node[draw, circle, black, fill=black, inner sep=1pt] (v2) at (-2,0) {};
    \node[left = 0pt of v1] {$v_1$};		
    \node[right = 0pt of v2] {$v_2$};
    \draw (v1) to (v2);
    
    \node at (-.5,0) {$\leadsto$};
    
    \foreach \x / \y / \n / \nn / \a / \p in {
      1/0/si/{s_1}/agentU/below, 5/0/ti/{t_1}/agentW/below,
      2/-1/wj/{w_1}/agentW/below, %
      2/1/w2/{w_2}/agentW/above,
      2/-0.4/yj/{y_1}/agentW/below, %
      2/0.4/y2/{y_2}/agentW/above, %
      4/-1/xj/{x_1}/agentU/below, %
      4/1/x2/{x_2}/agentU/above, %
      4/1/x2/{x_2}/agentU/above, %
      4/-0.4/uj/{u_1}/agentU/below, %
      4/0.4/u2/{u_2}/agentU/below%
    } {
      \node[\a] (\n) at (\x,\y) {};
      \node[\p = -0.5pt of \n] {$\nn$};
    }

    \draw (si) to (wj);
    \draw (uj) to (wj);
    \draw (yj) to (xj);
    \draw (yj) to (uj);
    \draw (ti) to (xj);
    \draw (ti) to (uj);

    \draw (si) to (w2);
    \draw (u2) to (w2);
    \draw (y2) to (x2);
    \draw (y2) to (u2);
    \draw (ti) to (x2);
    \draw (ti) to (u2);
    
    \draw (yj) to (u2);
    \draw (y2) to (uj);

    \path[draw]  (si) .. controls ($(w2)+(-1,1.2)$) and ($(x2)+(1,1.2)$) .. (ti);
  \end{tikzpicture}
  \caption{The left part shows graph $H=(\{v_1,v_2\},\{\{v_1,v_2\}\})$. Then $(H,1)$ is an instance of \IS. We reduce this instance to the \LESMsnonfragile\ instance of which the acceptability graph is shown on the right.}
  \label{fig:localsearch0}
\end{figure}
\fi

Now, we define the preferences of the agents from $U\cup W$.
For notational convenience, we define two subsets of agents which shall encode the neighborhood of a vertex: For each vertex~$v_i\in V(H)$,
define $Y(v_i)\coloneqq \{y_z \mid \{v_i,v_z\}\in E(H)\}$ and
$U(v_i)\coloneqq \{u_z \mid \{v_i,v_z\}\in E(H)\}$.
{
  \begin{center}
\begin{tabular}{lll}
    $\forall j \in [h]\colon$ & $s_j : w_1 \succ \dots \succ w_n \succ t_j$, \quad $t_j : u_1 \succ \dots \succ u_n \succ x_1  \succ \dots \succ x_n \succ s_j$,\\
    $\forall i \in [n]\colon$ & $x_i : t_1 \succ \dots \succ t_h \succ y_i$, \qquad \quad \qquad \qquad \qquad $y_i: u_i \succ x_i  \succ  [U(v_i)]$,	\\
   $\forall i \in [n]\colon$  & $u_i : w_i \succ [Y(v_i)] \succ y_i \succ t_1 \succ  \dots \succ t_h$,  \qquad $w_i : s_1 \succ \dots \succ s_h \succ u_i$ \text{.} \\
\end{tabular}
\end{center}
\par}

\iflong
Figure~\ref{fig:localsearch0} illustrates the underlying acceptability graph for a graph consisting of a single edge. %
\fi
\looseness=-1
Herein, $[Y(v_i)]$ (resp.\ $[U(v_i)]$) denotes the unique preference list where the agents in~$Y(v_i)$ (resp.\ $U(v_i)$) are ordered ascendingly according to their indices.

\noindent Observe that the acceptability graph $G(\ppp)$ includes the following edges:
\begin{compactitem}[--]
  \item For all $i \in \intervalI{1}{h}$ and $j \in \intervalI{1}{n}$, the edges~$\{s_i, t_i\}$, $\{s_i, w_j\}$, $\{t_i, x_j\}$, $\{t_i, u_j\}$, $\{w_j, u_j\}$, $\{y_j,x_j\}$, $\{y_j, u_j\}$ are in $E(G(\ppp))$.
  \item For all edges~$\{v_{i}, v_{i'}\}\in E(G)$,
  the edges~$\{u_i, y_{i'}\}$ and $\{u_{i'}, y_{i}\}$ are in~$E(G(\ppp))$.
\end{compactitem}

We define an initial matching $M_0$ on $G(\ppp)$ as $M_0 = \set{\{s_j, t_j\} \mid j \in \intervalI{1}{h}}$ $\cup \set{\{w_i, u_i\}, \{y_i, x_i\} \mid i \in \intervalI{1}{n}}$.
This completes the construction of~$I'$, which can clearly be done in polynomial time.
It is straight-forward to check that that~$\ppp$ is bipartite and the construction can be done in linear time.
\ifshort
The correctness proof is given in the appendix.
\else

Next, we show the correctness of the reduction, i.e.,
$I$ is a positive instance of \IS\ if and only if $I'$ is a positive instance of \LESMs. 
For notational convenience, define~$S=\{s_1,\ldots,s_h\}$,
$T=\{t_1,\ldots,t_h\}$, $X=\{x_1,\ldots, x_n\}$, $X=\{x_1,\ldots,x_n\}$, and $Y=\{y_1,\ldots, y_n\}$:

For the forward direction, assume that $(H,h)$ is a yes-instance of~\IS.
We need to show that we can reach an \es{} matching for the constructed instance of \LESMs\ in $2h$ swaps from $M_0$.
Let~$V'=\{v_{j_1}, \dots, v_{j_h}\}$ denote an $h$-vertex independent set.
For each~$z \in \intervalI{1}{h}$,
one can verify that $t_z$ and $w_{j_z}$ envy each other, implying that $\pair{t_z}{w_{j_z}}$ is \ebp{} of~$M_0$. 
Therefore, we swap their partners and eventually obtain $M_h$ with
$M_h(t_z) = u_{j_z}$ and $M_h(w_{j_z}) = s_z$ for all~$z\in [h]$.
The partners of the remaining agents remain unchanged, i.e., they are the same as in~$M_0$.
Now, for each~$z \in \intervalI{1}{h}$,
$\pair{x_{j_z}}{u_{j_z}}$ is an exchange-blocking pair of~$M_h$,
since $x_{j_z}$ prefers~$t_z$ to his partner~$y_{j_z}$
whereas $u_{j_z}$ prefers~$y_{j_z}$ to his partner~$t_z$.
We swap their partners and obtain the matching $M_{2h}$ where $M_{2h}(x_{j_z}) = t_z$ and~$M_{2h}(u_{j_z}) = y_{j_z}$ for all~$z\in [k]$.
The partners of the remaining agents remain unchanged, i.e., are the same as in~$M_h$.
It remains to be shown that $M_{2h}$ is \es, i.e., no agent is involved in an exchange-blocking pair of~$M_{2h}$:

\begin{itemize}[--]
  \item Each selector-agent~$s_z\in S$ is matched to~$w_{j_z}$. %
  Since all selector agents from~$S$ have the same preferences over~$\{w_i\mid i\in [n]\}$,
  no two agents from~$S$ can form an \ebp{}.
  Neither can an agent from~$S$ and an agent from~$\{u_i \mid i\in [n]\}$ jointly form an \ebp since~$u_i$ is either matched with its most-preferred agent or matched with~$y_i$ who is not acceptable to~$s_j$.
  Finally, no agent from~$S$ can form with an agent from~$X$ an \ebp{} since every agent from~$S$ either prefers its own partner or does not find the partner of the agents from~$X$ acceptable.

  \item Each vertex-agent~$x_i\in X$ is either matched to some selector-agent~$t_j$ or to~$y_i$.
  If $x_i$ is matched to some $t_j$.
  Then $x_i$ prefers all selector-agents~$t_{z}$ with $z < j$.
  Notice that $M_{2h}(t_{z})=x_{i'}$ for some vertex-agent~$x_{i'}\in X\setminus \{x_i\}$ other than~$x_i$.
  Since all vertex-agents from~$X$ have the same preferences over all selector-agents from~$T$,
  $x_{i'}$ does not envy~$x_i$, and will not form with~$x_i$ an \ebp{} of~$M_{2h}$.
  
  If $x_i$ is matched to~$y_i$, then $x_i$ envies every other agent~$x_{i'}\in X\setminus \{x_i\}$ which is matched to some selector-agent from~$T$.
  However, no agent from~$X\setminus \{x_i\}$ finds~$y_i$ acceptable
  and hence, will not form with~$x_i$ an \ebp{} of~$M_{2h}$.  
  
\item For each $i\in [n]$, vertex-agent~$u_i$ is either matched to~$w_i$ or to~$y_i$. 
  In the former case, $u_i$ does not prefer anyone over its current partner. 
  In the latter case, we know that $w_i$ is matched to some selector-agent~$s_z$.
  Agent~$u_i$ envies $s_z$, however $s_z$ is not involved in any \ebp{} (see above). 
  
  Agent~$u_i$ also prefers any~$y_{j} \in Y(v_i)$ to his partner~$y_i$.
  Suppose, for the sake of contradiction, that $u_i$ and $M_{2h}(y_j)$ form an \ebp of~$M_{2h}$ with $y_j\in Y(v_i)$.
  By our construction, either~$M(y_j)= x_j$ or $M(y_j)=u_j$ holds.
  In the former case, $x_j$ does not find~$y_i$ acceptable and will not form with~$u_i$ an \ebp{}, a contradiction.
  In the latter case, we infer that $v_j\in V'$, implying that $v_j$ and $v_i$ are non-adjacent, a contradiction too.

  \item Each vertex-agent~$y_i\in Y$ is either matched to~$u_i$ or to $x_i$.
  In the former case, $u_i$ is most acceptable to agent~$y_i$.
  Therefore, $y_i$ does not envy any other agent.	
  In the latter case, $y_i$ prefers only $u_i$ to~$x_i$.
  We know that $u_i$ must be matched to $w_i$.
  However, $w_i$ does not find $x_i$ acceptable.
  Therefore, $y_i$ is not involved in any exchange-blocking pair.

\item Each selector-agent~$t_z\in T$ is matched to some vertex-agent~$x_i$.
  Since all selector agents from~$T$ have the same preferences over~$X$,
  no two selector-agents will form an \ebp{} of~$M_{2h}$.
  Neither will any agent~$y_{i'}$ from~$Y$ form an \ebp{} with~$t_z$ by the previous reasoning.
  Neither will any agent~$w_i$ form an \ebp with~$t_z$ since he does not find~$x_i$ acceptable.

  \item  For each $i\in [n]$, vertex-agent~$w_i$ is either matched to some selector-agent~$s_z\in S$ or to~$u_i$.
  In the former case, $w_i$ envies~$M_{2h}(s_j)$ for all~$j<z$.
  However, for each~$j\in [h]\setminus \{z\}$,
  we know that the partner of~$s_j$ under $M_{2h}$ is a vertex-agent~$w_{i'}$ for some~$i'\in [n]\setminus \{i\}$ who also prefers~$s_j$ to~$s_z$,
  and will not form with~$w_i$ an \ebp.

  In the latter case, agent~$w_i$ prefers all agents from~$S$ to his partner~$u_i$.
  However, no partner of the agents from~$S$ finds~$u_i$ acceptable,
  and will not form with~$w_i$ an \ebp.
\end{itemize}

For the backwards direction, assume that $(\ppp, M_0, 2h)$ is a positive instance, meaning that there exists an \es{} matching $M_{\ell}$ that can be reached in at most $2h$ swaps from $M_0$ (and~$\ell \leq 2h$).
We show that the set $V' = \{v_i \mid M(w_i)\in S\}$ is an independent set of size~$h$, implying that $(H, h)$ is a positive instance of \IS.

First of all, we observe that $M_\ell$ is a perfect matching since it is reachable from~$M_{0}$ which is perfect.
\begin{claim}
  \label{loc:lem_sw}
  Each selector-agent~$s_j\in S$ has $M_{\ell}(s_j)\in W$.
\end{claim}

\begin{proof}  \renewcommand{\qedsymbol}{(of
    \cref{loc:lem_sw})~$\diamond$}
 
  Towards a contradiction,
  suppose that there is a selector-agent~$s_j$ with $M_{\ell}(s_j) = t_j$.
  Without loss of generality, assume that $h < n$ (as otherwise the instance is solvable in polynomial-time).
  This means there remains a vertex-agent~$w_i$ with $M(w_i)=u_i$.
  It is strait-forward to see that $w_i$ and $t_j$ form a \ebp, a contradiction.
\end{proof}

\begin{claim}
  \label{loc:lem_uy}
  No vertex-agent~$y_i\in Y$ is matched to some agent from~$U(v_i)$ in~$M_{\ell}$.
\end{claim}
\renewcommand{\qedsymbol}{(of
       \cref{loc:lem_uy})~$\diamond$}
\begin{proof}
  Suppose, for the sake of contradiction, that $M_{\ell}(y_i)\in U(v_i)$.
  Then, by the preferences of~$x_i$ we infer that $M_{\ell}(x_i)=t_j$ with $t_j\in T$ (since $M_{\ell}$ is perfect).
  However, we also infer that $t_i$ and $y_i$ envy each other, a contradiction.
\end{proof}

\begin{claim}
  \label{loc:lem_tu}
  No vertex-agent~$u_i$ has $M_\ell(u_i)\in T$.
\end{claim}

\begin{proof} \renewcommand{\qedsymbol}{(of
    \cref{loc:lem_tu})~$\diamond$}
  Suppose, for the sake of contradiction, that $M_{\ell}(u_i)\in T$.
  By \cref{loc:lem_uy}, we know that $M_{\ell}(y_i)\in U(v_i)$.
  Hence, $M_{\ell}(y_i)=x_i$.
  Now, $x_i$ and $u_i$ envy each other, a contradiction.
\end{proof}

\begin{claim}
  \label{loc:lem_selected}
  For each~$i\in [n]$, if $M_{\ell}(w_{i})\in S$, then $M_\ell(y_i) = u_i$.
\end{claim}

\begin{proof} \renewcommand{\qedsymbol}{(of
    \cref{loc:lem_selected})~$\diamond$}
  If $M_\ell(w_i) \in S$,
  then $u_i$ must be matched to $y_i$ since it cannot be matched to any agent from~$T$ (Claim~\ref{loc:lem_tu}) nor to any agent from~$Y(u_i)$ (Claim~\ref{loc:lem_uy}).
\end{proof}

\begin{claim}
  \label{loc:lem_exactlyk}
  $V'$ contains exactly $h$ vertices.
\end{claim}

\begin{proof} \renewcommand{\qedsymbol}{(of
       \cref{loc:lem_exactlyk})~$\diamond$}
     We know that every selector-agent~$s_j\in S$ is matched to some vertex-agent~$w_i$ (Claim~\ref{loc:lem_sw}).
     A vertex $v_i$ is selected into $V'$ if $w_i$~is matched to some $s_j$.
     Since $|S|=h$, we will select exactly $h$ vertices into $V'$.
\end{proof}

\begin{claim}
  \label{loc:lem_noNeigbors}
  No two vertices in $V'$ are adjacent in~$G$.
\end{claim}

\begin{proof} \renewcommand{\qedsymbol}{(of
       \cref{loc:lem_noNeigbors})~$\diamond$}
     Suppose we have selected two adjacent vertices $v_i, v_{i'}$ with $\{v_i,v_{i'}\}\in $.
     This means that $M_{\ell}(w_i),M_{\ell}(w_{i'})\in S$.
     By \cref{loc:lem_selected},
     we infer that  $M_{\ell}(u_i)=y_i$ and $M_{\ell}(u_{i'}) = y_{i'}$.
     Since $v_i$ and $v_{i'}$ are adjacent in~$G$,
     it follows $y_i\in Y(u_i)$ and $y_{i'}\in Y(u_{i'})$,
     meaning that $u_i$ and $u_{i'}$ envy each other, a contradiction.
   \end{proof}

   \cref{loc:lem_exactlyk,loc:lem_noNeigbors} immediately imply that $V'$ is an $h$-vertex independent set.
   Since the reduction is correct, it is a parameterized reduction.
\fi%
\end{proof}

\iflong
 \section{Conclusion}\label{sec:conclude}

 We conclude by mentioning that all obtained results transfer to the case when the preferences may contain ties (i.e., two agents are considered equally good).
 
 Regarding preference restrictions~\cite{BreCheFinNie2020-spscSM-jaamas}, it would be interesting to know whether deciding (coalitional) \esty for complete preferences would be become tractable for restricted preferences domains, such as single-peakedness or single-crossingness.

 Moreover, as it is unclear whether the number of exchange steps is polynomially bounded,
 the NP-containment of the problem of checking whether a given matching may reach an \es{} matching is open.

\paragraph{Acknowledgments.}
JC was supported by the WWTF, research project~(VRG18-012).
MS was supported by the Alexander von Humboldt Foundation.
 \fi

\bibliographystyle{abbrvnat}

\bibliography{bib.bib}

\end{document}